\documentclass[a4paper,11pt]{article}
\pdfoutput=1 

\usepackage{jcappub} 

\usepackage[T1]{fontenc} 

\usepackage{physics}
\usepackage{slashed}
\usepackage{subcaption}
\usepackage{float}
\usepackage[usenames,dvipsnames,svgnames]{xcolor}
\usepackage{multirow}
\usepackage{bm}
\usepackage[shortlabels,inline]{enumitem}
\usepackage[capitalise]{cleveref}
\PassOptionsToPackage{demo}{graphicx}
\newcommand*{\rom}[1]{\expandafter\ \romannumeral #1}
\usepackage{tikz-feynman}
\usepackage[left=2cm,right=2cm,top=2cm,bottom=2cm]{geometry}
\usepackage{blindtext}
\usepackage{enumitem}

\usepackage{framed}
\colorlet{shadecolor}{blue!20}
\usepackage[normalem]{ulem}

\makeatletter
\def\@fpheader{\relax}
\makeatother

\title{\boldmath Leptogenesis from a feebly interacting dark matter sector }

\author[a]{Suresh Chand$^{1,}$}
\author[b]{Mariana Frank$^{2,}$}
\author[a]{Poulose Poulose$^{3,}$}
\affiliation[a]{Department of Physics, Indian Institute of Technology Guwahati, Assam 781039, India}
\affiliation[b]{Department of Physics, Concordia University, 7141 Sherbrooke St. West, Montreal, Quebec, Canada H4B
1R6}

\emailAdd{$^1$sures176121102@iitg.ac.in}
\emailAdd{$^2$mariana.frank@concordia.ca}
\emailAdd{$^3$poulose@iitg.ac.in}

\abstract{We perform an analysis of leptogenesis in the context of a simple extension of the Standard Model by two fermions; one charged ($\chi $) and one neutral ($\psi$), in addition to three right-handed neutrinos, $N_i$, interacting through a charged  gauge singlet scalar $S$. The dark sector ($\chi$, $\psi$ and $S$) interacts feebly and produces a relic density consistent with measurements. The decay of right-handed neutrinos into the charged scalar $S$ and lepton provides an additional source  of  CP asymmetry, along with contributing through the virtual exchange of $S$ in the standard decay channel. With this the out-of-equilibrium decay of  right-handed neutrinos, combined with lepton number changing scattering processes
can generate the required baryon asymmetry of the universe even for right-handed neutrino masses in 10 TeV region, without  requiring neutrinos to have degenerate masses.}

\begin{document} 
\maketitle
\flushbottom

%
\section{Introduction}
\label{sec:intro}
The Standard Model (SM) of particle physics encapsulates our knowledge so far of the fundamental constituents of matter and their interactions. However, it suffers from theoretical inconsistencies and it is unable to explain some of the phenomena observed in the experimental data.
Among others, the SM is incompatible with the strong CP problem
, neutrino oscillations \cite{Casas:2001sr}, matter-antimatter asymmetry \cite{Cohen:1993nk,Cohen:1994ss}, and it is unable to provide a suitable candidate, or  explain the nature of dark matter   \cite{Akerib:2016vxi,Monteiro:2020wcb} and dark energy \cite{Aghanim:2018eyx}. The SM also does not incorporate gravity and is thus inconsistent with  general relativity. 

In particular, the SM lacks an explanation for the baryon asymmetry in nature, that is, the observation that the visible universe is made of baryonic matter and not antimatter. This matter antimatter asymmetry of the universe is  expressed in terms of the ratio of the difference between baryon density $n_B$ and antibaryon density to the number of photons in the universe. From the nucleosynthesis constraints and also from Cosmic Microwave Background (CMB) \cite{Planck:2019nip}, this ratio is determined to be \cite{Burles:1999ac}
\begin{equation}
\eta= \frac {n_B-n_{\bar B}}{n_\gamma}= \left \{  \begin{array} {cr} (5.8-6.6) \times 10^{-10}, & {\rm BBN} \\ (6.09 \pm 0.06) \times 10^{-10}, & {\rm CMB}\end{array} \right.
\label{eq:eta1}
\end{equation}

Sakharov \cite{Sakharov:1967dj} argued that the particle antiparticle asymmetry can arise dynamically from a charge symmetric or even from an arbitrary initial state. This mechanism required satisfying three physical conditions, under which   any physical process may generate the matter antimatter asymmetry, known as baryogenesis: baryon number (B)  violation,  Charge conjugation and Parity violation (CP), and departure from thermal equilibrium. In most particle physics models, the third condition of out-of-equilibrium is provided by the first-order phase transition of the electroweak symmetry breaking. In particular, in the SM, such first-order phase transition requires the Higgs boson to be lighter than its observed mass. On the other hand, many extension of the SM could accommodate this in natural way.  
A new mechanism for producing matter-antimatter asymmetry  was proposed in \cite{Fukugita:1986hr}, where the authors realized that the lepton number violating processes can lead the cosmological baryon asymmetry. The lepton number excess in the early stages of the Universe can be efficiently transformed into the baryon number excess through the so-called sphaleron processes.  Through a lucky coincidence, most neutrino-mass models require the presence of a heavy Majorana neutrino, the decay of which violates lepton number. Necessary CP violation is generated through the quantum corrections of the decay process, and the decay process goes out of equilibrium at temperature or the order of the mass the decaying particle.

Coming to the  dark matter (DM) problem \cite{Walker:1991ap,Fields:2019pfx,McGaugh:2009mt,Cyburt:2001pq}, while evidence for dark matter is well established through a variety of cosmological and astrophysical observations, the nature of dark matter remains to be understood.
The primary candidate for DM is a new kind of elementary particle. The most popular choice are weakly interacting massive particles (WIMPs), which can explain the observed value of the DM relic density by a mechanism  called freeze-out \cite{Bauer:2017qwy}. 
The assumption underlying thermal freeze-out is that the DM particle is a WIMP that was once in thermodynamic equilibrium with the hot plasma of SM particles  created after inflation. During this period, the universe was so hot that the DM particle was
highly relativistic. As the universe expanded and cooled down below the DM particle mass, the WIMP became
non-relativistic, and its abundance started to decrease due to DM self-annihilation 
into SM particles. This continues till the temperature reaches so that the reaction rate is smaller than the Hubble expansion rate, and from then on the relic density remains the same.  In such WIPM scenarios, the  required relic density is achieved by adjusting the dynamics through the coupling constant and the mass of the dark matter. 
However, extensive analyses of models with WIMPs ran into difficulties in trying to satisfy both relic density constraints and constraints from direct detection of dark matter experiments \cite{Billard:2021uyg}. In direct detection, WIMP DM particles could scatter off an atomic nucleus, with observable signal resulting from the recoil of the nucleus. In simple WIMP set ups, this scattering is also controlled by the same coupling and mass parameters that enter the processes controlling the relic density. To avoid WIMP difficulties, other possibilities like the feebly
interacting massive particles (FIMPs) are proposed. Unlike the WIMP, couplings of FIMP are too weak to have produced in abundance in the early Universe to be in thermal equilibrium. Rather, they are slowly produced, in most viable scenarios, through the decay of a partner particle, which itself is in thermal equilibrium to start with. The relic density slowly gets saturated to the presently observed value, in this so-called freeze-in mechanism \cite{Bernal:2017kxu}.  The advantage in this scenario is that the couplings are too weak to be of any significance in the direct detection experiments, thus evading the limits arising from those.

In this work we  construct a simple model where required dark matter relic density is generated by the freeze-in mechanisms. We do so by a simple extension of the SM, to include a dark sector. The model also includes a Majorana neutrino, thus presenting possibility of leptogenesis, along with addressing the neutrino mass generation through Type-I seesaw mechanism. We show that, unlike most other models, this scenario can achieve successful  leptogenesis with TeV scale Majorana neutrinos, without any fine-tuning as required in the resonant leptogenesis. A comprehensive study of freeze-in mechanism was presented in \cite{Belanger:2018sti,Goudelis:2021lra,Klasen:2013ypa,Hall:2009bx,Hall:2010jx,Elahi:2014fsa,Co:2015pka}, and related mechanisms for leptogenesis analysed in  \cite{Davidson:2008bu,Borah:2021qmi,Frigerio:2006gx,Drewes:2012ma,,Fong:2012buy,Falkowski:2011xh,An:2009vq,DeSimone:2007gkc,Pascoli:2006ci,Pascoli:2006ie}.

Our work is organized as follows.   In Section \ref{sec:Model} we begin with the description of our model, and theoretical and experimental constrains. This is followed by a brief discussion of the dark matter scenario with the FIMP mechanism corresponding to this framework in Section \ref{sec:FIMP}.  In Section \ref{sec:leptogenesis},  details of the leptogenesis arising in the proposed scenario are presented,
and  the influence of the new scalar degree of freedom in achieving TeV scale leptogenesis is established.  
 Detailed numerical study and discussion are presented in Section \ref{sec:NumAnalysis}. Finally we summarize our findings and conclude in Section \ref{sec:conclusions}.

\section{A model with FIMP dark matter}
\label{sec:Model}

The standard leptogenesis scenario with three right-handed Majorana neutrinos added to the SM requires these neutrinos to be heavier than $10^6$ GeV \cite{Giudice:2003jh,Buchmuller:2004nz,Buchmuller:2002rq,Davidson:2008bu}, when the masses are hierarchical. The required CP-asymmetry is provided through the quantum effects in the decay of the lightest of the heavy neutrinos to the charged leptons and Higgs boson. The quantum effects include  self-energy corrections for the heavy decaying neutrino, and  vertex corrections for the decay process, both involving the heavy neutrino of a flavour different than the decaying one in the loop. CP violation is introduced through the presence of complex Yukawa couplings, and the CP-asymmetry is proportional to the imaginary part of a combination of these couplings.  On the other hand, if the lightest of the additional neutrinos is degenerate in mass with at least one more heavy neutrino, the standard resonant mechanism can bring in large enough effects even with TeV scale masses for the neutrinos \cite{Pilaftsis:2003gt}.  

Here we consider a novel scenario with heavy neutrinos and with additional  particles added to spectrum \cite{Chakraborti:2019ohe}. In addition to addressing the dark matter problem successfully, this model provides a mechanism to generate leptogenesis with TeV scale heavy non-degenerate heavy neutrinos, enabled through their interaction with the dark sector particles, within the hierarchical mass scenario.
The extension to the SM particle content includes a gauge singlet charged scalar field $S^{+}$, plus a charged $(\chi^+$) and a neutral ($\psi$) singlet fermions. With an additional $Z_2$ symmetry under which both $\chi^+$ and $\psi$ are odd, while all other particles even, $\psi$ is a stable dark matter candidate. The additional particle spectrum along with their hypercharges and $Z_2$ charge are given in Table \ref{tab:tablabel}. 

\begin{table}[th]
	\begin{center}
		\begin{tabular}{ c|c|c|c}
			\hline \hline
			Fields & Spin&Y & $Z_2$ \\ 
			\hline
			$S^{+}$&0&+2&+\\
			\hline 
			$N_1$, $N_2$,$N_3$ &$\frac{1}{2}$&0&+\\[1mm]
			\hline 
			$\chi^{+}$&$\frac{1}{2}$&+2&$-$\\[1mm]
			\hline 
			$\psi$&$\frac{1}{2}$&0&$-$\\[1mm]
			\hline 
			\hline
		\end{tabular}
	\end{center}
	\caption{Additional fields in the model, together with their hypercharges and $Z_2$ charges. All fields are $SU(2)_L$ singlets, $N_i$ ($i=1,~2,~3$) are Majorana fermions, $\chi$, $ \psi$ are vectors-like fermions, and $S$ is a scalar.}
	\label{tab:tablabel}
\end{table}
With the above particle content, the Lagrangian of the model is given by
\begin{eqnarray}
\mathcal{L}_m&=& \mathcal{L}_{SM}+(D_{\mu}S)^{\dagger} D_{\mu}S+\bar{\chi}\, \imath \gamma^{\mu}D_{\mu}\chi+\bar{\psi}\, \imath \gamma^{\mu} \partial_{\mu} \psi +\sum_i \bar{N_i}\, \imath \gamma^{\mu} \partial_{\mu} N_i -m_{\chi} \bar{\chi} \chi -m_{\psi} \bar{\psi} \psi - \sum_{ij} m_{Nij} \bar{N_i} N_j  \nonumber\\
&-&(y_1 \bar{\chi} S \psi+ \sum_{ij} y_{2ij} \bar{N_i} S l_j+\sum_{ij} Y_{Nij} \bar{L}_i \tilde{\phi} N_j +h.c)-(\mu^{2}_{S} S^{\dagger} S+\lambda(S^{\dagger} S)^2+\lambda_1 S^{\dagger} S \phi^{\dagger} \phi ),
\label{eq:Lagrangian}
\end{eqnarray}
where $\phi$ represent the SM Higgs doublet, and $L_i$ and $l_i$ denote the SM left-handed lepton doublet and right-handed charged lepton singlet, respectively.
The summation indices ${i,~j}$ run from 1 to 3, indicating the three flavors of leptons. 
For simplicity, in our analysis we have considered diagonal $m_{Nij}=m_{N_i}\delta_{ij}$, however, this can be extended to a more general case. In the electroweak symmetry broken phase, the vacuum expectation value (VEV) of $\phi$, $v$ generates neutrino masses with the mass matrix 
\[  \left( {\begin{array}{ccc}
	0& \frac{1}{\sqrt{2}}vY_N\\
	\frac{1}{\sqrt 2} vY^{T}_N& m_N
	\end{array} } \right),
\]
where $Y_N$ and $m_N$ are $3 \times 3$ matrices with  $m_N={\rm diag}( m_{N_1},m_{N_2},m_{N_3})$.  Light neutrino masses generated through Type-I seesaw mechanism with  \( m_\nu = Y_N^Tm_N^{-1}Y_Nv^2\) limits $Y_N\sim 10^{-8}$ for $m_N$ in the GeV-TeV range.
The VEV of $\phi$ also contributes to the mass of the charged singlet scalar with
\begin{equation}
m^2_S=\mu^2_S+\frac{\lambda_1 v^2}{2}.
\end{equation}
On the other hand, the masses of the dark vector-like fermions, $\chi$ and $\psi$, arise purely through the parameters in the Lagrangian, $m_{\chi}$ and $m_{\psi}$, respectively.

\section{Fermionic FIMP Dark Matter}
\label{sec:FIMP}

The neutral $Z_2$-odd fermion, $\psi$ is a possible dark matter candidate, interacting with other particles solely through its Yukawa interaction with the  
vector-like gauge singlet charged fermion $ \chi$ and the newly introduced charged scalar $S$. The charged scalar, on the other hand, decays primarily through $S\to N\ell$ channel, which is controlled by the Yukawa coupling $y_2$, by requiring the mass hierarchy, $m_S> m_N$ for on-shell neutrino $N$. This, however, can be relaxed to accommodate $m_S< m_N$ so that $S$ decay goes through an off-shell neutrino $N$, $S\to\ell N^*\to\ell\nu H$. The $\chi$ decay has two possibilities. In the kinematic region with $m_\chi >m_S+m_\psi$, it decays through $\chi\to\psi S$, whereas for $m_\chi < m_S+m_\psi$  it proceeds through, $\chi\to\psi S^*\to\psi N^*\ell\to \psi\ell\nu H$, further requiring $m_\chi > m_H+m_\psi+m_\ell$.  While the 2-body decay is controlled by the Yukawa coupling $y_1$ alone, the other decay is dictated by the combination of $y_1y_2$. 
With the corresponding Yukawa couplings combination sufficiently large, the dark matter fermions can be produced copiously as to be in thermal equilibrium in the early stages of the Universe. Then, through the non-equilibrium processes near temperature $ T\sim m_\psi$,  the annihilation process ($\psi \psi\to SS$) brings down their number density to eventually satisfy the relic abundance at the decoupling.  This needs, however, the mass hierarchy $m_\psi > m_S$.  On the other hand, if the relevant coupling combination is much weaker, the production process $\chi\to\psi+\cdots$ can be sufficiently slow to enable the FIMP mechanism to generate the required dark matter abundance.   

Viability of all the above scenarios are studied in  Ref.  \cite{Chakraborti:2019ohe},  treating the Yukawa couplings $y_2$ as well as $Y_N$ diagonal. We shall relax this to include non-zero off-diagonal couplings in both cases, and explore the parameter space that could accommodate the required relic density. This generalisation is necessary to generate leptogenesis, the study of which is the main focus of this work.
The off-diagonal couplings  induce lepton flavour violation (LFV) through $Y_N$ as well as $y_2$. However, since $Y_N$ is involved in the seesaw mechanism, it is naturally required to be  small. Further, we shall restrict to the case of $y_2\ll 1$ so that it does not induce large LFV. Thus LFV in our model is negligibly small.

\subsection{Freeze-in via two body decay of $\chi$} 
\label{subsec:FIMP2body}	

When the couplings involving the dark matter particle are very small, we may envisage a scenario where the initial number density of dark matter particle is negligible, and the observed abundance of dark matter is produced by the slow decay of the partner particles. In this  case,  the process could occur through the slow decay of $ \chi \rightarrow S + \psi $ if kinematically allowed ($m_\chi>m_S+m_\psi$), and will continue until the Universe cools down to temperature $T<m_\chi$. Below this temperature, due to the Boltzmann suppression ($n_{\chi} \propto \exp (-m_{\chi}/T)$) of  the number density of $\chi$, there is no further addition to the number density of dark matter, $\psi$, leading to a constant co-moving number density. This mechanism of generating the dark matter relic density is known in the literature as the {\em freeze-in mechanism}.  Since the initial DM number density is negligible,  the inverse decay is irrelevant, and the Boltzmann equation satisfied by the dark matter particle is given by
	\begin{align}
		 \frac{dY_{\chi}}{dz} = & \frac{2Y^{eq}_{\chi}}{zH} \frac{K_{1}(z)}{K_{2}(z)} ~\Gamma_{\chi \rightarrow S\psi}
		\label{Boltzmann equation for dark matter}
	\end{align}
This leads to the relic density, \cite{Hall:2009bx} 
	\begin{equation}
		\Omega h^2 =\frac{2.19 \times 10^{27} g_{\chi} }{g^{S}_{*}\sqrt{g^{\rho}_{*}} } \frac{ m_{\psi \Gamma_{\chi \rightarrow S \psi}}}{ m^2_{\chi} }
\label{eq:Omega}
	\end{equation}
With the Lagrangian in Eq.  \ref{eq:Lagrangian}, the decay width in terms of the masses of the particles involved and the coupling $y_1$, becomes

	\begin{equation}
		\Gamma_{\chi \rightarrow S\psi}= \frac{y^2_1}{16\pi m^3_{\chi}} ~\left[(m_{\chi}+m_{\psi})^2-m^2_{S}\right] \left[(m^2_{\chi}-m^2_{S}-m^2_{\psi})^2-4m^2_{S}m^2_{\psi}\right]^{\frac{1}{2}} 
\label{eq:chiwidth12}
	\end{equation}
Putting Eq.~\ref{eq:chiwidth12} back into Eq.~\ref{eq:Omega}, and demanding $\Omega h^2$ to satisfy the observed relic density, the coupling $y_1$ can be constrained for given masses of the dark matter and the partner particles. Notice that, in this scenario, the dark matter sector is decoupled from the neutrino sector and from leptogenesis.
With this scenario, selecting the relic density to satisfy the observed value of $\Omega h^2 = 0.118\pm 0.001$ \cite{Aghanim:2018eyx}, we scanned the parameter space spanned by the masses $m_{\chi}, \, m_{\psi}$ and $m_{S}$ to obtain the value of the coupling, $y_1$, obeying the kinematic restriction for the decay, $m_\chi> m_S+m_\psi$. Focusing on light dark matter, and sub-TeV partner particles, we considered the scan range as given in Table~\ref{tab:tableScan12}.

\begin{figure}[h]
		\begin{center}
			\includegraphics[width=2.2in]{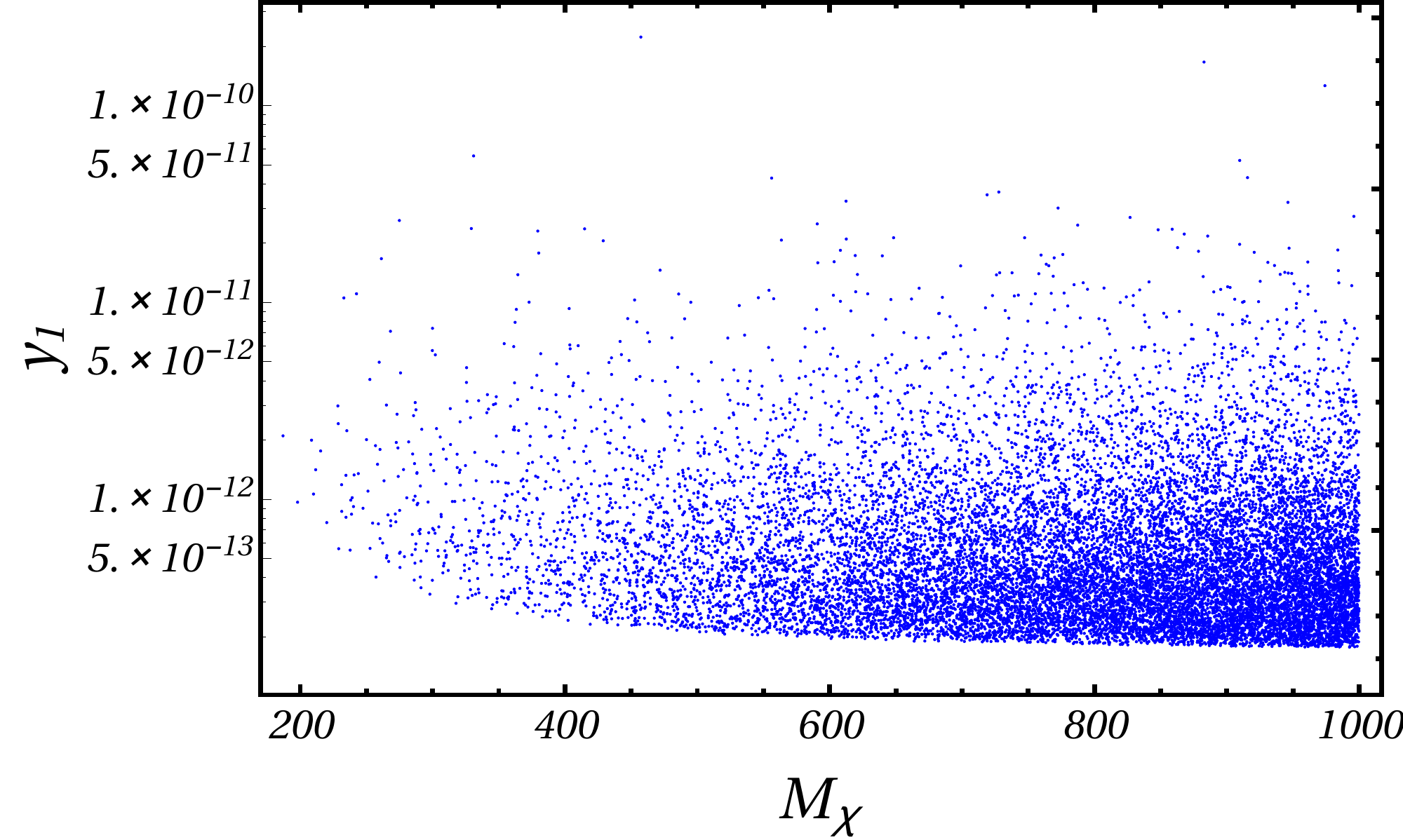}
			\includegraphics[width=2.2in]{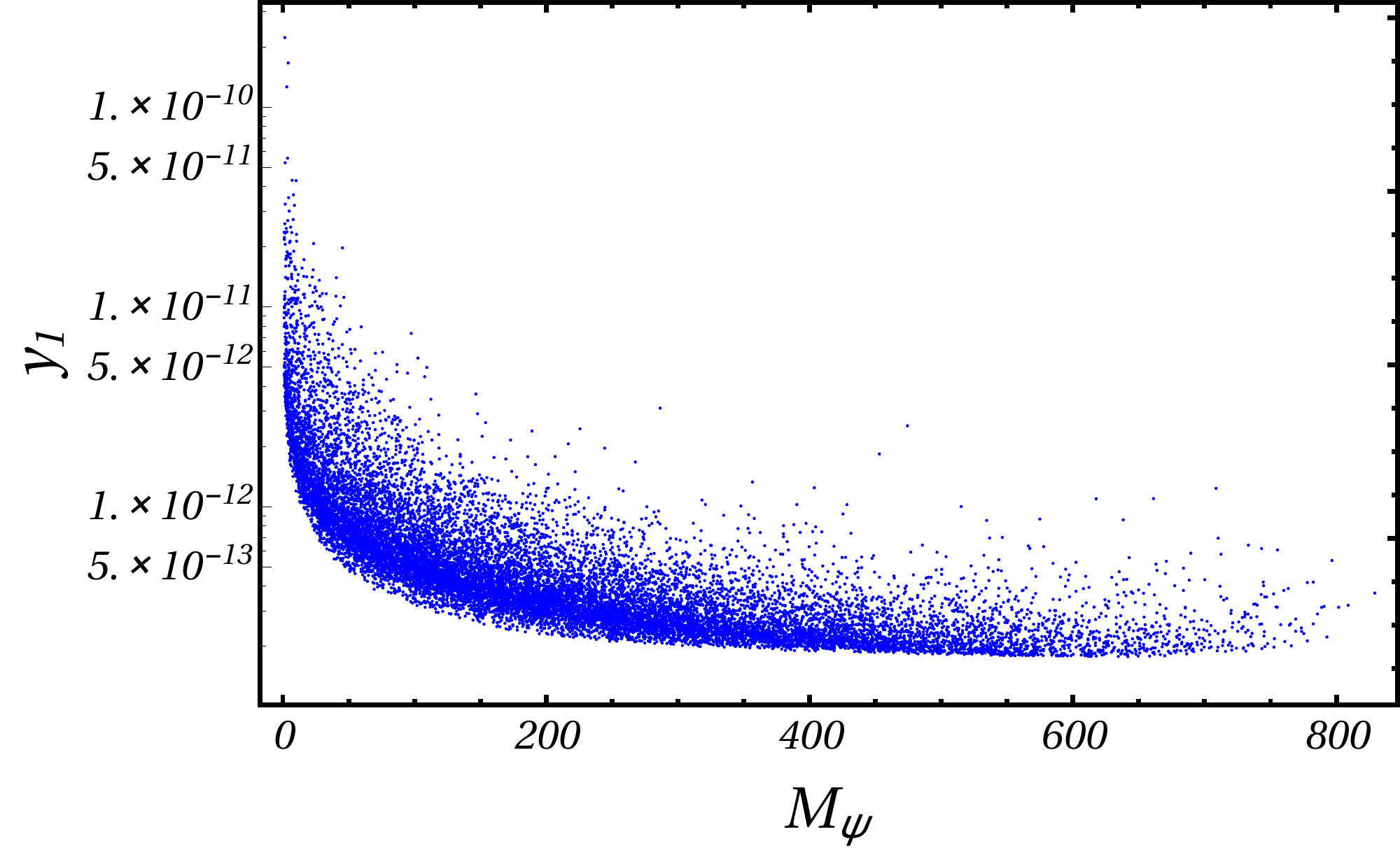}
			\includegraphics[width=2.2in]{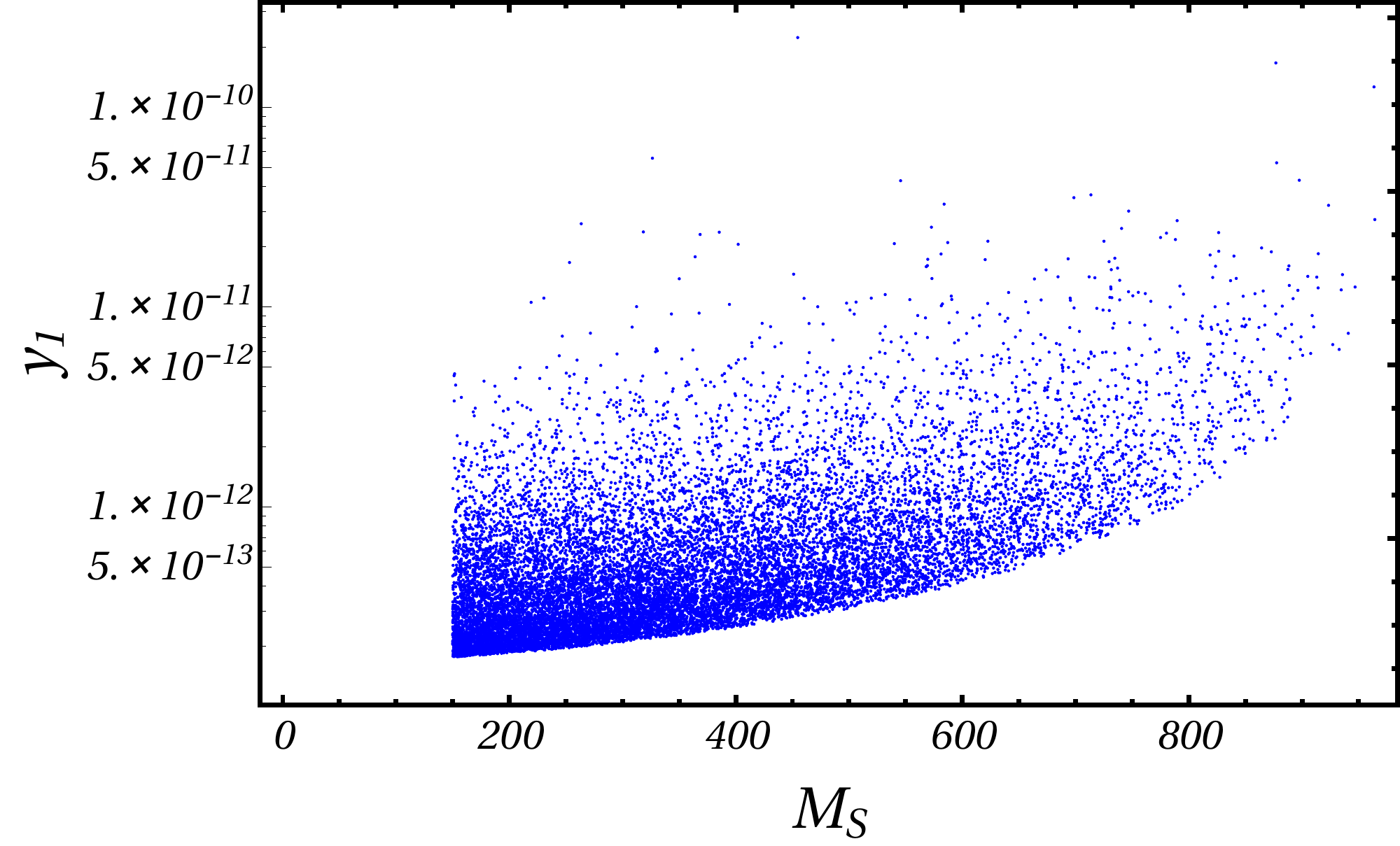}
			\caption{Yukawa coupling $y_1$ as a function of the masses of the dark matter ($m_\psi$), the charged fermionic partner particle ($m_\chi$) and the 
			charged scalar ($m_S$).}
			\label{fig:yvsm12}
		\end{center}
	\end{figure}

\begin{table} [h]
	\begin{center}
		\begin{tabular}{  c  | c }
			\hline\hline
			$m_{\chi}$ &  $150  - 1000$ \\
			$m_{S} $ &  $ 150  - 1000$ \\
			$m_{\psi}$ &  $1-  850 $\\
			\hline\hline
		\end{tabular}
\caption{Scan range of the masses in GeV, used to fix the coupling $y_1$ satisfying the observed relic density.}
\label{tab:tableScan12}
	\end{center}	
\end{table}
Allowed values of the parameter space arising from the scan are presented in different planes in Fig.~\ref{fig:yvsm12}.
	The parameter $m_S$ affects the relic density purely through the phase-space factor, and thus has a clear correlation with the couplings. On the other hand, $m_\chi$ is expected to have an anti-correlation effect, considering the dependence of the coupling and the mass of the decaying particle. This is indeed as shown in Fig. ~\ref{fig:yvsm12}. However, the effect is somewhat subdued owing to the additional inverse dependence of the relic density on $m_\chi$.   In the case of the dark matter mass, the direct dependence of the relic density on the combination $m_\psi y_1^2$, where the coupling is coming from the decay width, yields a parabolic dependence between the two parameters. The influence coming from the phase-space factor is subdued and indicated in the spread of the points. 
We checked that the width of $\chi$ in this parameter range  varies between $10^{-25}$ and  $10^{-21}$ GeV. The charged scalar decays through $S\to N\ell$, if kinematically allowed. However, as we shall consider the mass of the heavy neutrinos $m_N$ to be in the range of 10 TeV or more, this disallows the two-body decay. The three-body decay that follows through the heavily off-shell neutrino $N$, along with 
the Yukawa coupling required for the seesaw mechanism being in the range of $10^{-8}$ slows down the decay of $S$. For the parameters considered here, the decay width is in the range of $10^{-24}$ to $10^{-19}$ GeV for $m_S$ between 100  GeV and 1 TeV.

\subsection{Freeze-in via four body decay of $\chi$}
	\label{subsec:FIMP4body}

\begin{figure}[h!]
	\begin{center}
		\begin{tikzpicture}
		\begin{feynman}
		\vertex (a1){\(\chi\)};
		\vertex[  right=1.8cm of a1] (a2) ;
		\vertex[below right=1.4cm of a2] (a3) ;
		\vertex[above right=1.4cm of a2] (a4) {\(\psi \)} ;
		\vertex[above right=1.4cm of a3] (a5) {\({\ell} \)} ;
		\vertex[below right=1.4cm of a3] (a6) ;
		\vertex[below right=1.4cm of a6] (a7) { \(H \) } ;
		\vertex[above right=1.4cm of a6] (a8) { \(\nu \) } ;
		\diagram* { (a1) --  [fermion](a2),(a2)  --[fermion,] (a4),(a2) --[ charged scalar , edge label= {\( S^{*} \)}] (a3), (a5) --[fermion] (a3),  (a3) --[ fermion, edge label= {\( N^{*} \)} ] (a6), (a6) --[ scalar ] (a7), (a6) --[ fermion] (a8)
		};
		\end{feynman}
		\end{tikzpicture}
	\end{center}
	\caption{ Feynman  diagrams for the $\chi$ decay when $m_H+m_\psi<m_\chi < m_S+m_\psi$ }
	\label{fig:FD4bodydecay}
\end{figure}
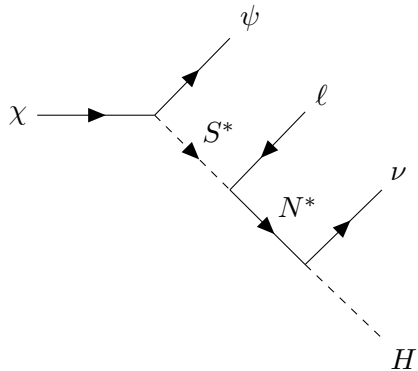

\begin{figure}[h]
	\begin{center}
		\includegraphics[width=2in]{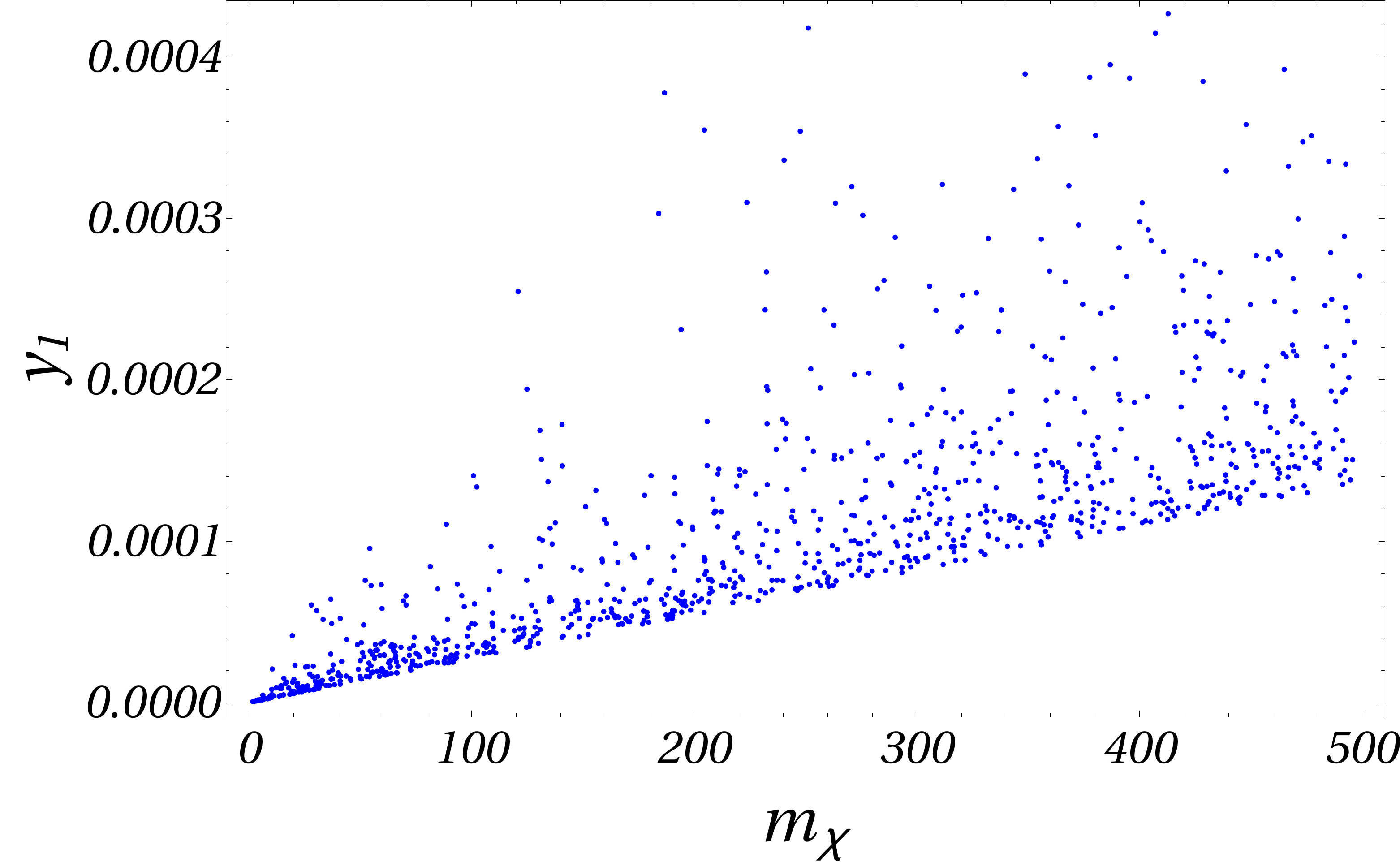}
		\includegraphics[width=2in]{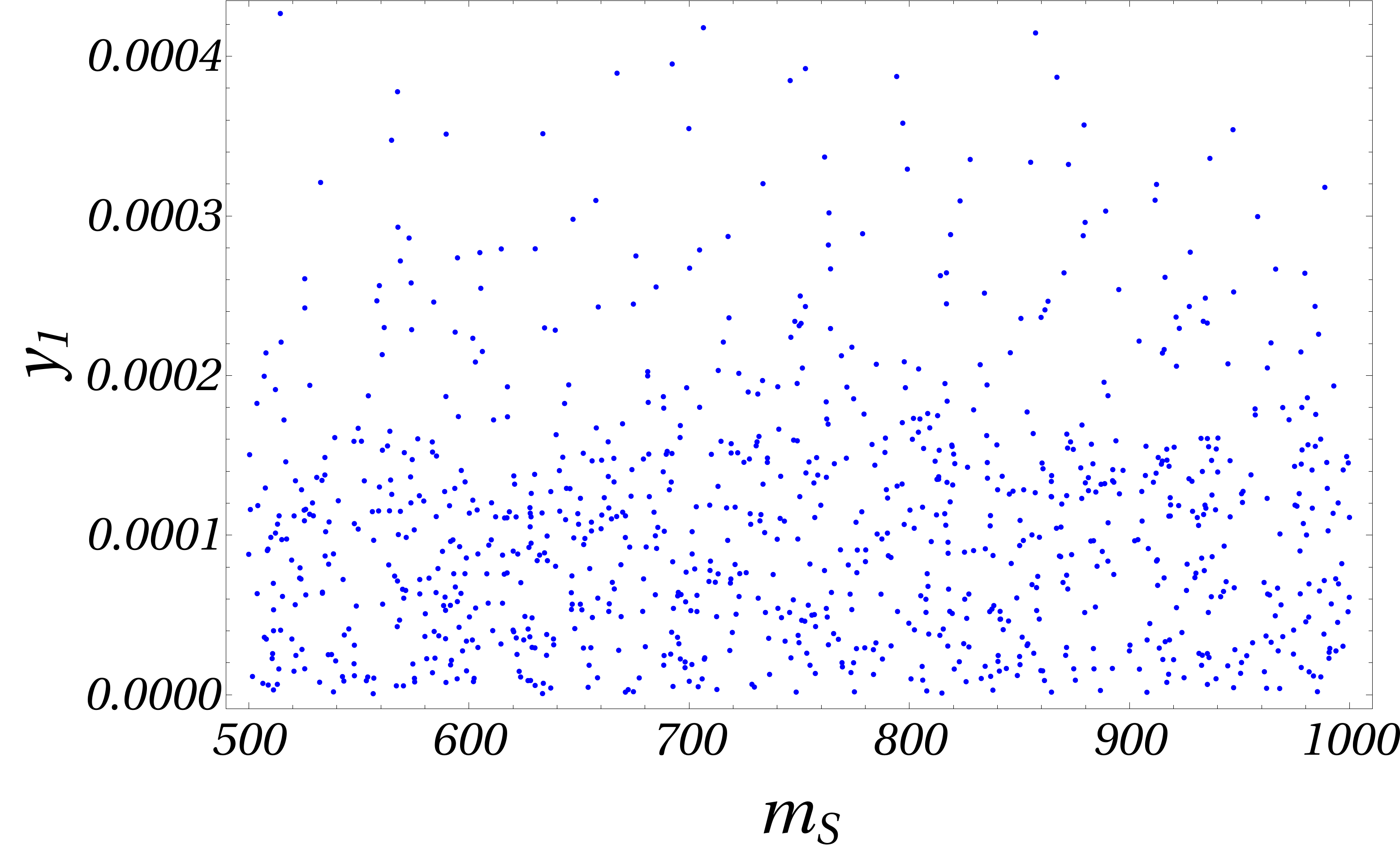} 
		\includegraphics[width=2in]{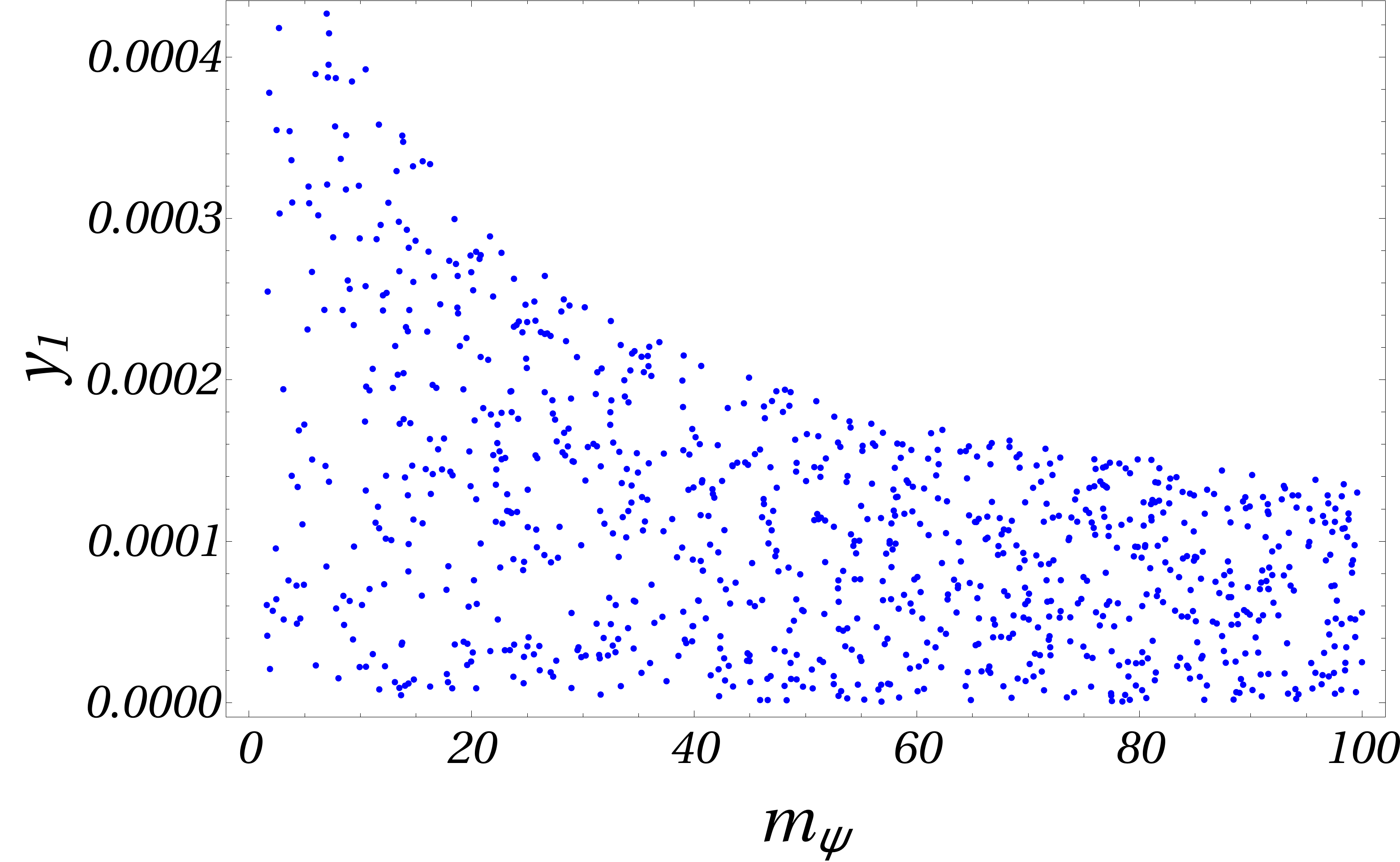}
	\end{center}
		\caption{Yukawa coupling $y_1$ versus the mass of non-SM particles, from the requirement that they satisfy the relic density bound.}
\label{fig:y1vsM4body}
\end{figure}

In the kinematic region where  $m_{\chi}< m_{S} +m_{\psi}$, the two body decay discussed above is not kinematically allowed. Noting that we are working with $m_S\ll m_N$, the leading allowed decay channel is the four-body decay shown in Fig.~\ref{fig:FD4bodydecay}. This involves the couplings, $y_1,~y_2$ and $Y_N$.   With $m_N\sim 10$ TeV, $Y_N$ in the range of $10^{-8}$  to satisfy the seesaw condition the cross section for this process will require larger $y_1$ values.
Consistent with parameters chosen from leptogenesis, as discussed in Section~\ref{sec:leptogenesis}, we restrict $y_2$ in the range of $10^{-3}-10^{-1}$.  With these restrictions, we scanned over the masses ($m_\chi,~m_S,~m_\psi$) and used Eq.~\ref{eq:Omega} replacing $\Gamma_{\chi\to S\psi}$ by the four-body width $\Gamma_{\chi\to\ell\nu H\psi}$, to obtain $y_1$ corresponding to the observed relic density. The scan range is as in Table~\ref{tab:tableScan12} with the condition that $m_\chi<m_S+m_\psi$. Resulting parameter space points projected on to $y_1-m_i$ plane, where $i=\chi,~S,~\psi$, are presented in Fig.~\ref{fig:y1vsM4body}.
Clearly, $y_1$ in the range of $10^{-4}$ is compatible with the dark matter observations.

\section{Leptogenesis}
\label{sec:leptogenesis}

So far, we insured that our model satisfies dark matter constraints within the range of parameters chosen. We move on to the analysis of leptogenesis in the scenario considered here, where the heavy Majorana neutrino can decay to leptons as well as to anti-leptons. Similar to the standard leptogenesis, CP-violation arising through the interference of the tree-level processes with the one-loop level self-energy and vertex corrections could lead to lepton number asymmetry, enabled by the non-equilibrium condition arising in the cosmic evolution equation, when the temperature of the Universe is of the order of mass of the decaying neutrino. The lepton number asymmetry thus generated is converted into baryon number asymmetry through non-perturbative sphaleron processes connecting two possible vacuum configurations,  which differ in their lepton and baryon numbers at the electroweak phase transition \cite{Fukugita:1986hr}. With hierarchical mass for the heavy neutrinos ($m_{N_1}\ll m_{N_2}\ll m_{N_3}$), any leptogenesis generated through the decay of $N_2$ and $N_3$ is washed out through the same sphaleron processes at high temperature, and thus the relevant leptogenesis is generated purely through $N_1$ decays.   For canonical thermal leptogenesis with hierarchal right-handed neutrinos, an upper limit on the CP asymmetry exits, providing a lower limits on the mass of lightest right-handed neutrino $m_{N_1}$  \cite{Giudice:2003jh}
\begin{equation}
m_{N_1} \geq 5 \times 10^8~{\rm GeV}~ \left( \frac{v}{246 ~{\rm GeV}}\right)^2.
\end{equation}
However, we notice that the same Yukawa coupling responsible for the seesaw mechanism and leptogenesis could add radiative corrections to the Higgs boson mass ($m_H$). Based on naturalness arguments requiring such corrections not  to exceed $\Delta m_H^2\sim 1~{\rm TeV}^2$ leads to an upper bound  \cite{Giudice:2003jh} 
\begin{equation}
m_{N_1} \leq 3 \times 10^7~{\rm GeV}~ \left( \frac{v}{246 ~{\rm GeV}} \right)^{\frac{2}{3}}
\end{equation}
clearly in conflict with the bound from leptogenesis above. It was pointed out in \cite{Pilaftsis:2003gt} and further studied in detail in \cite{Pilaftsis:2005rv}, that a resonant enhancement of the CP-asymmetry is possible when the mass difference between two  right-handed neutrinos is very small, and comparable to their decay widths ($m_{N_2}-m_{N_1}\sim \Gamma_{1,2}$), low energy leptogenesis is enabled, with $m_{N_{1,2}}\sim 1-10$ TeV.

In our model,  additional possibilities open through the Yukawa interaction $y_{2ij}~\bar \ell_i  N_jS+{\rm h.c.}$. First, $N\to S\ell$ and its CP-conjugate process provide new decay channels. Second, the self-energy and vertex corrections in each of the decays receive additional contributions, as explained below. We now investigate possibilities in the low-energy leptogenesis enabled by these new interactions.

\subsection{CP Asymmetry}
\label{subsec:CPAsymmetry}

\begin{figure}[h!]
\begin{center}
    	\begin{tikzpicture}  
	\begin{feynman}
	\vertex (a1){\(N_i\)};
	\vertex[right=1.4cm of a1] (a2) ;
	\vertex[above right=1.4cm of a2] (a3) ;
	\vertex[below right=1.4cm of a2] (a4) ;
	\vertex[ right=1cm of a4] (a5){\( \phi/S \)};
	\vertex[ right=1cm of a3] (a6) {\(\ell_l \)} ;
	\diagram* { (a2) --  [majorana](a1),(a4)  --[scalar ] (a5),(a3) --[fermion] (a6), (a3) --[majorana,edge label= {\( N_j \)}] (a4),  (a2) --[ scalar, edge label= {\( \phi/S\)}] (a3),(a4) --[anti fermion, edge label= {\(\ell_{k}\)} ] (a2),    
	};
	\end{feynman}  
	\end{tikzpicture} \hskip 10mm
		\begin{tikzpicture}  
	\begin{feynman}
	\vertex (a1){\(N_i\)};
	\vertex[right=1.4cm of a1] (a2) ;
	\vertex[right=1.4cm of a2] (a3) ;
	\vertex[right=1.1cm of a3] (a4) ;
	\vertex[above right=1cm of a4] (a5){\( \phi / S \)};
	\vertex[below right=1cm of a4] (a6){\(\ell_l \)} ;
	\diagram* { (a2) --  [majorana](a1),(a4)  --[scalar ] (a5),(a4) --[fermion] (a6), (a3) --[majorana,edge label= {\( N_j \)}] (a4),  (a2) --[ scalar,half left, edge label= {\( \phi/S\)}] (a3),(a2) --[ fermion,half  right, edge label= {\( \ell_{k}\)} ] (a3),    
	};
	\end{feynman}  
	\end{tikzpicture}       
	\caption{ Feynman diagram corresponding to one loop vertex and self-energy corrections to $N_i \rightarrow \ell_l ~\phi/S$.}
	\label{fig:FDcpv}
\end{center}
\end{figure}
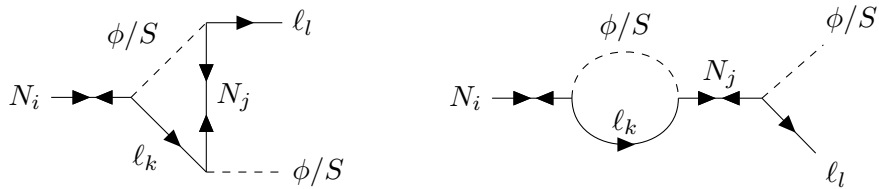

As in  standard leptogenesis, in the present case  CP asymmetry also arises through the interference of the tree-level process and the one-loop process. 
The decay processes relevant to our analysis are $ N_1 \rightarrow L \phi $ and $ N_1 \rightarrow \ell S $. 
CP violation leads to asymmetric decays to leptonic and anti-leptonic channels. We  denote by $\epsilon_1$ the CP-violating parameter arising from the standard process involving the Higgs boson 
\begin{equation}
\epsilon_1=\frac{\Gamma(N_1 \rightarrow L \phi)-\Gamma( N_1 \rightarrow \bar{L} \bar{\phi})}{\Gamma(N_1 \rightarrow L \phi)+\Gamma( N_1 \rightarrow \bar{L} \bar{\phi})},
\label{eq:cpvphi}
\end{equation}
and by $ \epsilon_2$ the corresponding parameter arising from the decay involving the new charged singlet scalar 
\begin{equation}
\epsilon_2=\frac{\Gamma( N_1 \rightarrow \ell S)-\Gamma( N_1 \rightarrow \bar{\ell} \bar{S})}{\Gamma( N_1 \rightarrow \ell S)+\Gamma( N_1 \rightarrow \bar{\ell} \bar{S})}.
\label{eq:cpvS}
\end{equation}
In Fig.~\ref{fig:FDcpv} we show the relevant one-loop diagrams showing the vertex corrections as well as the self-energy corrections.
The CP-violating parameters in Eq.~\ref{eq:cpvphi} and Eq.~\ref{eq:cpvS} can be written explicitly in terms of the self-energy ($\epsilon^{s\xi}$) and vertex contributions ($\epsilon^{v\xi}$), with $\xi=\phi,~S$, as
\begin{align}
\epsilon_1= \epsilon^{s\phi}+\epsilon^{v\phi}
\label{epa}
\end{align}
\begin{align}
\epsilon_2= \epsilon^{sS}+\epsilon^{vS},
\label{eq:epb}
\end{align}
where
\begin{align}
\epsilon^{s\phi}=& \frac{1}{8 \pi K_{11}}\sum_{j=2,3}\left[  \frac{m_{N_1}}{m^2_{N_1}-m^2_{N_j}} \Im\left(m_{N_j} ~K^2_{1j}+m_{N_1}~ \kappa_{1j}K_{1j}+m_{N_j} ~\kappa_{j1} K_{1j}\right)\right] \, , 
\label{eq:CPh}\\
\epsilon^{sS}=& \frac{1}{8 \pi \kappa_{11}}\sum_{j=2,3}\left[  \frac{m_{N_1}}{m^2_{N_1}-m^2_{N_j}} \Im\left(m_{N_j} ~\kappa^2_{j1}+m_{N_1}~K_{1j} \kappa_{j1}+m_{N_j} ~K_{j1} \kappa_{j1}  \right)\right]  \, ,
\label{eq:CPs} \\
\epsilon^{v\phi}=&\frac{1}{8 \pi K_{11}} \sum_{j=2,3} \Im \left(K^2_{1j}\right) \mathcal{F}\left( \frac{m^2_{N_j}}{m^2_{N_1}} \right) \, ,
\label{eq:CPV1}\\
\epsilon^{vS}=&\frac{1}{8 \pi \kappa_{11}} \sum_{j=2,3} \Im \left(\kappa^2_{j1}\right) \mathcal{F}\left( \frac{m^2_{N_j}}{m^2_{N_1}} \right) \, ,
\label{eq:CPV2}
\end{align}
where $\mathcal{F}(x)= \sqrt{x} \left[1+(1+x)~\ln\frac{x}{1+x}\right]$, and we have defined the relevant product of the Yukawa couplings as 
\(K_{ij}=\left(Y_{N}^\dagger Y_N\right)_{ij}\) and \(\kappa_{ij}=\left(y^\dagger_{2}y_2\right)_{ij}. \)  
Unlike $Y_{N}$, the Yukawa couplings $=\left(y^\dagger_{2}y_2\right)_{ij}. \)  
Unlike $Y_{N}$, the Yukawa couplings $\kappa_{ij}$ connecting $N\ell S$ do not play any role in the seesaw mechanism. However, they may influence the lepton flavour violating processes, which is somewhat tamed by the heaviness of the right-handed neutrino. We shall, therefore, keep these couplings somewhat unrestricted, but not larger than of the order of 0.1.  The $Y_{N}$ couplings, on the other hand, are constrained from the light neutrino sector through the seesaw mechanism. They, together with $m_i$, control the masses and mixings of the light neutrinos. Note that the standard leptogenesis is obtained in the limit $\kappa_{ij} \to 0$, when only the first term in Eq. \ref{eq:CPh}, and  Eq. \ref{eq:CPV1} survive. Conversely, from the experimental results on the masses and mixings of the light neutrinos, we can reconstruct $Y_N$ for given heavy neutrino masses. We adopt the Casas-Ibarra parametrisation (CI) \cite{Xing:2009vb,Casas:2001sr} to represent $Y_N$  as
\begin{equation}
y_{N}=  D_{\sqrt{M}} R D_{\sqrt{\kappa}} U^{\dagger},
\label{CIeqn}
\end{equation}
where $U $ is light neutrino mixing matrix, which we take to be  the PMNS matrix \cite{ParticleDataGroup:2020ssz}, $R$ is any arbitrary complex orthogonal matrix, $(D_{\sqrt{M}})_{ij}=\sqrt{m_{N_i}}~\delta_{ij}$,  $(D_{\sqrt{\kappa}})_{ij}=\frac{\sqrt{m_i}}{v}~\delta_{ij}$, where $m_{N_i}$ and $m_i$ are the masses of the heavy right-handed neutrinos and the light neutrinos, respectively, and $v$ is the VEV of the Higgs field.

\subsection{The  Boltzmann equations}
\label{subsec:BoltzmannEqn}
 The time evolution of the number density in non-thermal equilibrium  is studied using the Boltzmann equation. As mentioned earlier, in the hierarchical heavy neutrino mass case, any lepton asymmetry generated by the decay of a massive Majorana particle is washed out before the electroweak symmetry breaking. Thus, the surviving lepton asymmetry is generated by the decay of the lightest right-handed neutrino ($N_1$). The number of density of $N_1$  depends on its decay, inverse decay and scattering process. In the present set-up  the following processes fit in these categories:
 \begin{enumerate}
 \item Decay: $N_1\to \phi L$, ~~$N_1\to \bar\phi \bar L$, ~~$N_1\to S\ell$, ~~$N_1\to \bar S \bar \ell$;
 \item Inverse decay: $\phi L\to N_1,~~~\bar \phi \bar L\to N_1,~~~S \ell\to N_1,~~~\bar S \bar \ell\to N_1$;
 \item Scattering: \\
 standard $\Delta L=1$ $s$-channel processes: $\ell N_1 \rightarrow  d\bar u$, $\bar \ell N_1 \rightarrow  \bar d u$; \\ 
 standard $\Delta L=1$ $t$-channel processes: $N_1 u\to d\bar\ell$, $N_1 \bar u\to\bar d \ell$, $N_1 d\to u\ell$, $N_1 \bar d\to \bar u\bar \ell $; \\
 standard $\Delta L=1$ processes involving gauge boson: $N_1\phi\to A L$,  $N_1A \to \phi L$; \\
 new $\Delta L=1$ processes involving $S$: $N_1\ell \to S\phi$, $N_1\bar \ell \to \bar S\bar \phi$; \\
 new  $\Delta L=1$ processes involving $\chi,\psi$: $N_1\ell\to \chi\psi$,  $N_1\bar\ell\to \bar\chi\bar\psi$.
 \end{enumerate}
A lepton number asymmetry is induced by all the above processes, and in addition, induced also by the $\Delta L=2$ processes below.
\begin{enumerate}[start=4]
 \item standard $\Delta L=2$ processes:  $\ell\ell\to \bar\phi\bar\phi$, $\phi\ell\to\bar\phi\bar\ell$; 
 \item new  $\Delta L=2$ processes:  $S\ell\to \bar S\bar\ell$, $\phi\ell\to \bar \ell \bar S$, $\bar \phi\bar\ell\to \ell S$.
  \end{enumerate}
We considered only those processes in which the number density of $N_1$, and that of leptons and antileptons change. We study the evolution of the number density normalised by entropy $Y_{N_1}$, $Y_L=y_{l}-y_{\bar{l}}$. Within this minimal framework the Boltzmann equations can be written in the following form,
\begin{equation}
Hsz\frac{dY_{N_1}}{dz}= D_d+S_{ns}+S_s\, ,
\label{eq:BE1}
\end{equation} 
\begin{equation}
Hsz \frac{dY_L}{dz}= \left[ \frac{Y_{N_1}}{Y^{eq}_{N_1}}-1\right] \{ \epsilon_{n2} \Gamma_{D_2} +\epsilon_{n1} \Gamma_{D_1} \} -S Y_L\, ,
\label{eq:BE2}
\end{equation}
where decay and inverse decay processes $D_d$ is given in Eq.~\ref{eq:Dd}, while $S_s$ and $S_{ns}$ contributions come from the scattering terms that affect the number density of $Y_{N_1}$, are given  in Eq.~\ref{eq:Ss} and Eq.~\ref{eq:Sns} below. \begin{equation}
D_d=\left[-\frac{Y_{N_1}}{Y^{eq}_{N_1}}+1 \right](\Gamma_{D_1}+\Gamma_{D_2})
\label{eq:Dd}
\end{equation}
\begin{equation}
S_{ns}=\left[  1-\frac{Y_{N_1}}{Y^{eq}_{N_1}}    \right]\left[2\Gamma_{s_{new1}}+2\Gamma_{s_{new2}} \right]
\label{eq:Sns}
\end{equation}
$S$,  given in Eq.~\ref{eq:S} below, is the contribution of decays, inverse decays, and scattering process.
\begin{equation}
S_s=\left[  1-\frac{Y_{N_1}}{Y^{eq}_{N_1}}    \right]\left[\Gamma_{s_{1}}+\Gamma_{s_{2}}+\frac{\Gamma_{s_3}}{2}+ \frac{\Gamma_{s_4}}{2}+\frac{\Gamma_{s_5}}{2}+\frac{\Gamma_{s_6}}{2} \right]
\label{eq:Ss}
\end{equation}
\begin{equation}
S=\frac{1}{2 Y^{eq}_L} \left[\Gamma_{D_1}+\Gamma_{D_2} +2\Gamma_{st}+ \frac{Y_{N_1}}{ Y^{eq}_{N_1}} (2\Gamma_{s_{new1}}+2\Gamma_{s_{new2}}+\Gamma_{s_1}+\Gamma_{s_2}) + \frac{\Gamma_{s_3}}{2}+\frac{\Gamma_{s_4}}{2}+\frac{\Gamma_{s_5}}{2}+\frac{\Gamma_{s_6}}{2}\right]
\label{eq:S}
\end{equation}
Where $\Gamma_{D}$ is defined for a thermal average of decay widths. The thermal average of a general decay process $X_i\rightarrow a_{\alpha} b_{\beta}$ is
\begin{equation}  
\Gamma_{D_i}= \Sigma_i \Sigma_{\alpha, \beta}\int d\Pi_X e^{\frac{-E_x	}{T}} \int d\Pi_a d\Pi_b (2 \pi)^4 \delta^4(P_x-P_a-P_b) |M(X_i\rightarrow a_{\alpha} b_{\beta})|^2.
\label{eq:OE}
\nonumber
\end{equation}
Where $\Gamma$ represents the thermal average of scattering cross-section of a process:
\begin{equation}
\Gamma_{s}=\Gamma(x+a\rightarrow 1+2)=\frac{T}{(32 \pi^4)}\int s^{\frac{3}{2}} ds K_1(\frac{s}{T})\lambda(1,\frac{m^2_1}{s},\frac{m^2_2}{s}) \sigma\, .
\label{eq:BS17}
\nonumber
\end{equation}
Here $\Gamma_{D_1}$ and $\Gamma_{D_2}$ are thermal decay widths for $N_1\to \phi L$ and $N_1\to S\ell$, respectively. The contributions $\Gamma_{s_{new1}}$, $\Gamma_{s_{new2}}$, $\Gamma_{s_{1}}$, $\Gamma_{s_{2}}$, $\Gamma_{s_{3}}$, $\Gamma_{s_{4}}$, $\Gamma_{s_{5}}$, $\Gamma_{st}$  and $\Gamma_{s_{6}}$ are thermal average of scattering cross-section of $N_1\ell\to \chi\psi$, $N_1\ell \to S\phi$,  $\ell N_1 \rightarrow  d\bar u$, $\ell N_1 \to \phi A$, $N_1 u\to d\bar\ell$, $N_1 d\to u\ell$,  $N_1\phi\to A L$, $\ell\ell\to \bar\phi\bar\phi$  and  $N_1A \to \phi L$, respectively. The $\Delta L=2$ scattering with both $s$-channel and $t$-channel contributions are represented by $\Gamma_{st}$.

\subsection{Lepton Number Asymmetry}
\label{subsec:LeptonAsym}
In the previous subsection we have given the Boltzmann equations for the  number density distribution functions relevant for neutrinos, leptons and anti-leptons in our model.   To evaluate the asymmetry, we explore the evolution of the phase space distribution of the lepton sectors. The number density of lepton sectors varies with the number density of $N_1$. Therefore, we solve Boltzmann equation for $N_1$ and for the  lepton sectors. Boltzmann Eqs. \ref{eq:BE1} and \ref{eq:BE2}  can be written as,
\begin{equation}
\frac{dY_{N_1}}{dz}=D(-Y_{N_1}+Y^{eq}_{N_1})\, ,
\label{eq:BE3}
\end{equation}
\begin{equation}
\frac{dY_{L}}{dz}=\left( \epsilon_{1} \frac{\Gamma_{D_1}}{HszY^{eq}_{N_1} }+\epsilon_{2} \frac{\Gamma_{D_2}}{Hsz Y^{eq}_{N_1}} \right)\left(-Y_{N_1}+Y^{eq}_{N_1}\right)-\frac{S Y_L}{Hsz}\, .
\label{eq:BE4}
\end{equation}
The rate change of the number density per unit entropy with energy levels, as defined in Eq. \ref{eq:BE3},  affects the number density of daughter particles. At equilibrium,  $Y_{N_1}=Y^{eq}_{N_1}$, there is no change in the number density of $Y_{N_1}$ for that energy level. Deviations of $Y_{N_1}$ from $Y^{eq}_{N_1}$ determine the changes of number density in the  particles it decays into. Here we want to study change of number density of leptons,  given in Eq. \ref{eq:BE4}, the effect of the coefficient of $(-Y_{N_1}+Y^{eq}_{N_1})$ conversion factor into the leptonic sector. This conversion factor is also energy dependent. If the conversion factor is large at an energy level, that is when $Y_{N_1}$ deviates from $Y^{eq}_{N_1}$,  it can generate more matter antimatter asymmetry at that time. The last term in Eq. \ref{eq:BE4} is the wash-out term. The analytic solutions of Eqs. \ref{eq:BE3}  and \ref{eq:BE4} are
\begin{equation}
Y_{N_1}=e^{-\int D dz} \int_{z_{min}}^{Z_{max}} Y^{eq}_{N_1} D  e^{\int D dz} dz,
\end{equation}
\begin{equation}
Y_{L}=\epsilon_{1} \zeta_{1}+\epsilon_{2} \zeta_{2}.
\end{equation}
Where  $\zeta_1$ and $\zeta_2$ are  efficiency factors and given below in Eqs. \ref{eq:kappa1}, \ref{eq:kappa2}. There are three possibilities for generating matter antimatter asymmetry. The first one is that term $\epsilon_{1} \zeta_{1}> \epsilon_{2} \zeta_{2}$ so $\epsilon_{1} \zeta_{1}$ generates matter antimatter asymmetry. In second scenario $\epsilon_{2} \zeta_{2}> \epsilon_{1} \zeta_{1}$ so $\epsilon_{2} \zeta_{2}$  generates required matter antimatter asymmetry. The third possibility is that $\epsilon_{1} \zeta_{1}$  and $\epsilon_{2} \zeta_{2}$  {\it both} play an important role in generating matter antimatter asymmetry,  if $\epsilon_{1} \zeta_{1}$  and $\epsilon_{2} \zeta_{2}$  are  of the the same order, ${\cal O}(10^{-10})$.
\begin{equation}
\zeta_{1}=e^{-\int S dz} \int_{z_{min}}^{z_{max}} \frac{\Gamma_{D_1}}{HSz Y^{eq}_{N_1}} \left[\left( e^{-\int D dz} \int_{z_{min}}^{Z_{max}} Y^{eq}_{N_1} D  e^{\int D dz} dz \right)-Y^{eq}_{N_1} \right] e^{\int S dz} dz
\label{eq:kappa1}
\end{equation}
\begin{equation}
\zeta_{2}=e^{-\int S dz} \int_{z_{min}}^{z_{max}} \frac{\Gamma_{D_2}}{HSz Y^{eq}_{N_1}} \left[\left( e^{-\int D dz} \int_{z_{min}}^{Z_{max}} Y^{eq}_{N_1} D  e^{\int D dz} dz \right)-Y^{eq}_{N_1} \right] e^{\int S dz} dz
\label{eq:kappa2}
\end{equation}
The above equations indicate that the efficiencies $\zeta_{1}$ and $\zeta_{2}$ are different. In $\zeta_{1}$ the thermal average is decay width  $\Gamma_{D_1}$ while for $\zeta_{2}$  the  thermal average factor is the decay width $\Gamma_{D_2}$. While thermal average of decay width $\Gamma_{D_1}$ depends on the Yukawa coupling constant $Y_{N_i}$, the thermal average of decay width $\Gamma_{D_2}$ depends on Yukawa coupling $y_2$, which   is much larger than Yukawa coupling $Y_{N_i}$, and so the efficiency factor $\zeta_{2}$ is greater than  $\zeta_{1}$.

\section{Numerical Analysis}
\label{sec:NumAnalysis}

In this section, we perform the numerical analysis of the 
 lepton asymmetry generated through a combination of CP-asymmetry arising through the standard decay, and the decay to the newly introduced charged scalar, as well as considering the influence of the wash-out arising through different scattering processes  mentioned in Section \ref{subsec:BoltzmannEqn}. The CP parameters depend on the coupling constants $Y_{N}$ and $y_2$, where, as we noted before, $Y_{N}$ is constrained by light neutrino masses and mixings, whereas $y_2$ is largely unconstrained. Considering the CI parametrisation as in Eq. \ref{CIeqn}, the $Y_{N}$ depends on the masses and mixing matrix elements of light neutrinos,  on the masses of the heavy neutrinos, and on the elements of an arbitrary orthogonal matrix, $R$. We consider normal hierarchy in the light neutrino sector, with the lightest neutrino considered to be massless, and the other two masses set in agreement with constraints from the the neutrino oscillation experiments \cite{ParticleDataGroup:2020ssz, Esteban:2020cvm}. Accordingly, we chose   $m_1=0$ GeV, $m_2=0.0083$ eV and $m_3=0.051 $ eV.  The complex orthogonal matrix, $R$ is chosen as
\[ R= \left( {\begin{array}{ccc}
	~\cos\theta& \sin\theta&0\\
	-\sin\theta&\cos\theta&0\\
	~0&0&1
	\end{array} } \right),
\]
We scan over the heavy neutrino mass keeping a hierarchy of 
$m_{N_1}\ll m_{N_2}\ll m_{N_3}$, with the lightest mass in the 10 to 100 TeV range, and the heavier ones differing from it by at least one order of magnitude. 
The dependence of $Y_N$ and $y_2$ enter the CP-asymmetries through their combination $K_{1j}$ and $\kappa_{1j}$ with $j=2,3$, respectively, as given in Eq.s \ref{eq:CPh}-\ref{eq:CPV2}.  The complex nature of these parameters is important in determining the amount of CP-violation.  Let us first consider $K_{1j}$. With the standard Yukawa couplings given by the CI parametrisation as in Eq. \ref{CIeqn}, 
\begin{equation}
K_{1j} = (Y_N)_{k1}^*(Y_N)_{kj} =\sum_{k,\alpha,\beta}m_{N_k}\frac{\sqrt{m_\alpha m_\beta}}{v^2}~R^*_{k\alpha}R_{k\beta}~U_{1\alpha}U^*_{j\beta}.
\end{equation}
With the assumed structure of $R$ and the choice of $m_1=0$, this  takes a simpler form,
\begin{equation}
K_{1j} = \frac{m_{N_1}m_2}{v^2}~|R_{12}|^2~U_{12}U^*_{j2}+ \frac{m_{N_2}m_2}{v^2}~|R_{22}|^2~U_{12}U^*_{j2}+ \frac{m_{N_3}m_3}{v^2}~U_{13}U^*_{j3}
\end{equation}
Thus, the phases of the elements of $R$ are irrelevant, and we can consider it a real orthogonal matrix for our analysis. Taking here $U$  as the PMNS matrix, the only phases that enters in $K_{12}$ are those of $U_{13}$ and  $U_{22}$, while in  $K_{13}$ the phases of $U_{13}$ and  $U_{32}$ give rise of the relevant CP phase. However, noticing that in the hierarchical case considered in the heavy neutrino sector, the third term, proportional to $M_{N_3}$  dominates over the others,  the only phase that is relevant in both $K_{12}$ and $K_{13}$ is that of $U_{13}$. We write this as
\begin{eqnarray}
K_{12} &\sim& \frac{m_{N_3}m_3}{v^2}~\sin\theta_{23}\cos\theta_{13}~\sin\theta_{13}~e^{-i\delta_{CP}} \nonumber \\
K_{13} &\sim& \frac{m_{N_3}m_3}{v^2}~\cos\theta_{23}\cos\theta_{13}~\sin\theta_{13}~e^{-i\delta_{CP}} 
\end{eqnarray}
The real matrix element $K_{11}$ can be written as
\begin{eqnarray}
K_{11} &\sim& \frac{m_{N_3}m_3}{v^2}~\sin^2\theta_{13}
\end{eqnarray}
We use the present experimental values of  the mixing angles and the $\delta_{CP}$ parameter \cite{Esteban:2020cvm}
\begin{eqnarray}
\theta_{13}={8.57^\circ}^{+ 0.13}_{0.12},~~\theta_{23}={49.0^\circ}^{+1.1}_{-1.4},~~\delta_{CP}={195^\circ}^{+51}_{-25}.
\label{numixing}
\end{eqnarray}
Taking $m_{N_3}\sim 10^3$ TeV, we have $K_{1j}\sim 10^{-8}$.  The standard contribution to the self-energy term in the CP-violating parameter can then be read from Eq. ~\ref{eq:CPh} as 
\begin{equation}
\epsilon^{s\phi}_{\rm std}= \frac{1}{8 \pi K_{11}}\left[  \frac{m_{N_1}}{m_{N_2}} \Im\left(K^2_{12}\right)+ \frac{m_{N_1}}{m_{N_3}} \Im\left(K^2_{13}\right)\right]\sim
\frac{1}{8 \pi K_{11}}~  \frac{m_{N_1}}{m_{N_2}} ~\Im\left(K^2_{12}\right)
\label{eq:epsphistd}
\end{equation}
and the new contribution to the standard decay channel is dominantly 
\begin{equation}
\epsilon^{s\phi}_{\rm new}\sim
\frac{1}{8 \pi K_{11}}~  \frac{m_{N_1}}{m_{N_2}} ~\Im\left(\kappa_{21}K_{12}\right).
\label{eq:epsphinew}
\end{equation}
With a suitably chosen $\kappa_{21}$, it is then possible to lift up the CP-violating parameter to the value required in leptogenesis, even when the standard contribution is a few orders of magnitude smaller.  In addition, there is a contribution from the new decay channel $N_1\to \ell S$, which contributes to the self-energy  (Eq.~\ref{eq:CPs}),
\begin{equation}
\epsilon^{sS}\sim
\frac{1}{8 \pi \kappa_{11}}~  \frac{m_{N_1}}{m_{N_2}} ~\Im\left(\kappa_{21}^2\right).
\label{eq:epsSi}
\end{equation}
Notice that the contribution to the standard channel from the virtual effects of the new scalar particle, $S$ is present even if $\kappa_{21}$ is real, whereas the contribution coming from the new channel is subdominant in this case, as the contribution will now be 
\begin{equation}
\epsilon^{sS}\sim
\frac{1}{8 \pi \kappa_{11}}~  \frac{m_{N_1}}{m_{N_2}} ~\Im\left(\kappa_{21}K_{21}\right).
\label{eq:epsSr}
\end{equation}
If $\kappa_{1j}$ is real, the effect of $S$ is completely absent in the vertex correction contribution to the CP-violating parameter, Eqs. \ref{eq:CPV1}-\ref{eq:CPV2}.  However, once $\kappa_{1j}$ has an imaginary part, it can be suitably chosen to get the required CP-asymmetry, which now has a dominant contribution from the new scalar $S$
\begin{equation}
\epsilon^{vS}\sim
\frac{1}{8 \pi \kappa_{11}}~  \frac{m_{N_3}}{m_{N_1}} ~\Im\left(\kappa_{31}^2\right).
\label{eq:epvS}
\end{equation}
As it is proportional to $m_{N_3}$, the vertex contribution dominates over the self-energy contribution for $\kappa_{1j}$ of the same order.
With these motivations, we shall now undertake the numerical analysis, scanning the parameter space to find regions  that are compatible with the observed baryon asymmetry.
We shall consider four distinct possibilities for $\kappa_{1j}$, {\it i. e.}
\begin{enumerate}
\item Real and positive $\kappa_{1j}$;
\item Real and negative $\kappa_{1j}$;
\item Imaginary and negative $\kappa_{1j}$;
\item Complex $\kappa_{1j}$ with negative imaginary part.
\end{enumerate}
The signs are chosen so that in each situation we get the right sign for the CP-asymmetry. In each of the four cases above, we shall consider two separate mass choices for the analyses, one with fixed $m_{N_j}$ and the other with varying $m_{N_j}$, but still keeping the normal mass hierarchy.  While the cases where the right-handed neutrino masses are constant are included in the more general cases where we allow them to vary, we perform  analyses with masses kept constant to clearly understand the influence of the couplings ($\kappa$) on the CP asymmetries $\epsilon_1$ and $\epsilon_2$. We then analyse the generated baryon asymmetry only for the cases where we vary the neutrino masses to decide if each scenario is compatible with the experimental data.

We perform the scans keeping the parameters in the range specified in Table~\ref{tab:summary} for each scenario. We shall consider specific cases with 
$\kappa_{1j}$ taken to be real, purely imaginary or complex and perform separate analyses for each case. 
\begin{table}[H]
	\centering
	\footnotesize
	{
					\begin{tabular}{l l|c|c|c|c}
		\hline\hline\\
					\multicolumn{6}{c}{$m_\chi=200~{\rm GeV},~~m_S=175~{\rm GeV},~~m_\psi=60~{\rm GeV},~~10^{-4}\le \kappa_{11}\le 10^{-1}$}\\[2mm]
			\hline\hline
			\multicolumn{2}{c|}{\multirow{2}*{\bf Scenario} }& {{\bf couplings}}  & \multicolumn{3}{c}{{\bf Mass of Heavy Neutrinos}}\\ \cline{3-6} 
			\multicolumn{2}{c|}{ }& {\bf $\kappa_{12}$},  {\bf $ \kappa_{13}$}  & $m_{N_1}$ TeV&$m_{N_2}$ TeV&$m_{N_3}$ TeV \\ 
			\hline \hline
			Case 1a&Real positive $\kappa$, constant mass	&\multirow{2}* {\bf $10^{-3}\leq\kappa_{1j}\leq 10^{-1}$} &  $10$&$10^3$ &$10^5$\\
			\cline{1-2} \cline{4-6}
			Case 1b&Real positive $\kappa$,  variable mass &    &$(10, 100) $&$(10^{3},  10^{4})$ &$(10^{5} , 10^{6} )$ \\
			\cline{1-6}
			Case 2a&Real negative $\kappa$, constant mass &  \multirow{2}* {\bf$ -10^{-1}\leq\kappa_{1j}\leq -10^{-3}$}  &10&$ 10^3$ &$  10^5$\\
			\cline{1-2} \cline{4-6}
			Case 2b&Real negative $\kappa$, variable mass & & $(10, ~100) $&$(10^3, ~ 10^4)$ &$(10^{5} , 10^{6} )$ \\
			\cline{1-6}
			Case 3a&Imaginary $\kappa$, constant mass	&  \multirow{2}* {\bf $  -\imath 10^{-3}\leq\kappa_{13}\leq -\imath 10^{-1}$}  & $10$&$10^3$ &$10^5$\\
			\cline{1-2} \cline{4-6}
			Case 3b&Imaginary $\kappa$, variable mass & & $(10, 100) $&$(10^{3},  10^{4})$ &$(10^{5} , 10^{6} )$ \\
			\cline{1-6}
			Case 4a&Complex $\kappa$, constant mass	& \multirow{2}* {\bf $ (1-\imath)10^{-3}\leq\kappa_{13}\leq (1-\imath )10^{-1}$}  &$10$&$10^3$ &$10^5$\\
			\cline{1-2} \cline{4-6}
			Case 4b&Complex $\kappa$, variable mass && $(10, 100) $&$(10^{3},  10^{4})$ &$(10^{5} , 10^{6} )$ \\
			\hline\hline
	\end{tabular}}
	\caption{The possible leptogenesis scenarios, yielding correct signs for the CP asymmetry, and the limits on the couplings, and neutrino masses, where applicable,  from the asymmetry parameters for each case.}	
	  \label{tab:summary}
\end{table}

Below we shall  analyse these cases separately. 
\vskip0.1in
\noindent
{\bf Case 1a: Real positive $\kappa$ and constant mass}\\[5mm]

\begin{figure}[H]
	\begin{center}
		\includegraphics[width=2.80in]{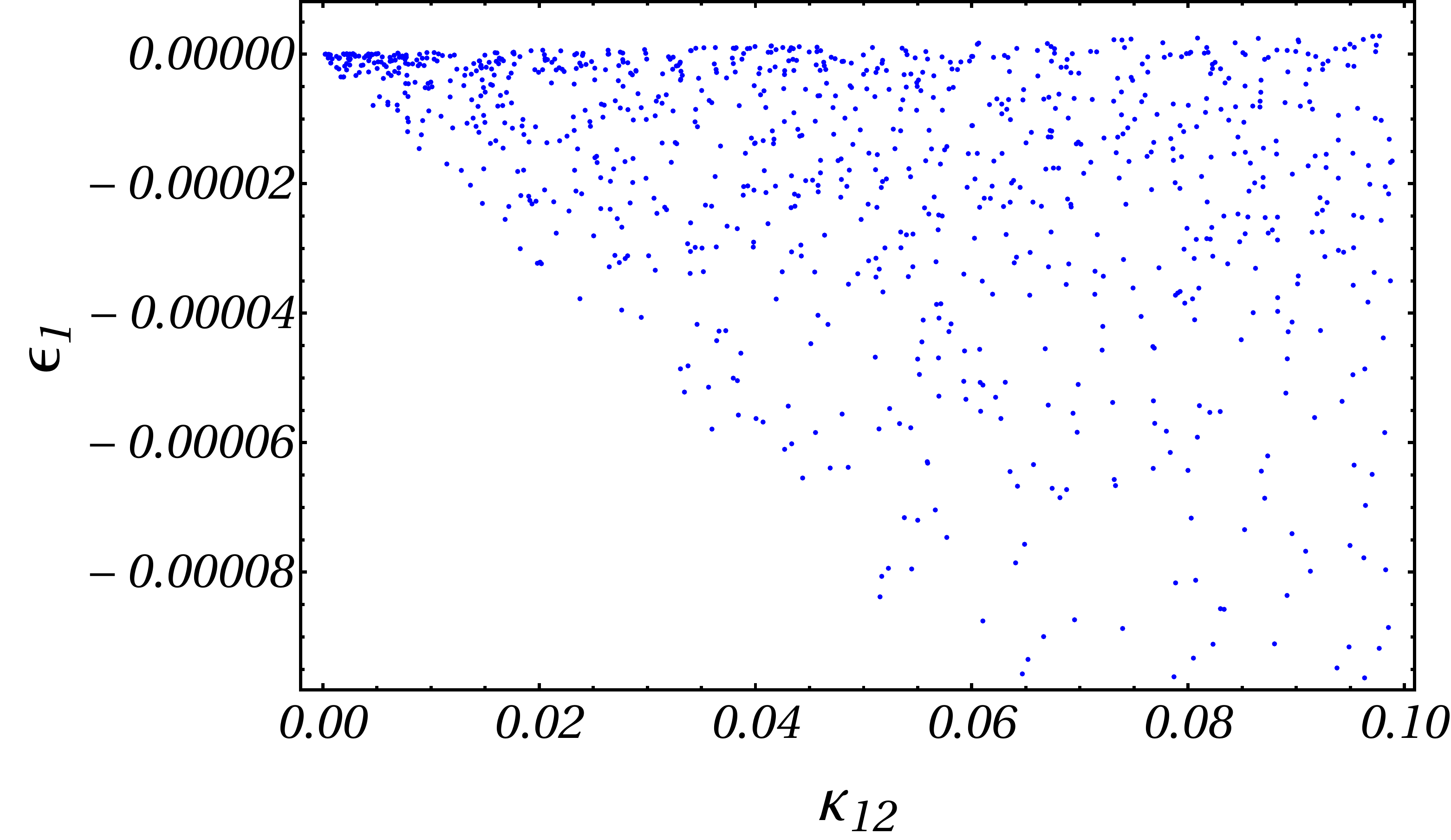}
		\includegraphics[width=2.90in]{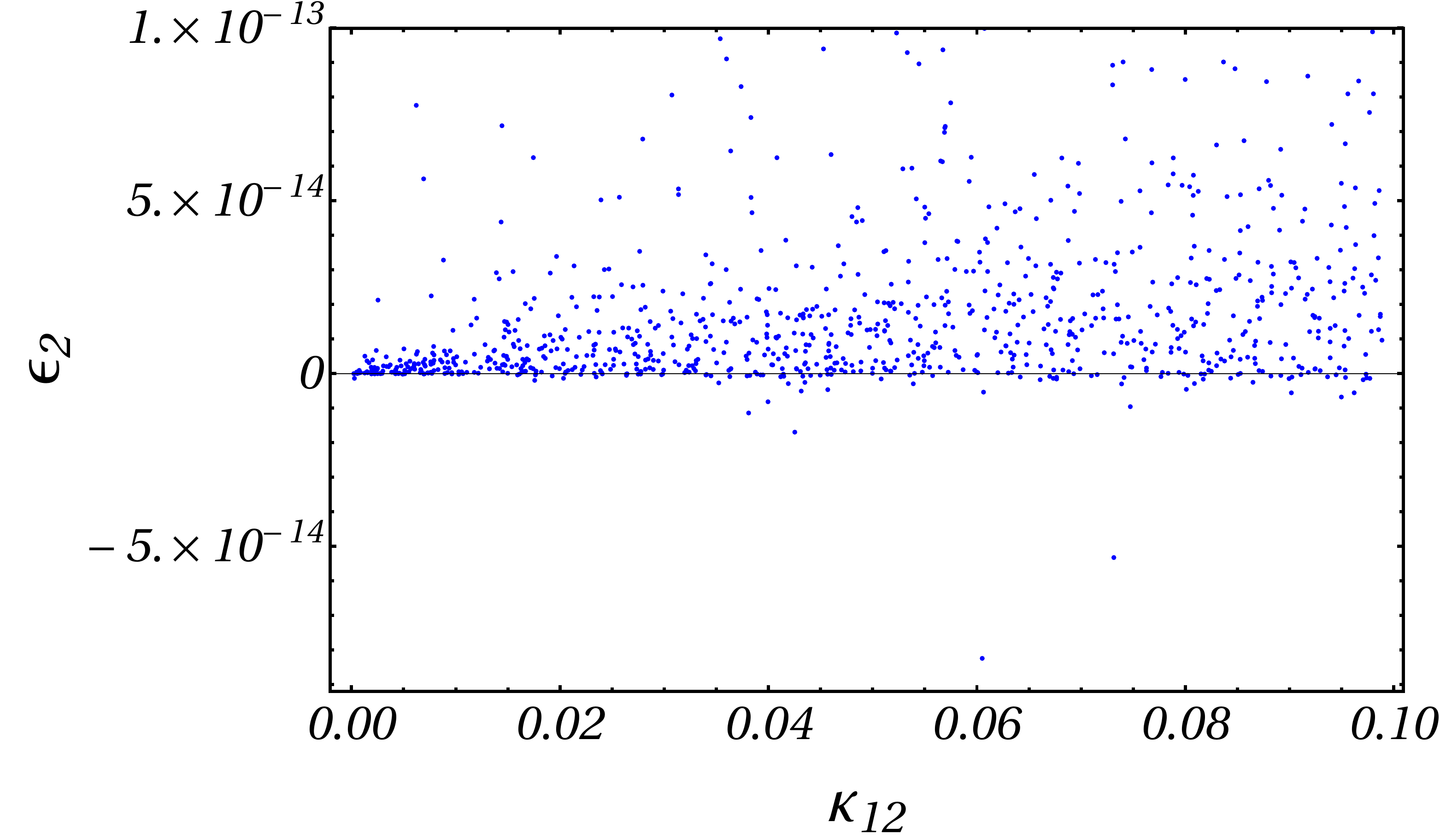}
		\caption{CP-asymmetry parameters $\epsilon_1$ and $\epsilon_2$ versus $\kappa_{12}$  for Case 1a, for real $\kappa$ and constant right-handed neutrino masses, $m_{N_j}$,  as given in Table \ref{tab:summary}.}	
\label{fig:ERek}	
	\end{center}
\end{figure}

We start our numerical analysis assuming $\kappa$ to be real and positive, and we fixe throughout the masses of right-handed neutrinos to be $m_{N_1}=10$ TeV, $m_{N_2}=10^3$ TeV and $m_{N_3}=10^5$ TeV, so that $\epsilon_{1}$ and $\epsilon_{2}$ depend only on $\kappa_{11}$, $\kappa_{12}$ and $\kappa_{13}$.  In this case, the vertex correction does not contribute to $\epsilon_2$, and thus the CP-asymmetry is dominated by $\epsilon^{sS}$ as in Eq.~\ref{eq:epsSr}. In the standard channel, the influence of $S$ to the self-energy correction will also be affected by $\kappa_{12}$ being real, as in Eq. \ref{eq:epsphinew}.
The scans of $\epsilon_{1}$ versus $\kappa_{12}$  and $\epsilon_{2}$ versus $\kappa_{12}$  
are shown for real $\kappa$ and constant mass in Fig. \ref{fig:ERek}.  Note that $\epsilon_1$ and $\epsilon_2$ depend  on $\kappa_{13}$ as well. However, this variation is very mild, so we do not show it here. It is  clear that both $\epsilon_{1}$ and $\epsilon_{2}$ depend strongly  on $\kappa_{12}$. 
Here the CP asymmetry parameter $\epsilon_{2}$ is very small due to its inverse dependence on $\kappa_{11}$, which is orders of magnitude larger than $K_{11}$. 
For this set of parameters, $\epsilon_{1}$ is negative while $\epsilon_2$ takes positive values, except for a very small fraction of points yielding negative but close to zero $\epsilon_2$.
\noindent
{\bf Case 1b: Real positive $\kappa$ and variable mass}\\

\begin{figure}[H]
	\begin{center}
		\includegraphics[width=2.8 in]{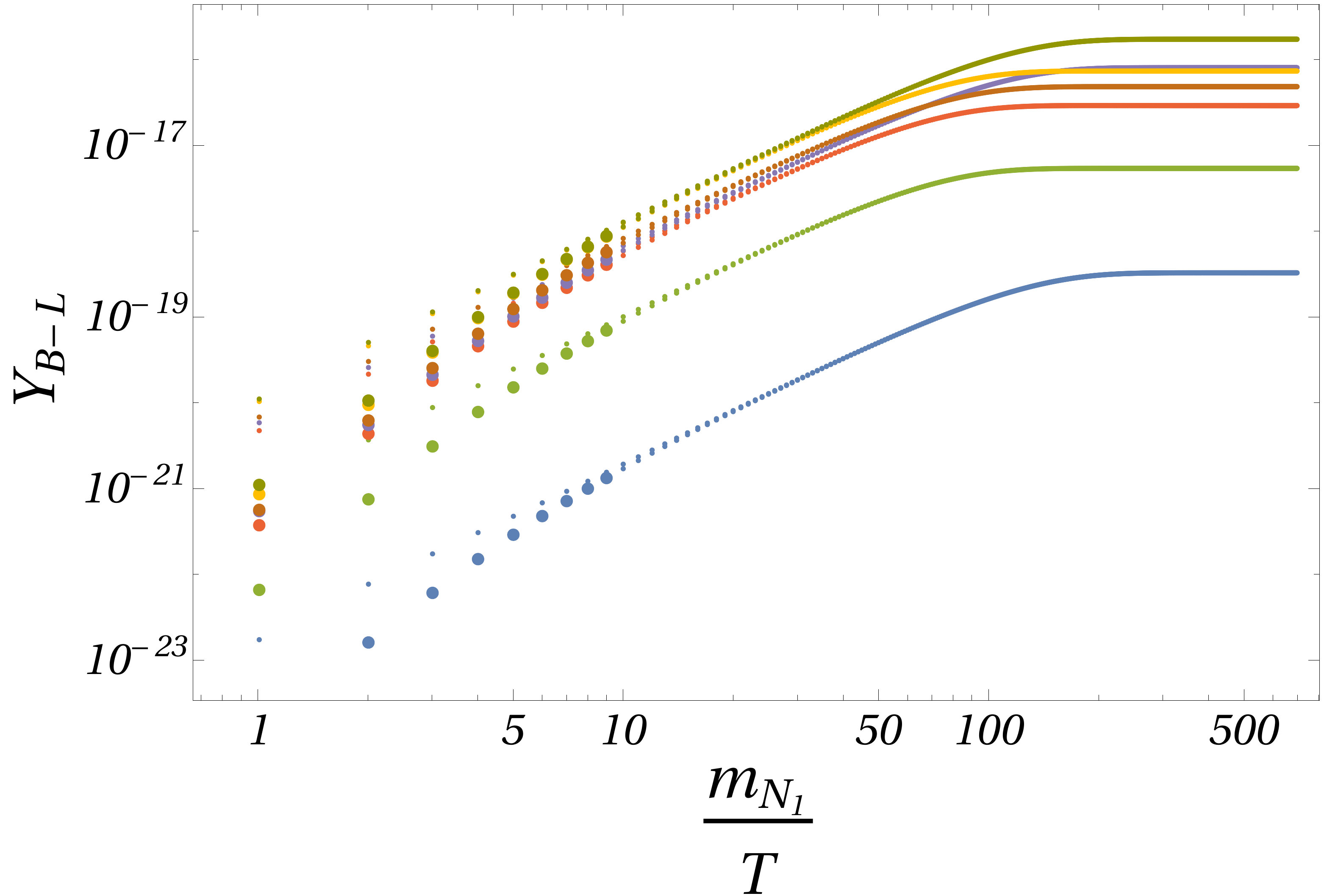}
		\includegraphics[width=2.8 in]{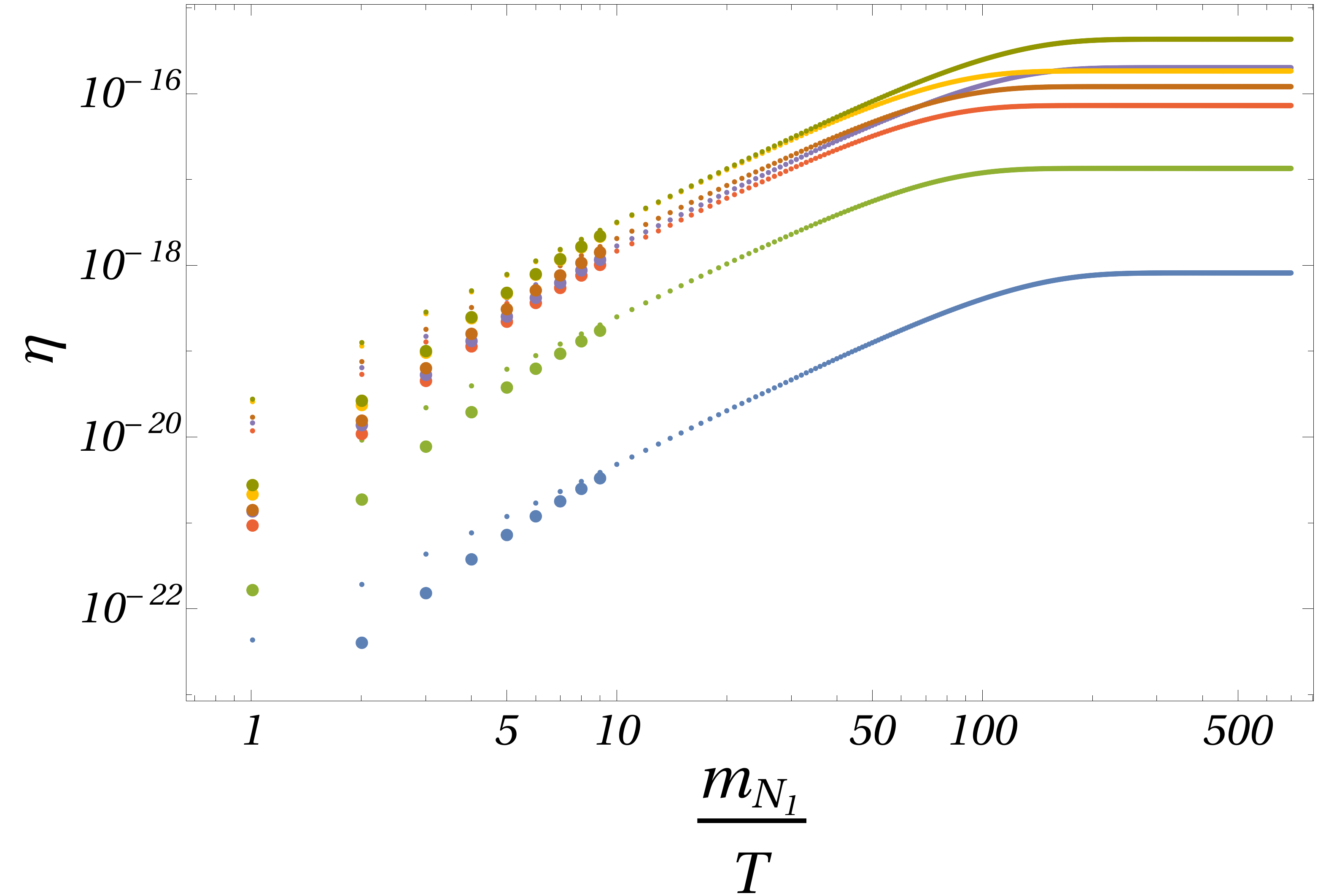}
		\caption{The lepton asymmetry, $Y_{B-L}$ (left) and the baryon asymmetry, $\eta$ (right) versus $z=\frac{m_{N_1}}{T}$ corresponding to real $\kappa$ for selected set of parameter values.  The ten curves in the plots correspond to parameter  values given in Table \ref{tab:etaRek}. }
		\label{fig:YBLetaRek}	
	\end{center}
\end{figure}

We next analyse the case where $\kappa$ is constrained to be real and positive, but the 
right-handed neutrino masses are not fixed.  The mass of the lightest one, $m_{N_1}$ is varied  from 10 TeV to $100$ TeV; the mass $m_{N_2}$ is varied from $10^3$ TeV to $10^4$ TeV and the mass $m_{N_3}$ is varied from $10^5$ TeV to $10^6$ TeV. As we have already seen, the vertex correction does not contribute to the CP asymmetry parameter $\epsilon_{2}$ for real $\kappa_{12}$ and $\kappa_{13}$. Only self-energy corrections yield a non-zero CP asymmetry parameter $\epsilon_{2}$, which is positive, while $\epsilon_1$ is negative. More points with reduced $\epsilon_2$ values appear with increasing $m_{N_2}$. As discussed in the beginning of this section, the CP-asymmetry is not significantly affected by values of $m_{N_3}$ and $\kappa_{13}$.
 The efficiency factors $\zeta_{1}$ and $\zeta_{2}$ also depend on the mass of right-handed neutrinos. 
Fig. \ref{ig:ERekM}	 shows the dependence of $\epsilon_{1}$  and $\epsilon_2$ on $\kappa_{12}$ and on the masses $m_{N_1}$ and $m_{N_2}$.  One can note that, varying the right-handed neutrino masses, $\epsilon_1$ remains negative, while $\epsilon_2$ is positive, showing that this is a robust prediction of real positive values of $\kappa$.  Notice that, compared to the plots for constant mass,  the plots of $\epsilon_1$ and $\epsilon_2$ versus $\kappa_{12}$ become more populated for at smaller values of $\kappa_{12}$,
but  their range does not change much. 
  \begin{figure}[H]
	\begin{center}
		\includegraphics[width=2.2in]{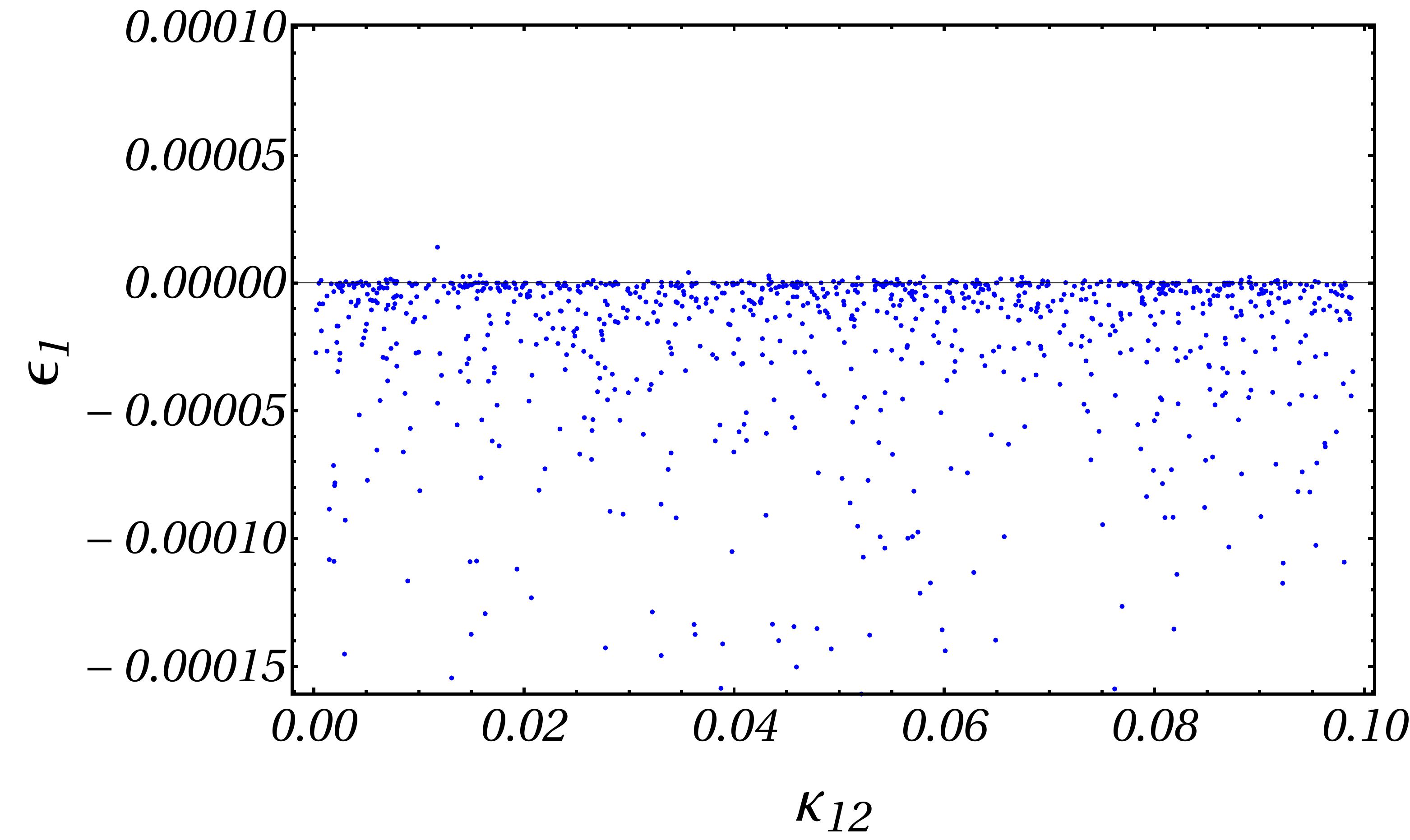}
		\includegraphics[width=2.2in]{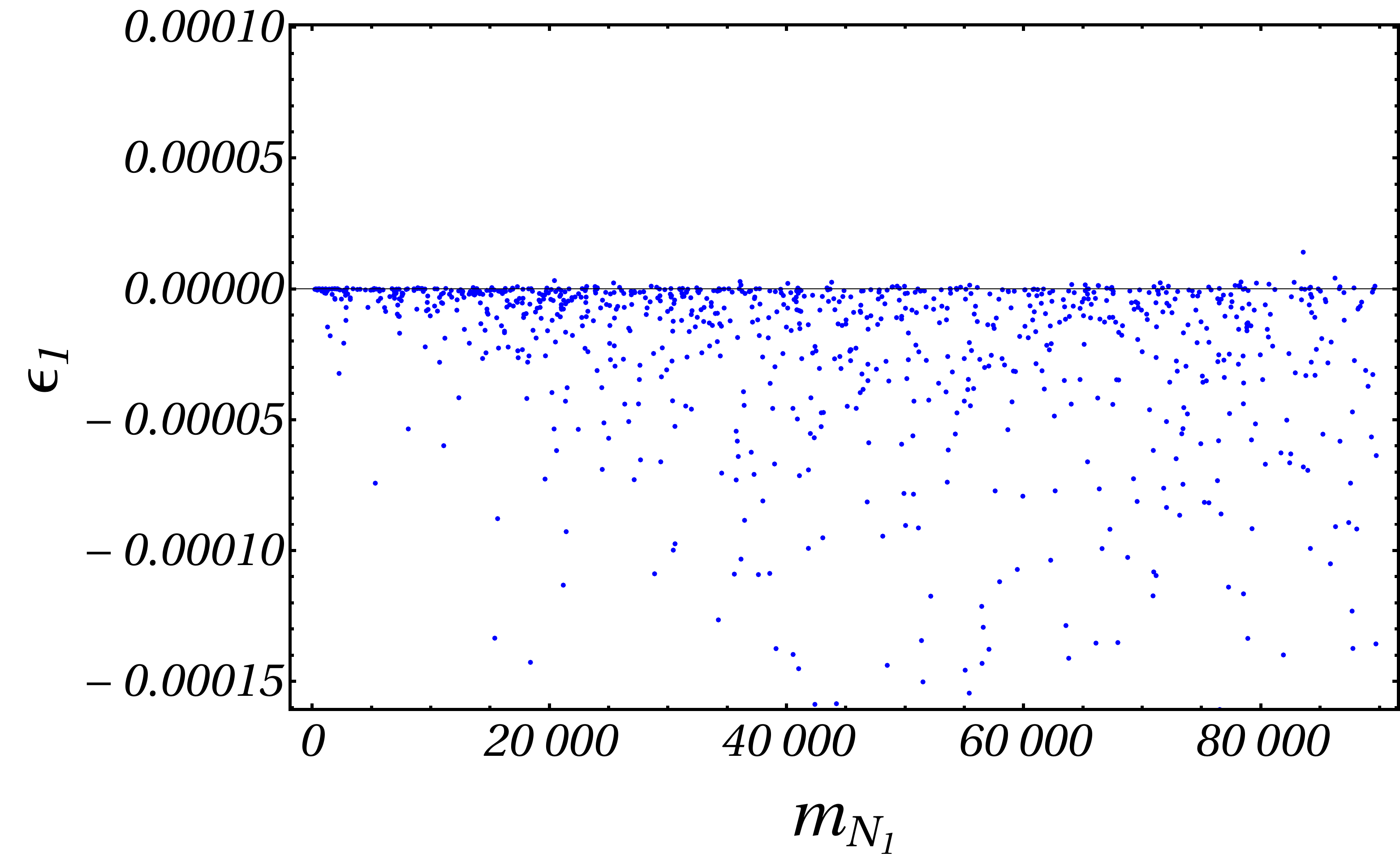}
		\includegraphics[width=2.2in]{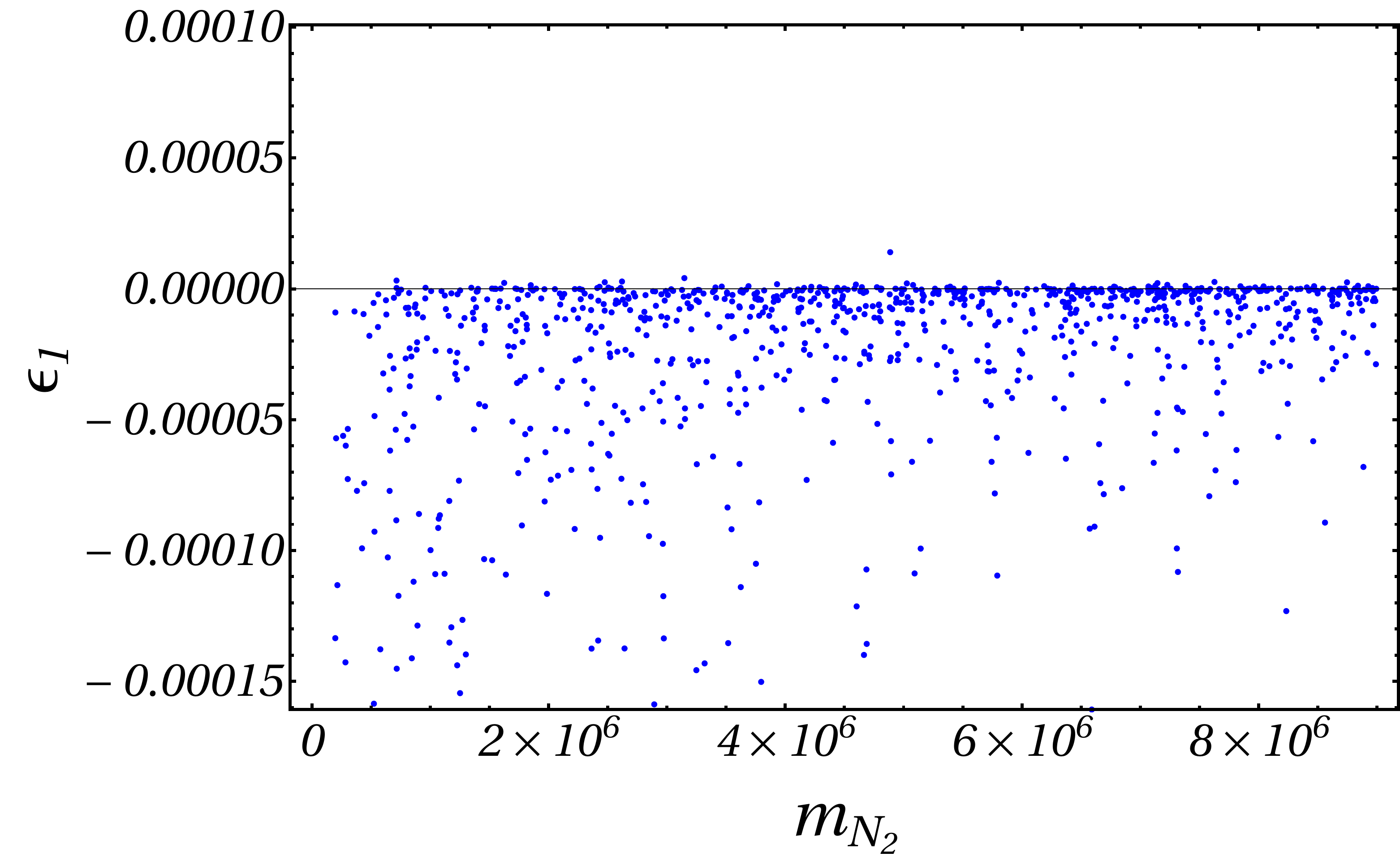}
		\includegraphics[width=2.2in]{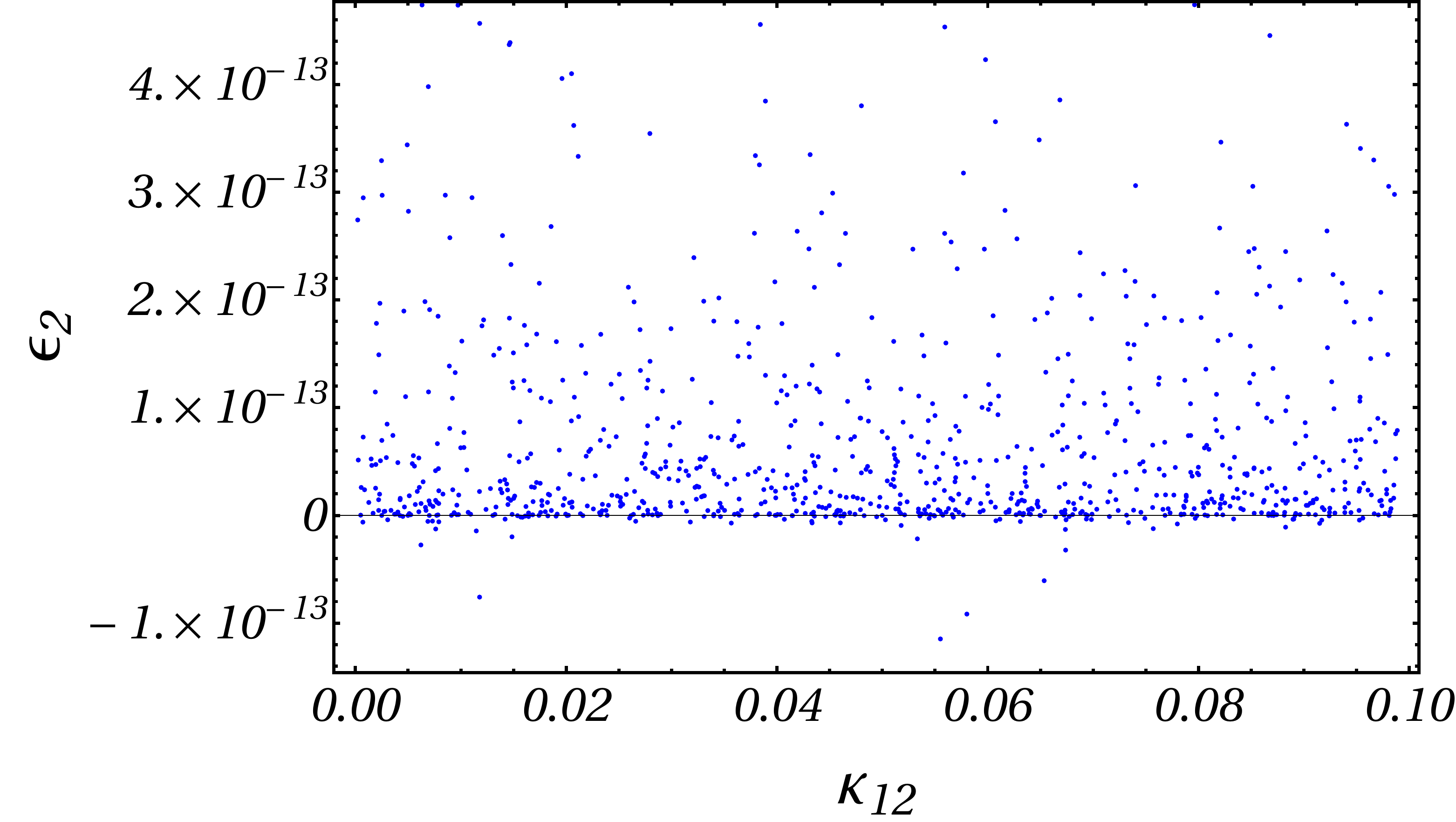}
		\includegraphics[width=2.2in]{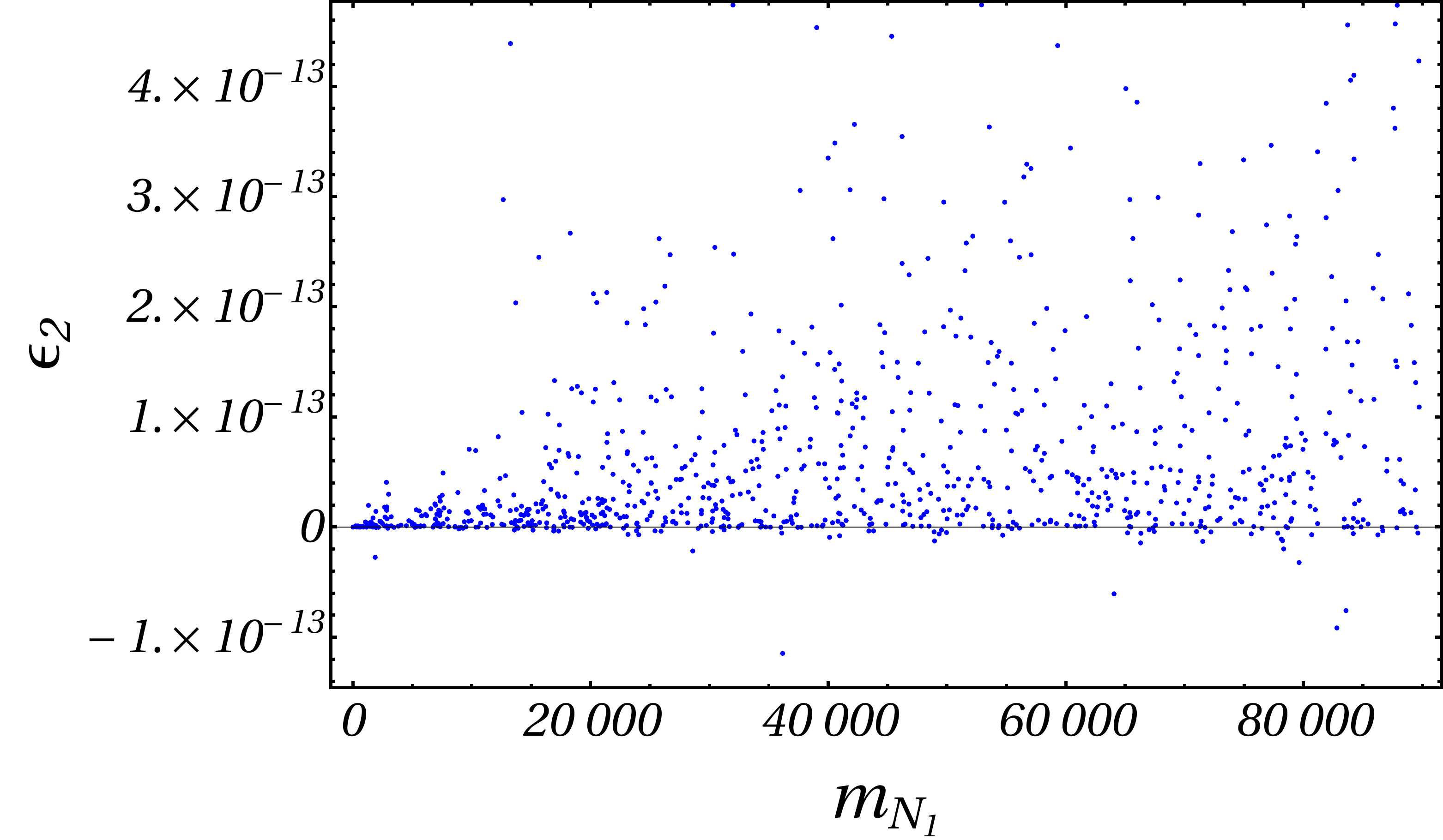}
		\includegraphics[width=2.2in]{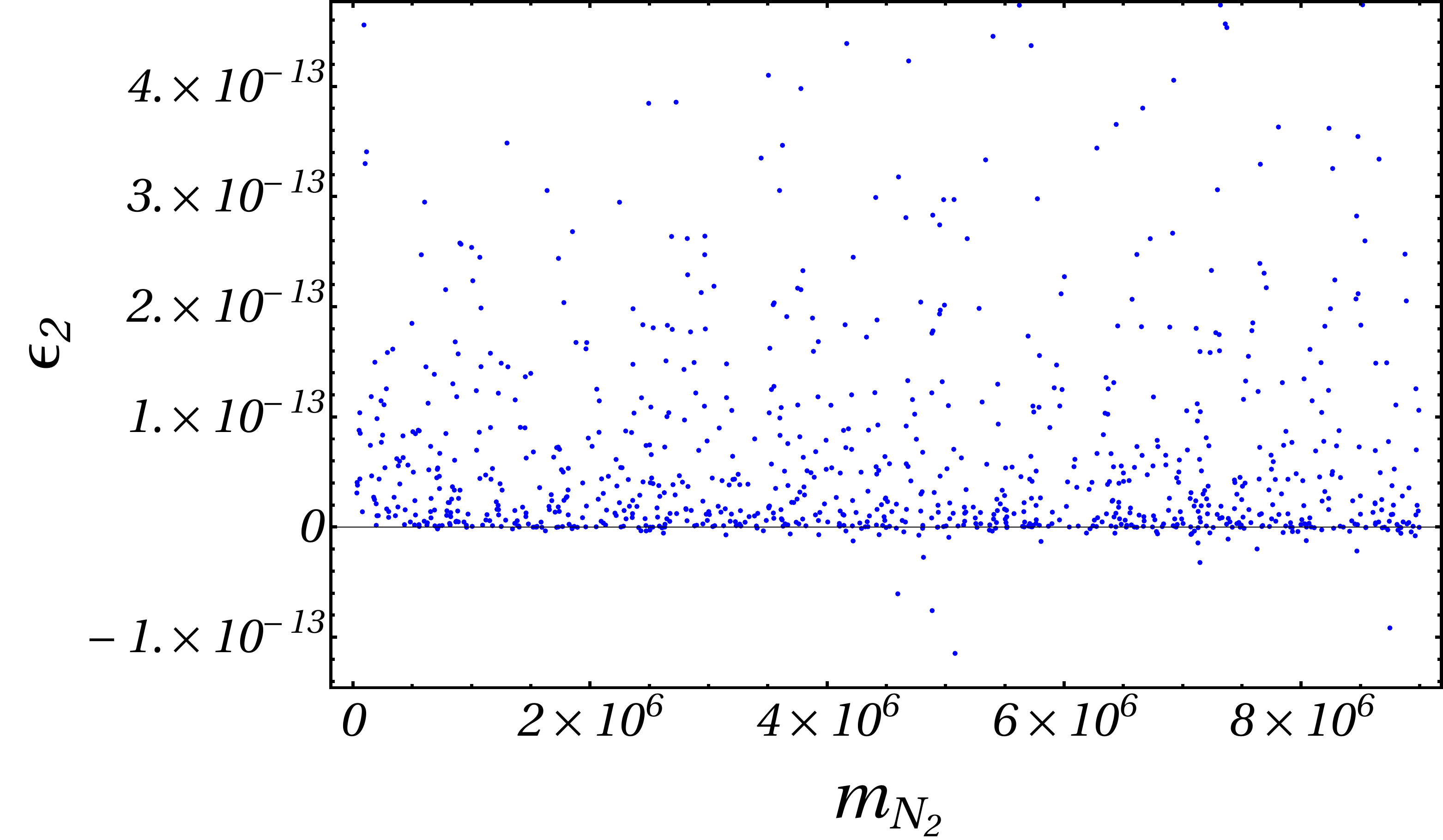}
		\caption{CP-asymmetry against the relevant coupling  $\kappa_{12}$ and the masses, $m_{N_1}$ and $m_{N_2}$ 
		for Case 1b with real $\kappa$ and variable masses $m_{N_j}$. Top panels: Variation of $\epsilon_1$ with $\kappa_{12}$,  $m_{N_1}$ and $m_{N_2}$. Bottom panels: same, but for $\epsilon_2$.  }	
\label{fig:ERekM}	
	\end{center}
\end{figure}
To further elucidate the resulting baryon asymmetry generated in this case, in  Fig. \ref{fig:YBLetaRek} we plot the lepton asymmetry and the corresponding baryon asymmetry $\eta$ against $z=\frac{m_{N_1}}{T}$ for some selected set of parameter values, as given in Table \ref{tab:etaRek}.  Clearly, the obtainable baryon-antibaryon asymmetry is many orders of magnitude smaller than the required value. 

\begin{table}[H]
	\begin{center}
		\small
		\begin{tabular}
			{p{0.6in} |p{0.6in} |p{0.8in} |p{0.8in}}
			\hline \hline
			$\kappa_{11}$ & $\kappa_{12}$& $m_{N_1}$ (TeV)& $m_{N_2} (TeV) $ 
			\\[1mm] \hline 
			0.0243446&$0.0972258$ &$29.8197$ & $6.9371\times 10^3$  \\
			\hline 
			0.0276601&$0.07091$ &$49.8281$ & $5.80265\times 10^3$  \\
			\hline 
			0.0245508&$0.07091$ &$61.8766$ & $4.56419\times 10^3$ \\
			\hline 
			0.0974737&$0.0861245$ &$83.6984$ & $7.58584\times 10^3$  \\
			\hline 
			0.0132375&$0.0867719$ &$19.5765,$ & $7.37307\times 10^3$  \\
			\hline 
			0.0217837&$0.089865$ &$75.7554$ & $1.4591\times 10^3$  \\
			\hline 
			0.0247151&$0.0218281$ &$19.7915$ & $7.84946\times 10^3$  \\
			\hline 
			0.0261621&$	0.00796258,$ &$88.6163$ & $1.06065\times
			10^3$  \\
			\hline	
			0.0305289&$0.00796258$ &$11.3217$ & $8.48216\times 10^3$  \\
			\hline	0.0247151&$0.0891052$ &$73.3373$ & $7.8375\times 10^3$  \\
			\hline \hline
		\end{tabular}
		\caption{Parameter values for the $Y_{B-L}$ and $\eta$ plots in Fig. \ref{fig:YBLetaRek}.}
	\label{tab:etaRek}
	\end{center}
\end{table}

\noindent
{\bf Case 2a: Real Negative $\kappa$ and Constant Mass}\\[5mm]

\begin{figure}[H]
	\begin{center}
		\includegraphics[width=2.80in]{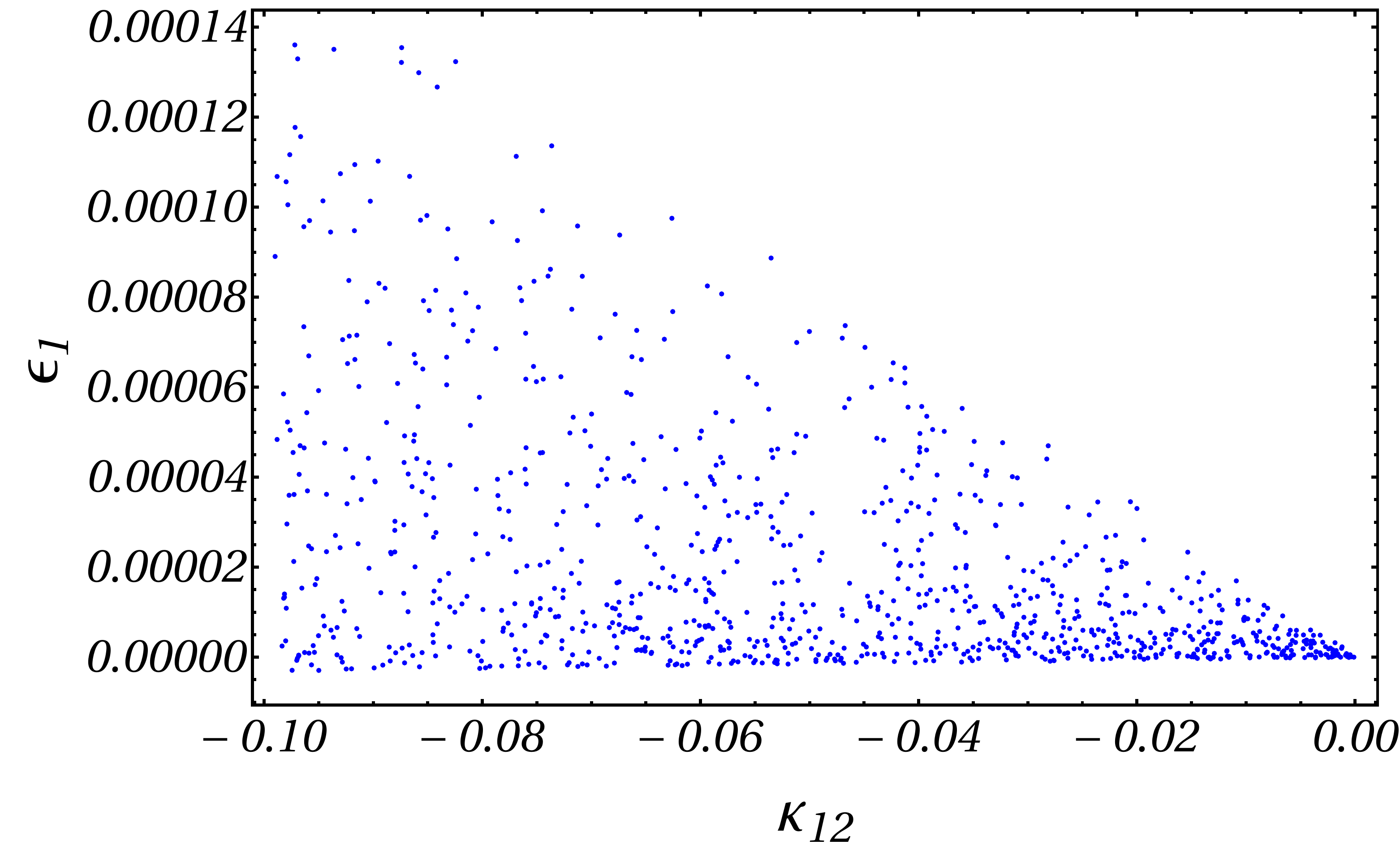}
		\includegraphics[width=3.0in]{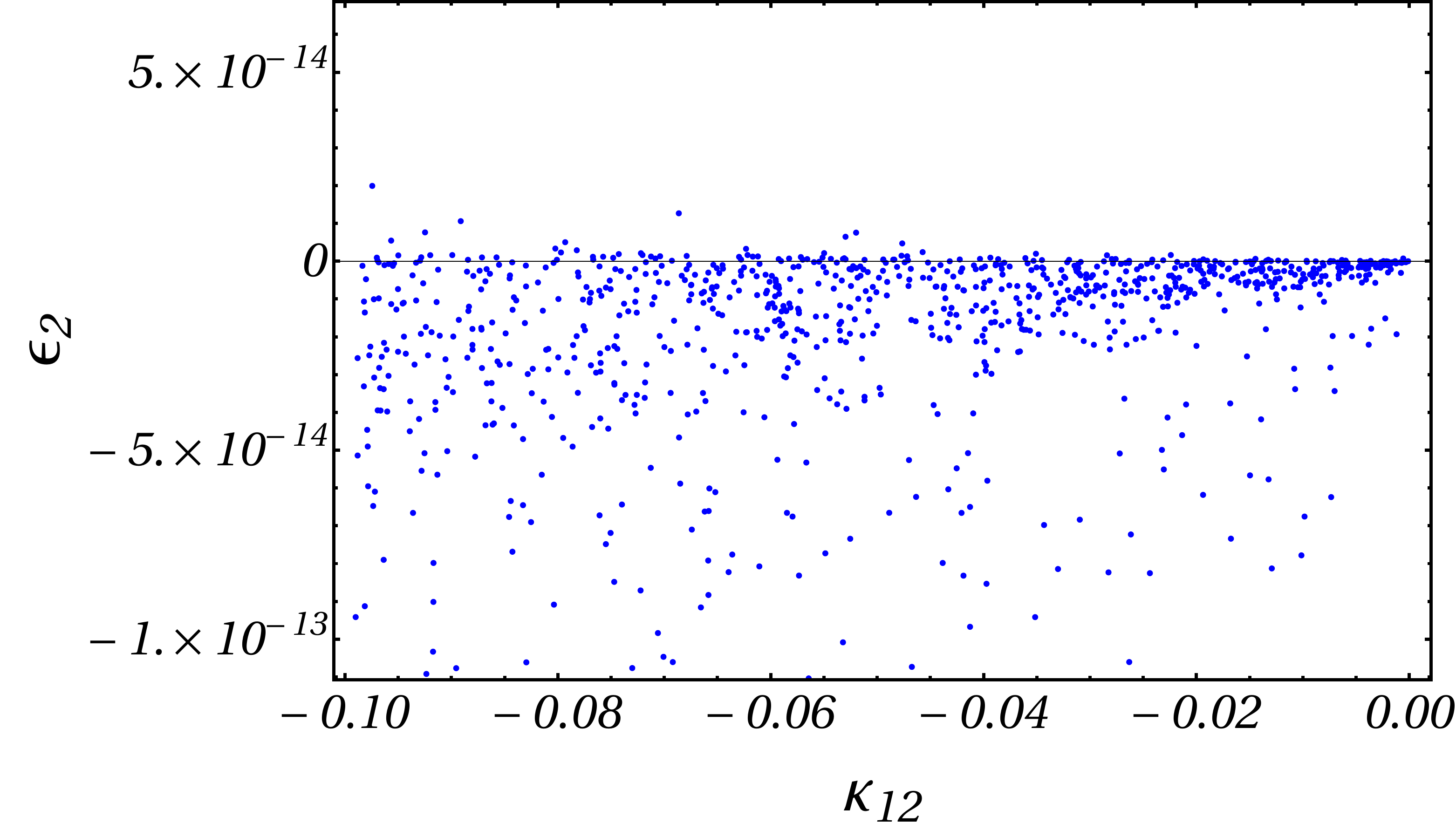}
		\caption{CP-asymmetry parameters $\epsilon_1$ (left) and $\epsilon_2$ (right) versus $\kappa_{12}$  for Case 2a, for real negative  $\kappa$ values and constant right-handed neutrino masses, $m_{N_j}$,  as given in Table \ref{tab:summary}. }	
		\label{fig:EReNk}
	\end{center}
\end{figure}

For negative values of $\kappa_{1j}$, the standard channel yields positive contributions ($\epsilon_1$) to the CP-asymmetry, while the contribution due to the new channels ($\epsilon_2$) is negative, as is clear from Eq. ~\ref{eq:epsphinew} and Eq.~\ref{eq:epsSr}.  However, $\epsilon_2$ being many orders smaller compared to $\epsilon_1$, the overall effect is such that the sign of $\eta$ is negative, indicating a non-acceptable case of excess of anti-baryons. These are plotted in Fig.~\ref{fig:EReNk}.
However, we notice that with the large uncertainty in the measurement of $\delta_{CP}$ (Eq.~\ref{numixing}), the sign of $K_{12}$ is flipped for the lower range of $\delta_{CP}$ compared to the central value and upper range. In particular, if we consider $\delta_{CP}< 180^\circ$, we get the correct sign for $\epsilon_1$.  Still, even in that case, the numerical values of the CP asymmetry still are too small to provide the required leptogenesis.

\vskip0.1in
\noindent
{\bf Case 2b: Real negative $\kappa$ and variable mass}\\[5mm]
Varying the mass does not improve the situation, as the sign of $\epsilon_1$ remain positive, and $\epsilon_2$ remains in the same range for the entire 
span of mass range considered.

\vskip0.1in
\noindent
{\bf Case 3a: Imaginary $\kappa$ and constant mass}\\[5mm]

\begin{figure}[H]
	\begin{center}
		\includegraphics[width=2.80in]{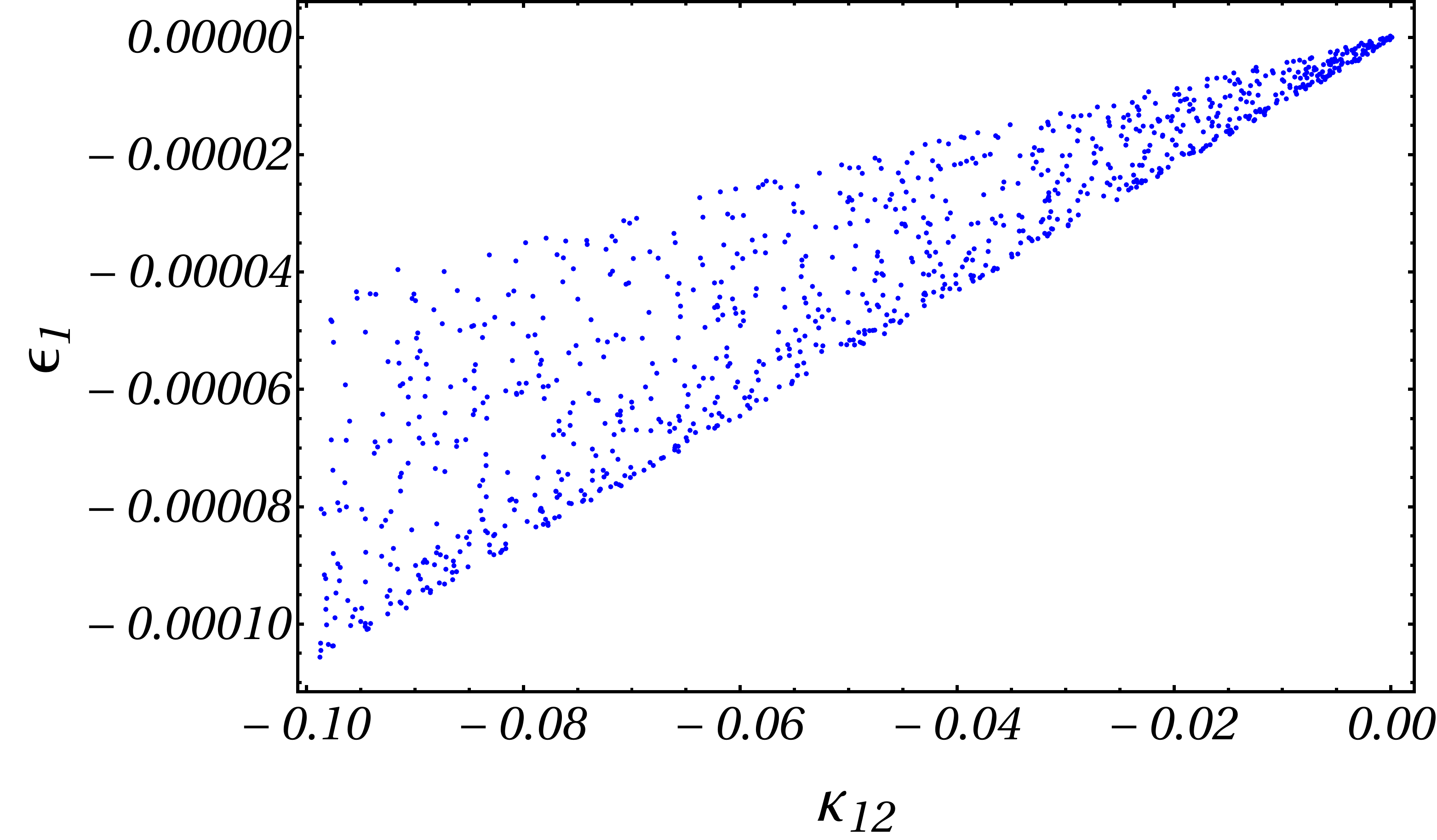}
		\includegraphics[width=2.80in]{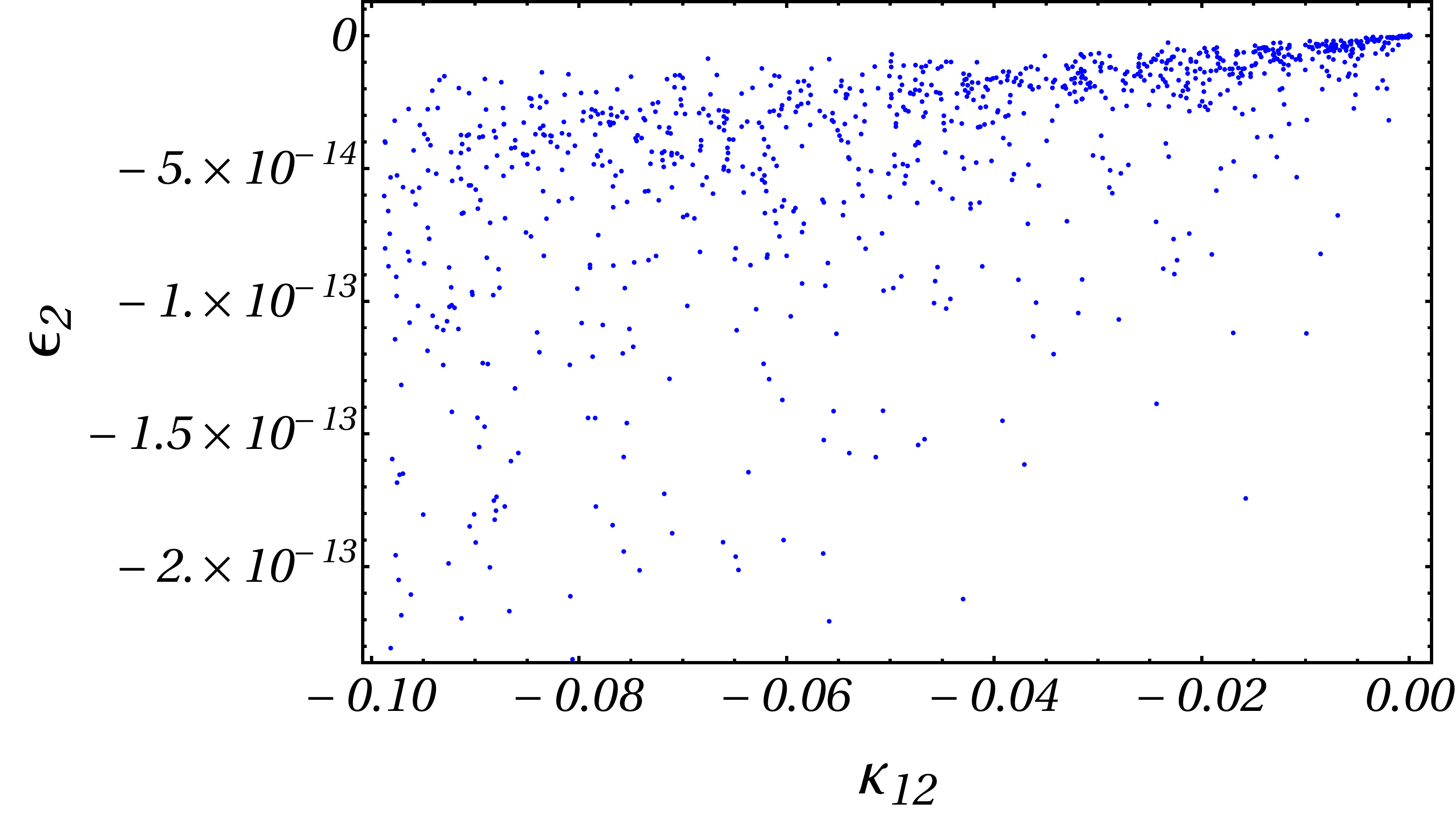}
		\caption{CP-asymmetry parameters $\epsilon_1$ (left) and $\epsilon_2$ versus $\kappa_{12}$ (right)  for Case 3a, with imaginary values for $\kappa_{12}$ and constant right-handed neutrino masses, $m_{N_j}$,  as given in Table \ref{tab:summary}.}		
		\label{fig:EImk}
	\end{center}
\end{figure}

We now study the case where $\kappa_{12}$ and $\kappa_{13}$ are purely imaginary. Since $\epsilon_{1}$ and $\epsilon_{2}$   depend more sensitively on $\kappa_{12}$ than on any other $\kappa_{ij}$'s,  we focus on variations with $\kappa_{12}$. The imaginary part of $K_{12}$ plays the role of the phase factor for real positive $\kappa_{12}$ and real negative $\kappa_{12}$, but in this case the real part of $K_{12}$ also contributes to the asymmetry. The order of the real and imaginary parts of $K_{12}$ are approximately same. Unlike the case where $\kappa_{12}$ was real, here $\epsilon_{1}$ and $\epsilon_{2}$ have the same sign, meaning that in this case their  numerical values are negative for  imaginary values of $\kappa_{12}$ and $\kappa_{13}$. The numerical values of $\epsilon_{2}$ are very small in comparison to $\epsilon_{1}$, so $\epsilon_{1}$ and the efficiency factor $\zeta_{1}$ are able to give required matter antimatter asymmetry. Scan plots of $\epsilon_{1}$ versus $\kappa_{12}$ and  $\epsilon_{2}$ versus $\kappa_{12}$  are shown in Fig. \ref{fig:EImk}.

\noindent
{\bf Case 3b: Imaginary $\kappa$ and variable mass}\\[5mm]

\begin{figure}[H]
	\begin{center}
		\includegraphics[width=2.2in]{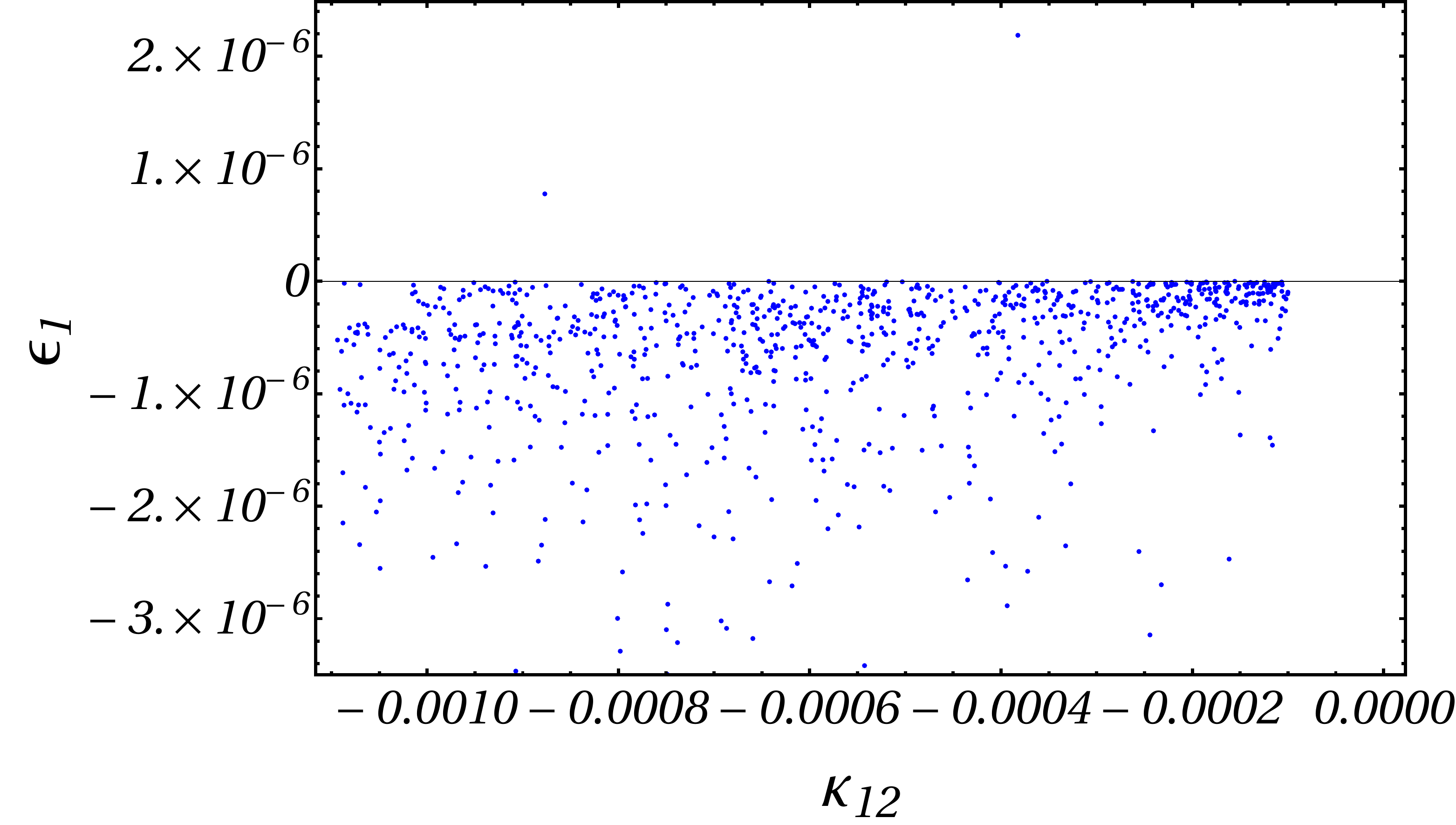}
		\includegraphics[width=2.2in]{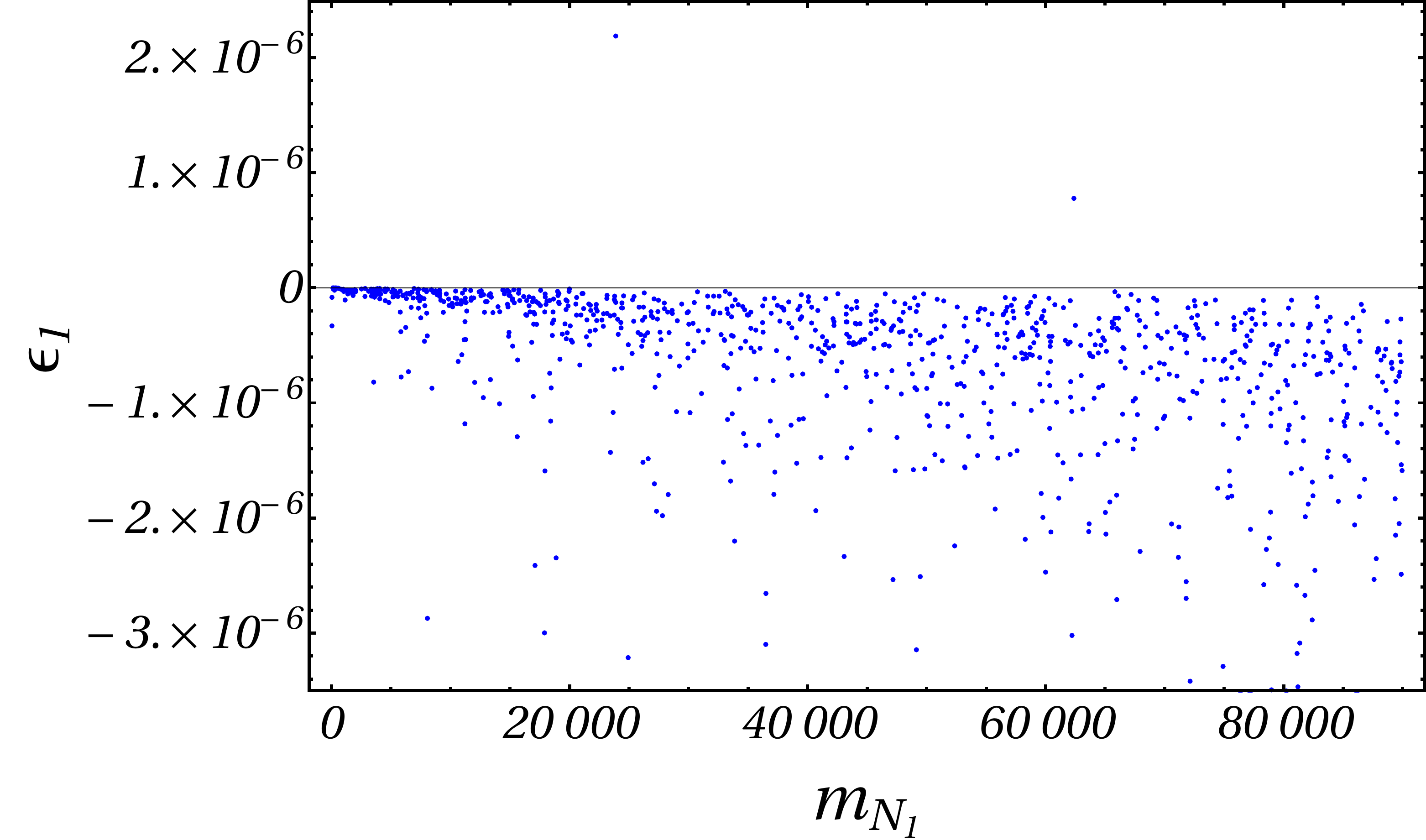}
		\includegraphics[width=2.2in]{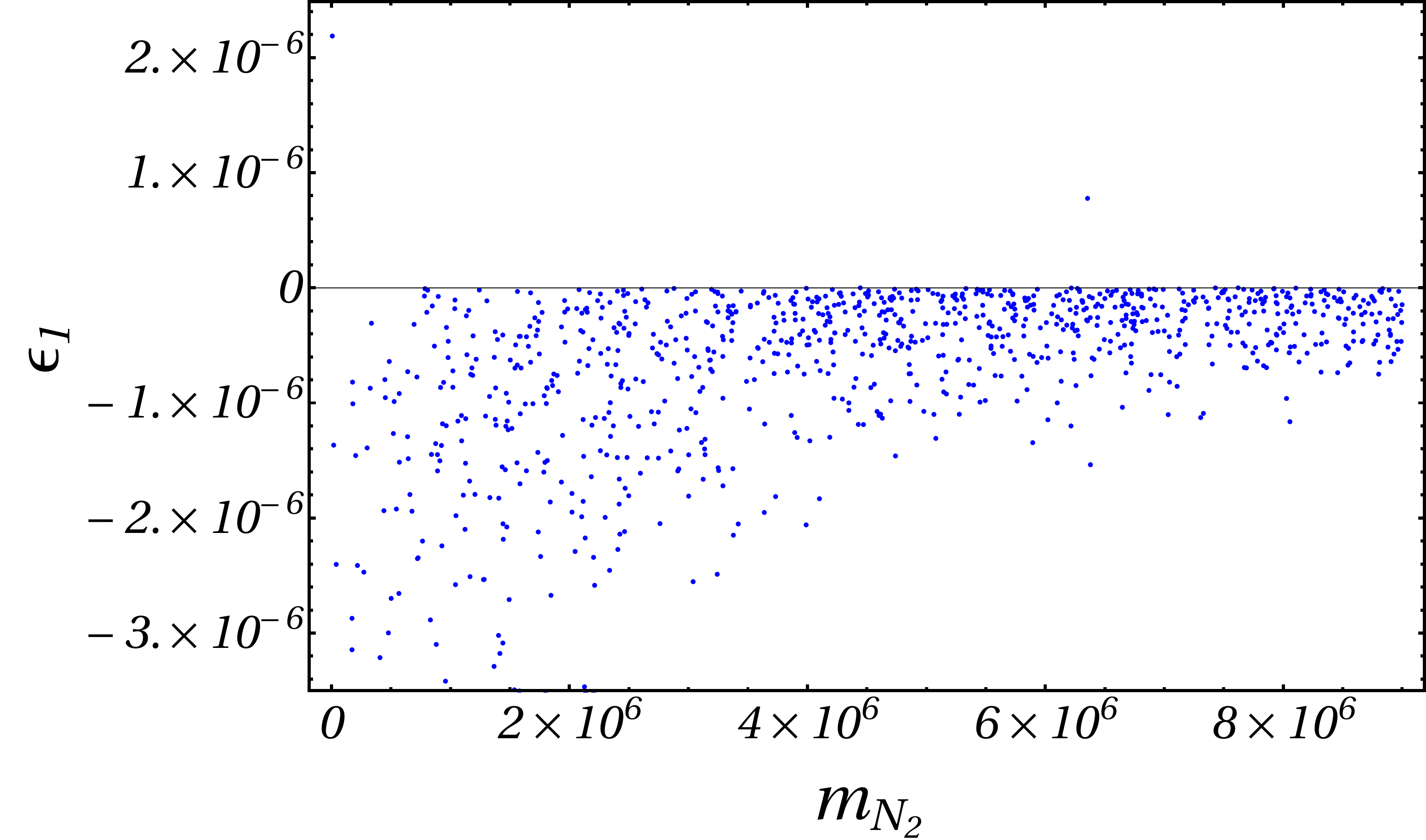}
		\includegraphics[width=2.2in]{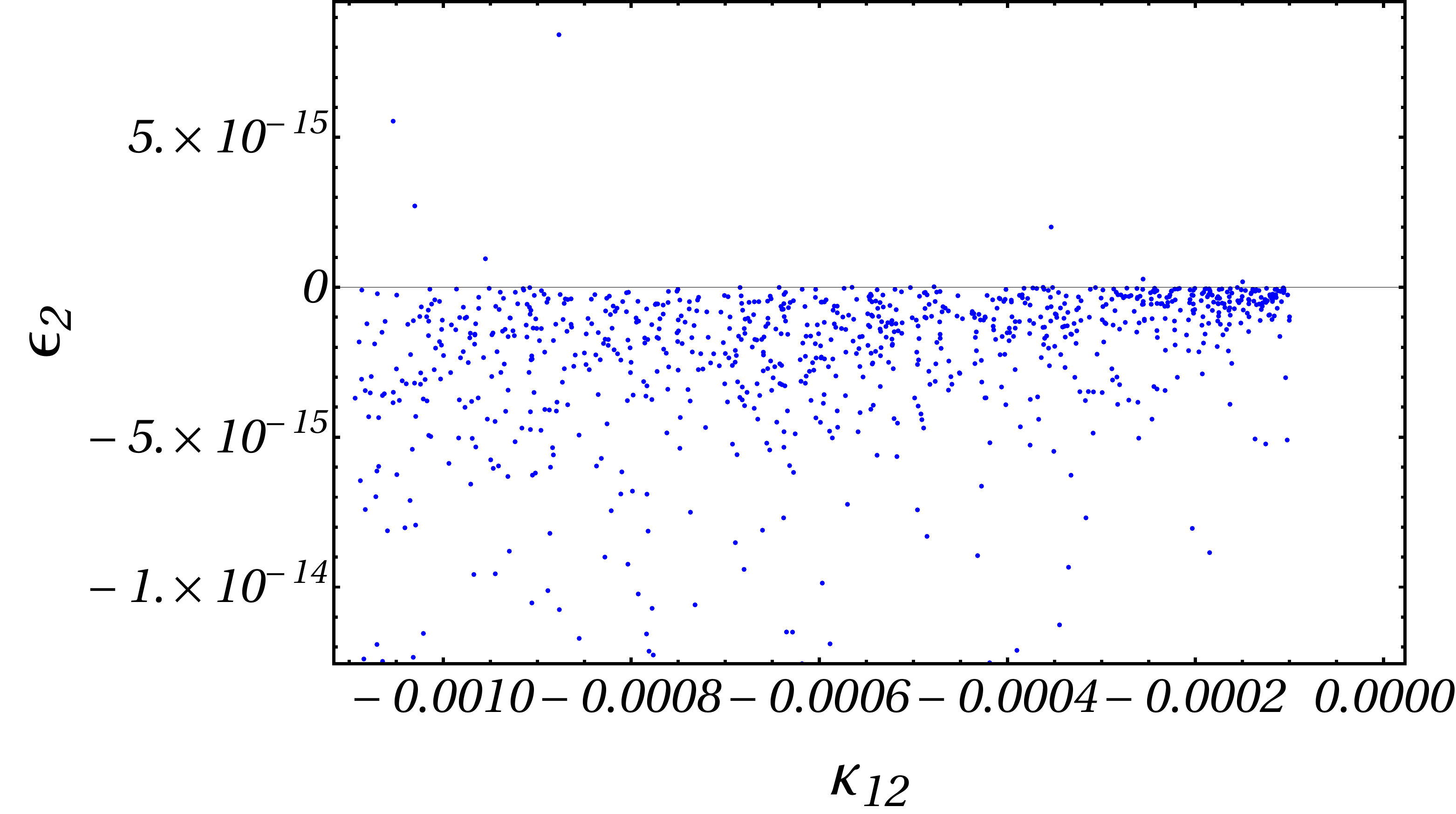}
		\includegraphics[width=2.2in]{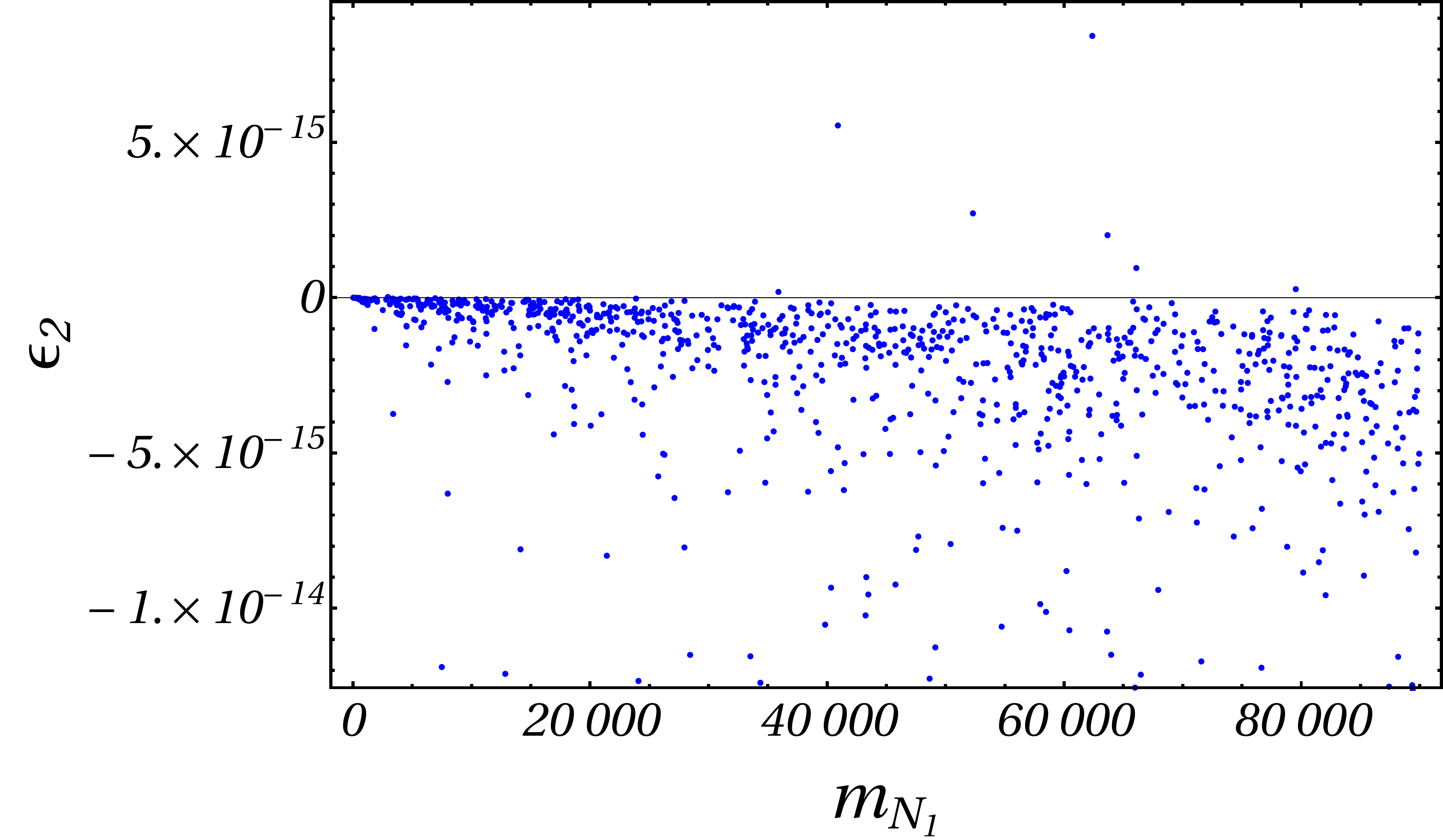}
		\includegraphics[width=2.2in]{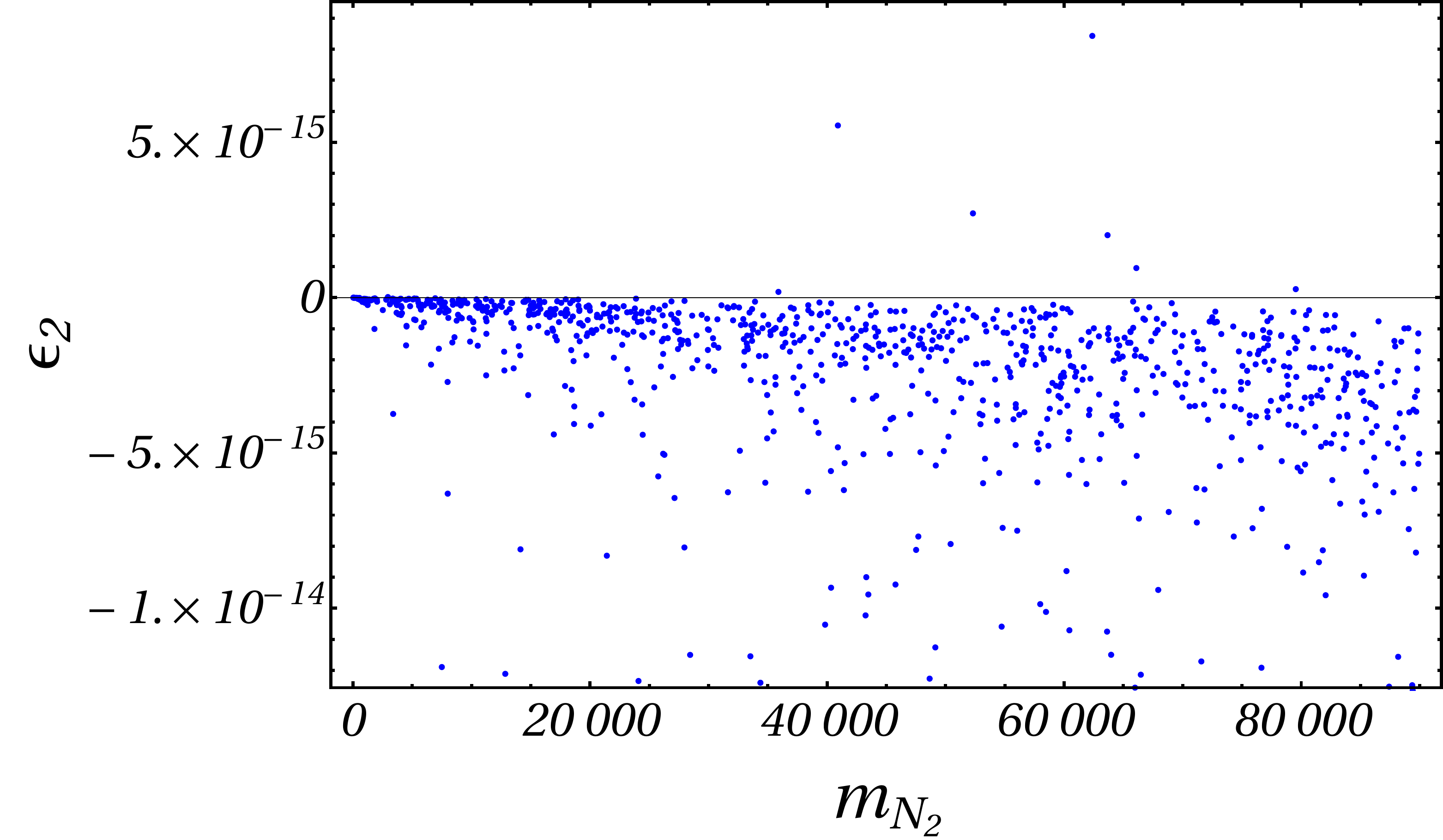}
		\caption{CP-asymmetry against the relevant coupling combination $\kappa_{12}$ and right-handed neutrino masses, $m_{N_1}$ and $m_{N_2}$ for $\epsilon_1$ (top panels) and $\epsilon_2$ (bottom panels), 
		for Case 3b, with imaginary values for $\kappa_{12}$ and varying $m_{N_j}$.}
		\label{fig:EImkM}
	\end{center}
\end{figure}

\begin{figure}[H]
	\begin{center}
		\includegraphics[width=2.6in]{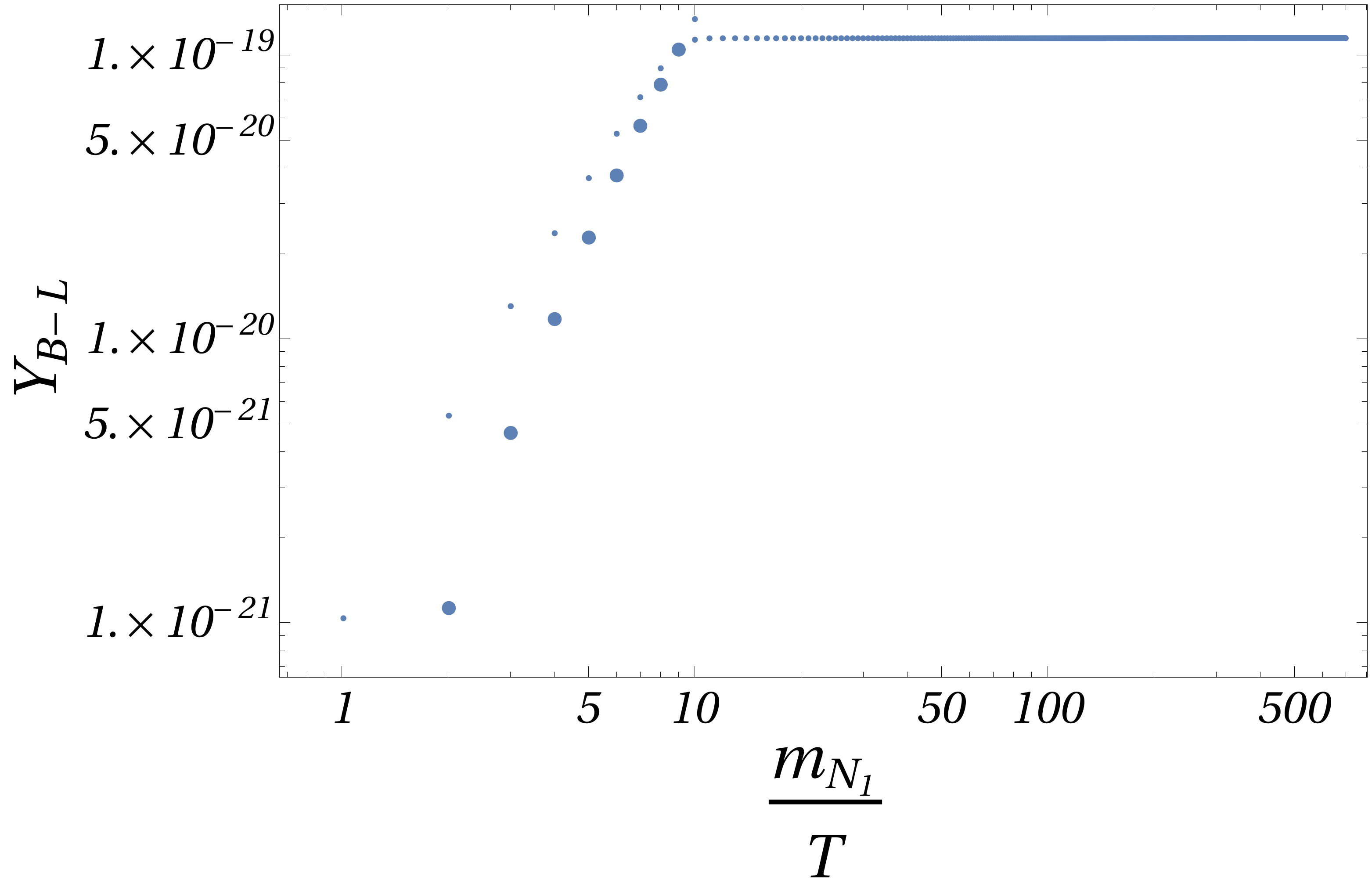}
		\includegraphics[width=2.6in]{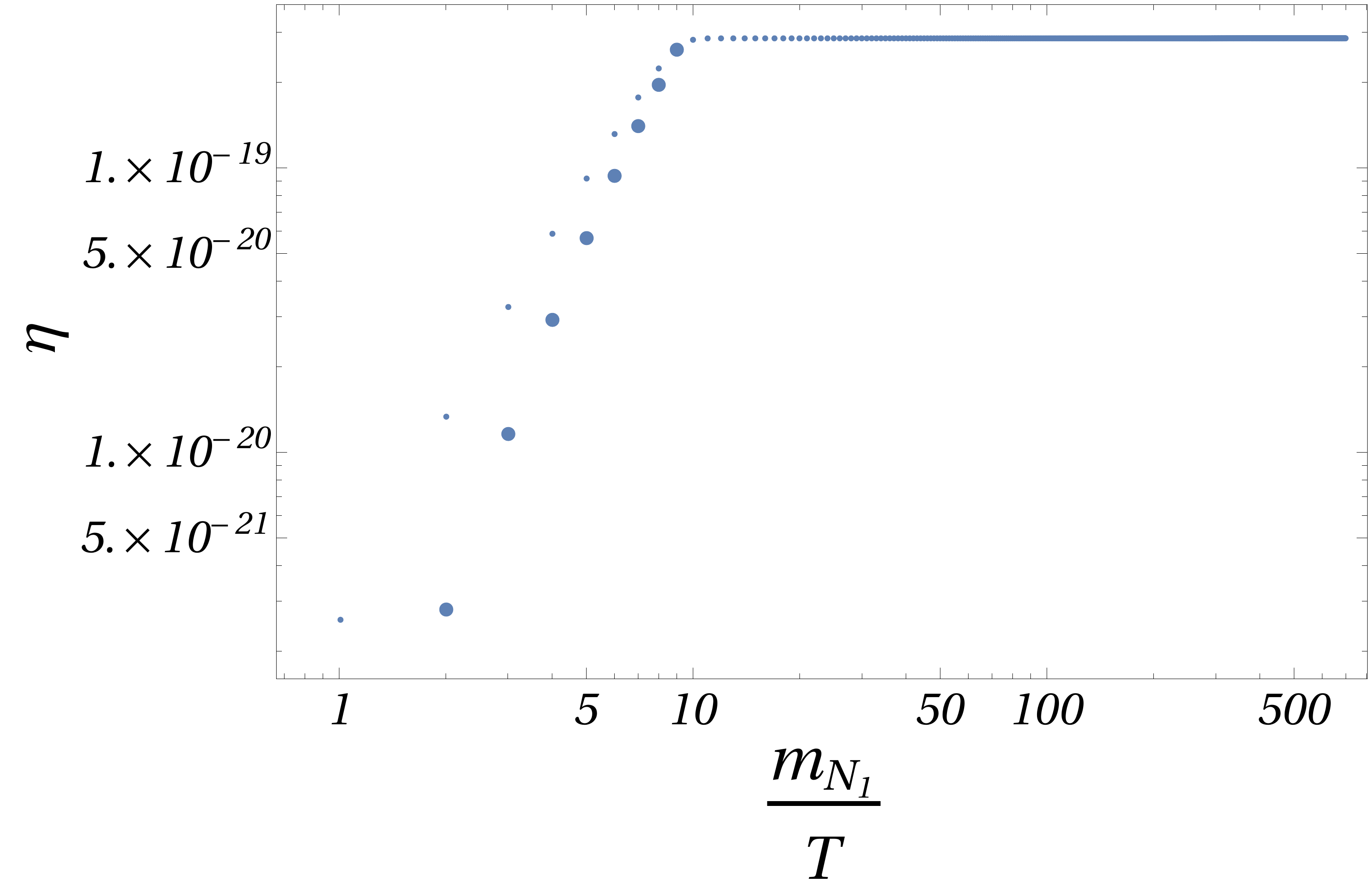}
		\caption{The lepton asymmetry, $Y_{B-L}$ (left) and the baryon asymmetry, $\eta$ (right) versus $z=\frac{m_{N_1}}{T}$ corresponding to imaginary $\kappa$ with parameter values as in Table \ref{tab:EImk}. }
		\label{fig:YBLetaImk}
	\end{center}
\end{figure}

The values for $\epsilon_{1}$ and $\epsilon_{2}$ for imaginary $\kappa_{12}$, $\kappa_{13}$ with variable mass are  similar to the case where  $\kappa_{12}$, $\kappa_{13}$ are imaginary and the right-handed neutrinos have constant mass.  Just as for the case where $\kappa$ is real, varying neutrino masses populates the plot of $\epsilon_1$ versus $\kappa_{12}$ for $\epsilon_1$ values closer to 0, while affecting $\epsilon_2$ somewhat less. The $\epsilon_1$ and $\epsilon_{2}$ are negative for this setup of parameters, and numerical values of $\epsilon_2$ are much smaller than those of $\epsilon_{1}$. Scan plots for $\epsilon_1$ and $\epsilon_2$ as functions of $\kappa_{12}$ and right-handed neutrino masses are show in Fig. \ref{fig:EImkM}. 
Different sets of $\kappa_{11}$, $\kappa_{12}$, and $\kappa_{13}$ give the different solutions for Boltzmann equations. In Fig. \ref{fig:YBLetaImk}	 we show the evolution of $Y_{B-L}$ and $\eta$ for selected sets of parameter values, as given in Table \ref{tab:EImk}. As for the case of real $\kappa$, it is not possible to generate the required BAU in this case.

\begin{table}[H]
	\begin{center}
		\small
		\begin{tabular}
			{ p{0.6in} |p{0.8in} |p{0.8in} |p{0.8in} }
			\hline \hline
			$\kappa_{11}$ & $\kappa_{12}$& $m_{N_1}$ (TeV)& $m_{N_2} (TeV) $ 
			\\[1mm] \hline 
			0.0848485&$-0.000631741 \iota$ &$41.8584$ & $3.20559\times 10^3$  \\
			[1mm]\hline
			0.0260666&$-0.000762459 \iota $ &$16.3979$ & $1.46463\times 10^3$ \\
			\hline \hline
		\end{tabular}
		\caption{Parameter values for the $Y_{B-L}$ and $\eta$ plots in Fig. \ref{fig:YBLetaImk}.}
		\label{tab:EImk}
	\end{center}
\end{table}

\noindent
{\bf Case 4a: Complex $\kappa$ and constant mass}\\[5mm]

\begin{figure}[H]
	\begin{center}
		\includegraphics[width=1.60in]{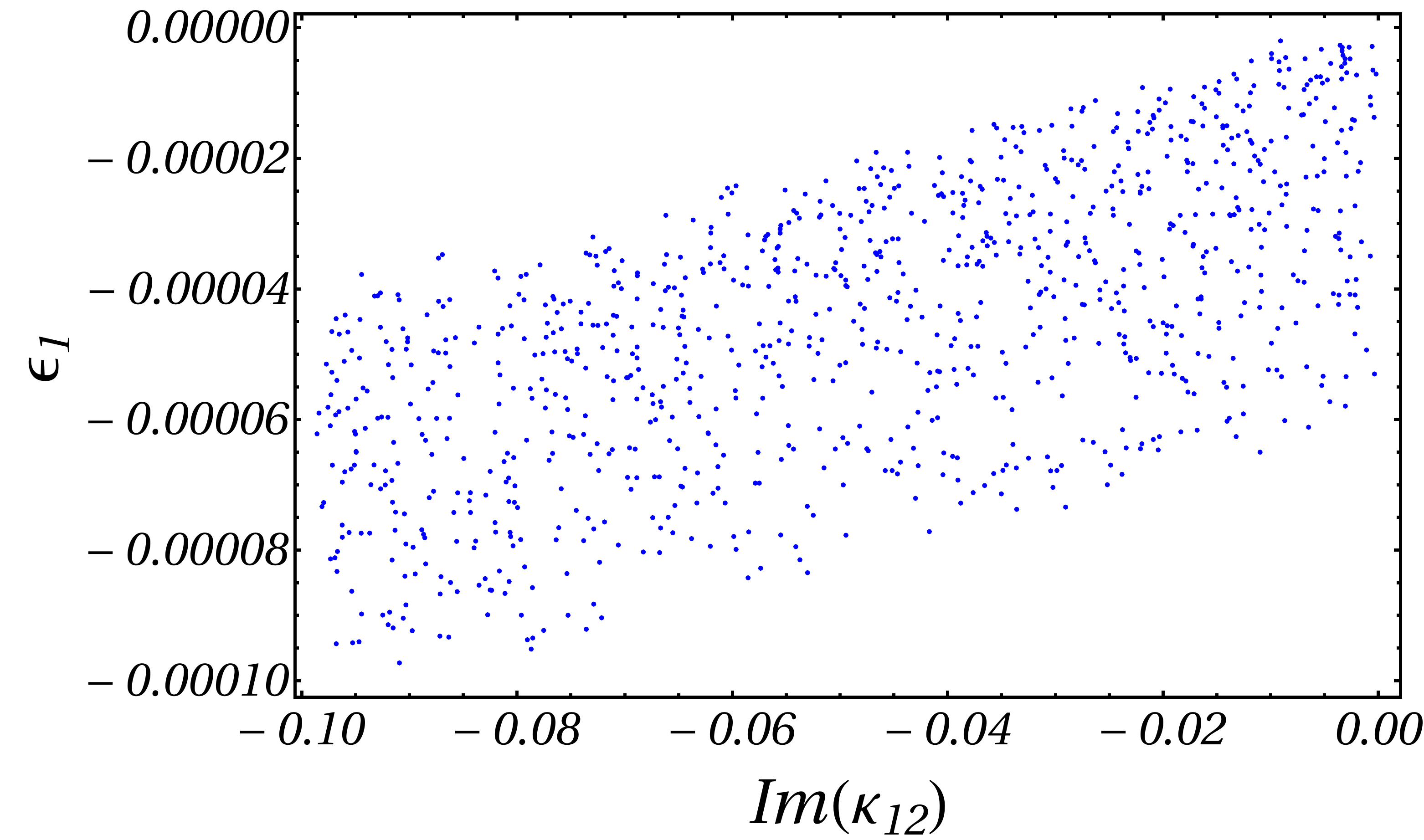}
		\includegraphics[width=1.60in]{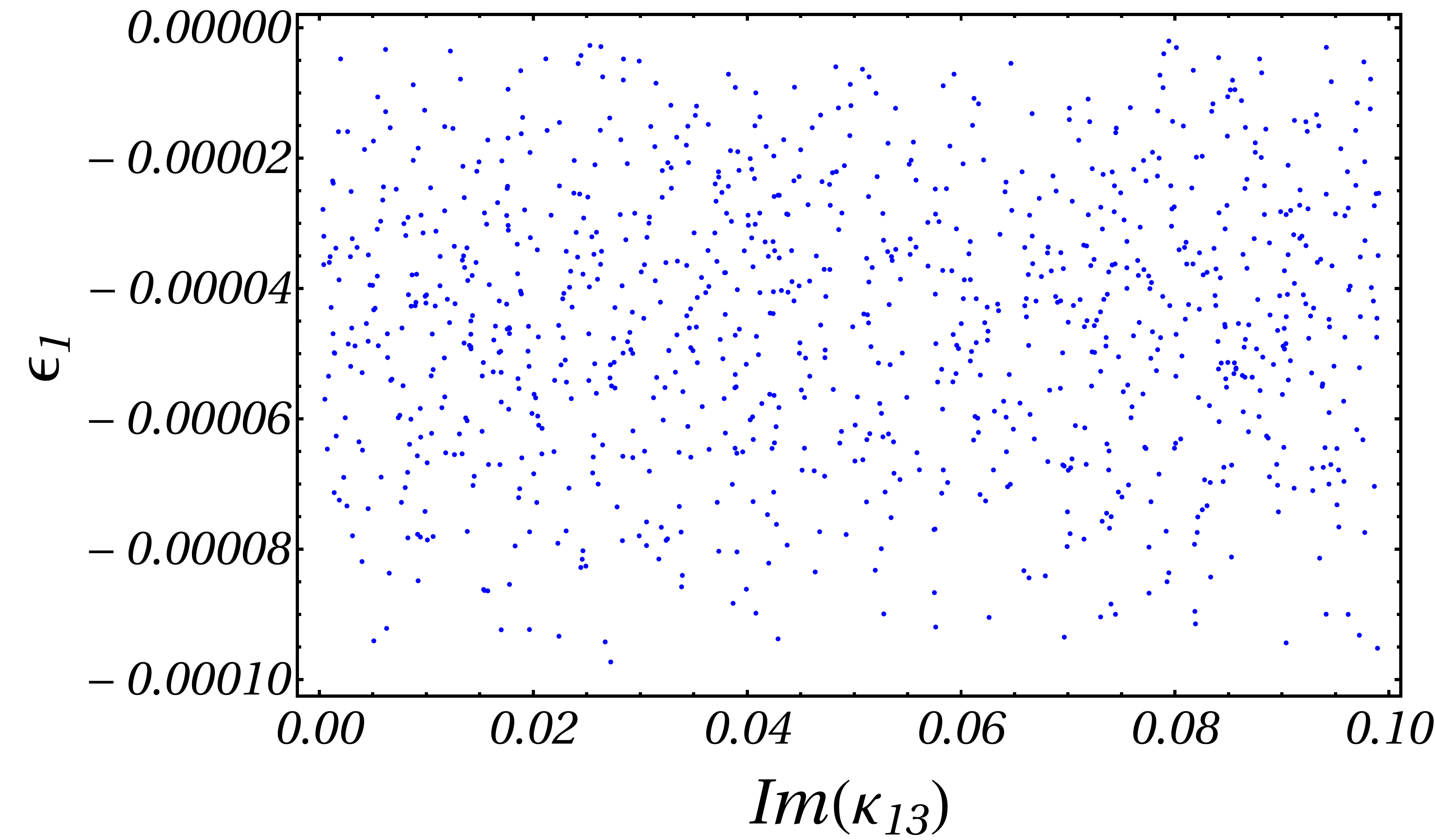}
		\includegraphics[width=1.60in]{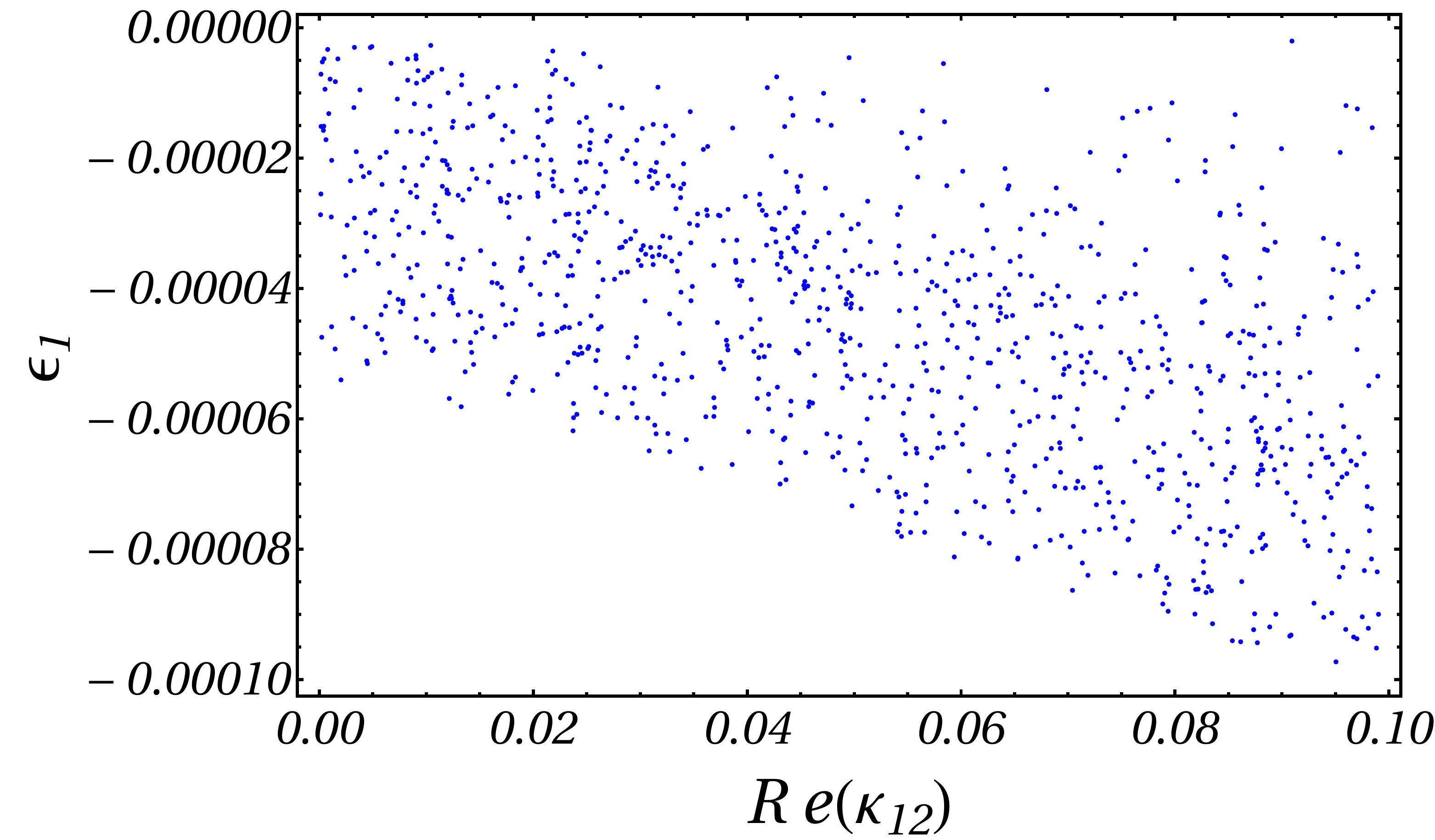}
		\includegraphics[width=1.60in]{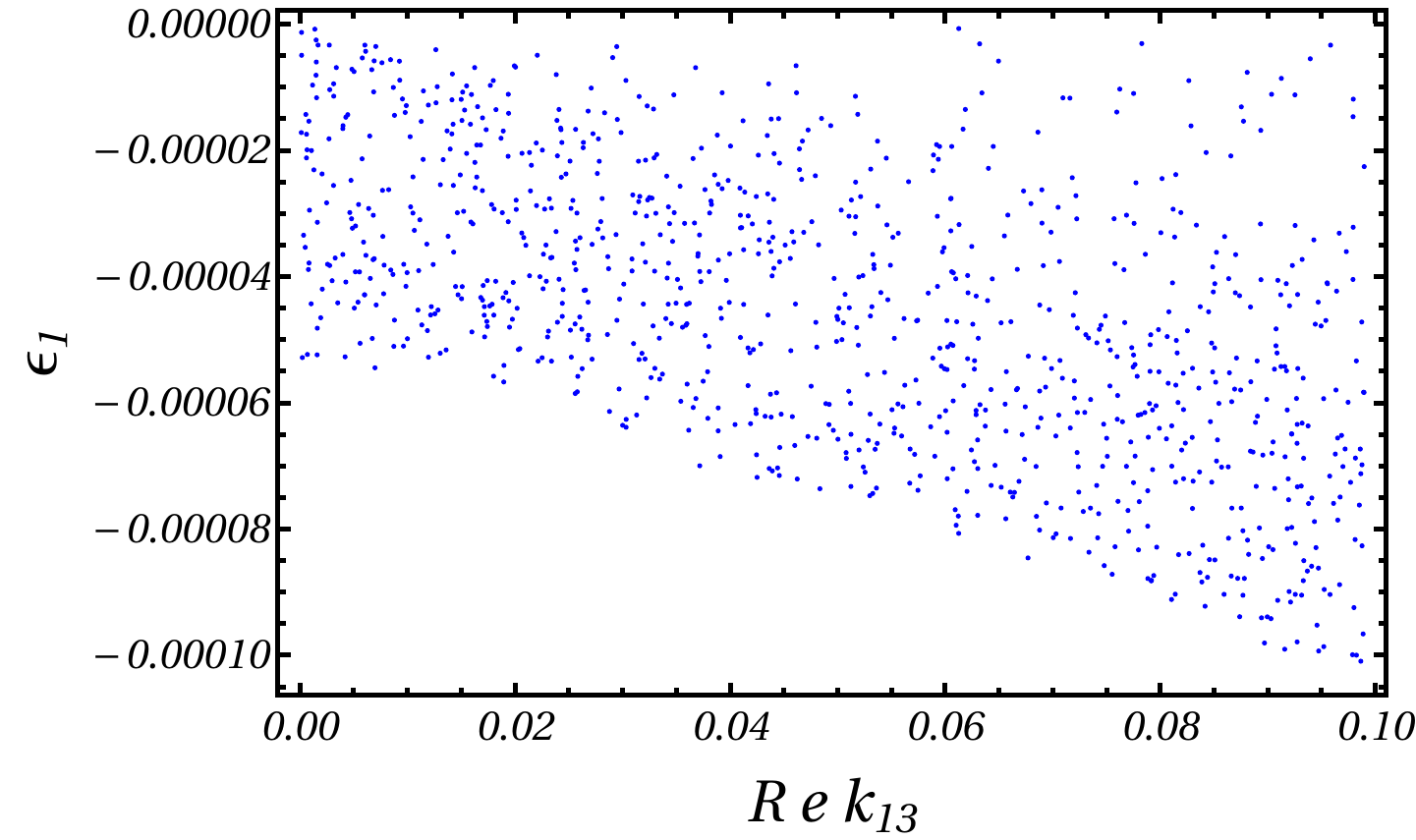}
           \includegraphics[width=1.60in]{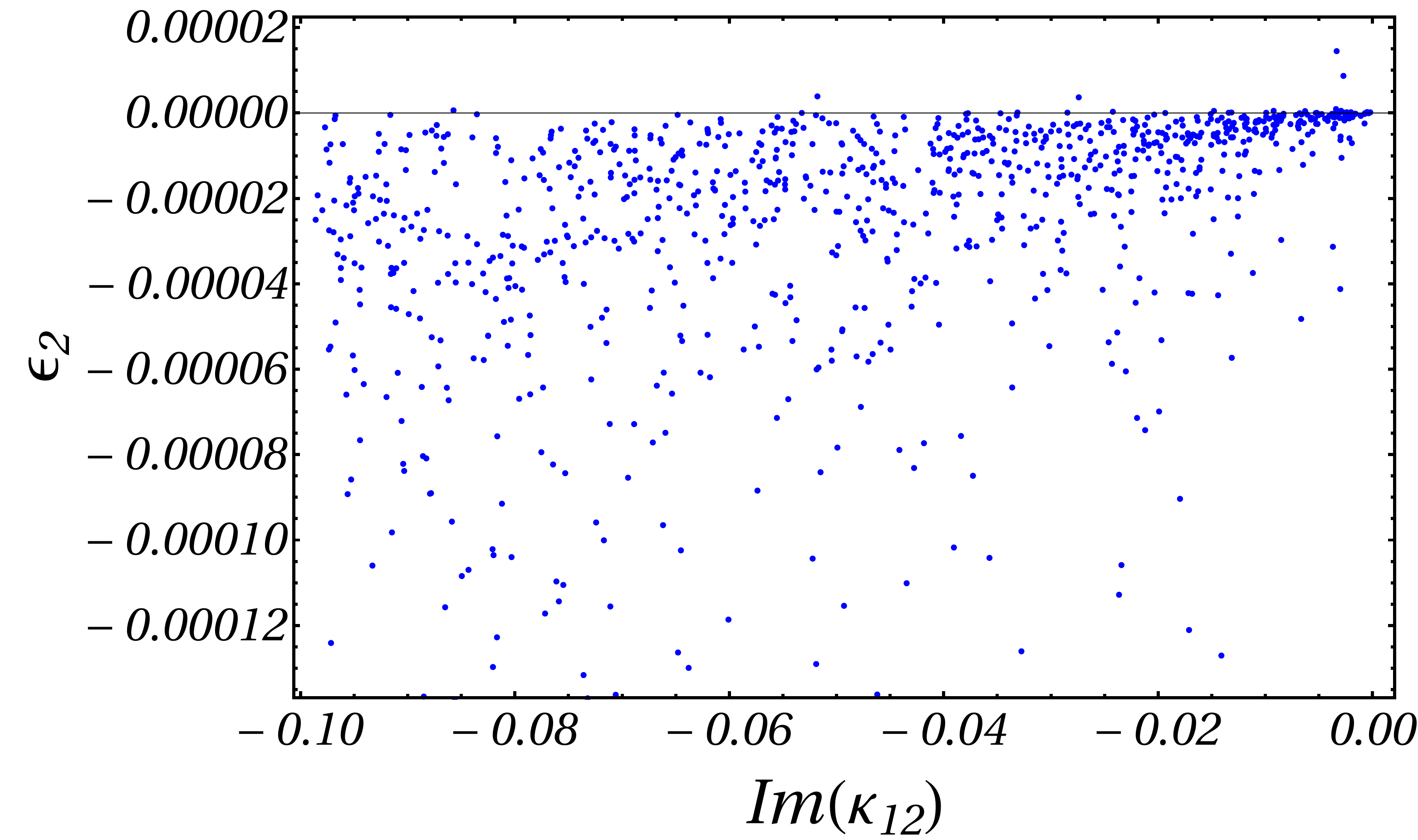}
		\includegraphics[width=1.60in]{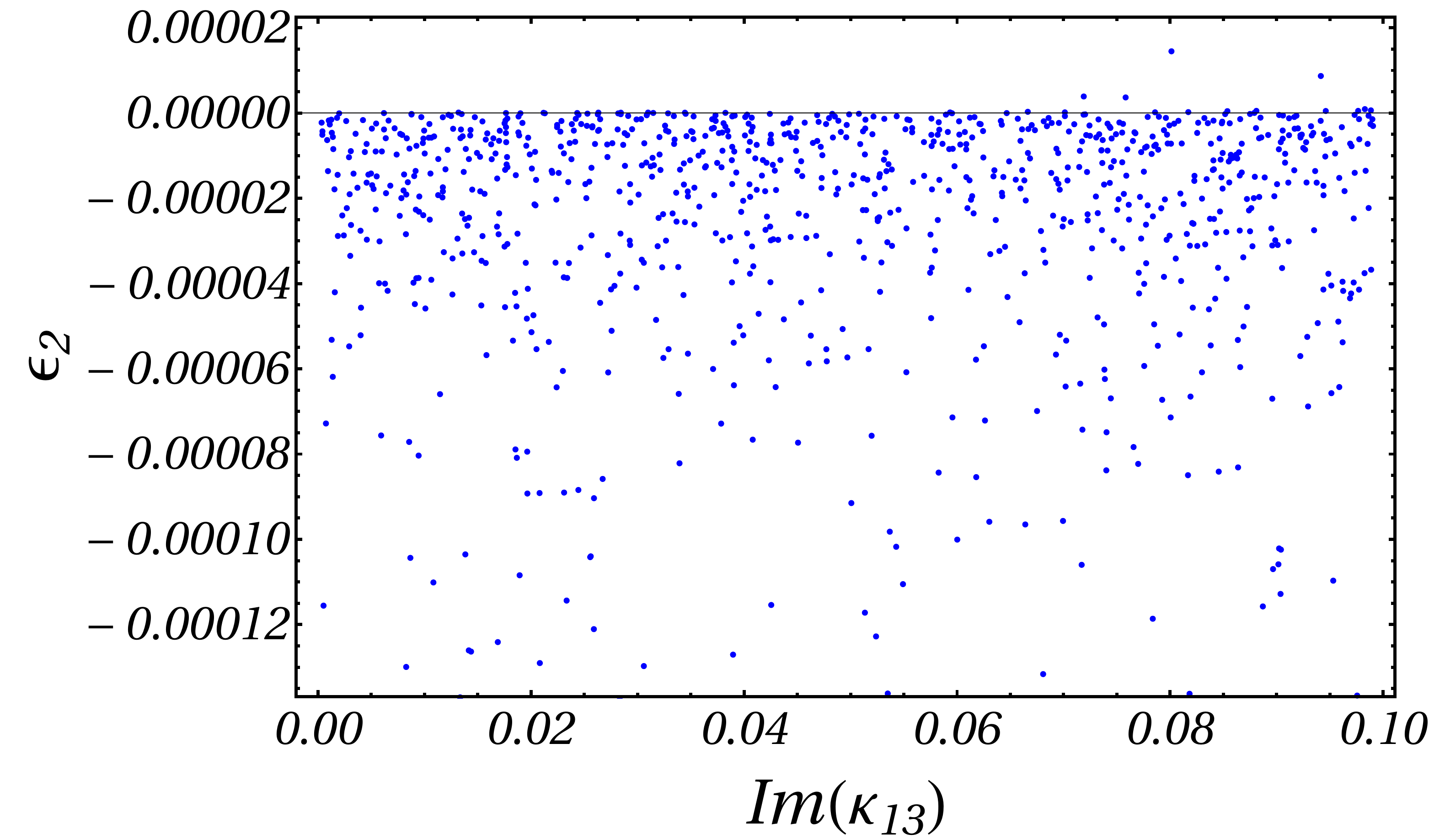}
		\includegraphics[width=1.60in]{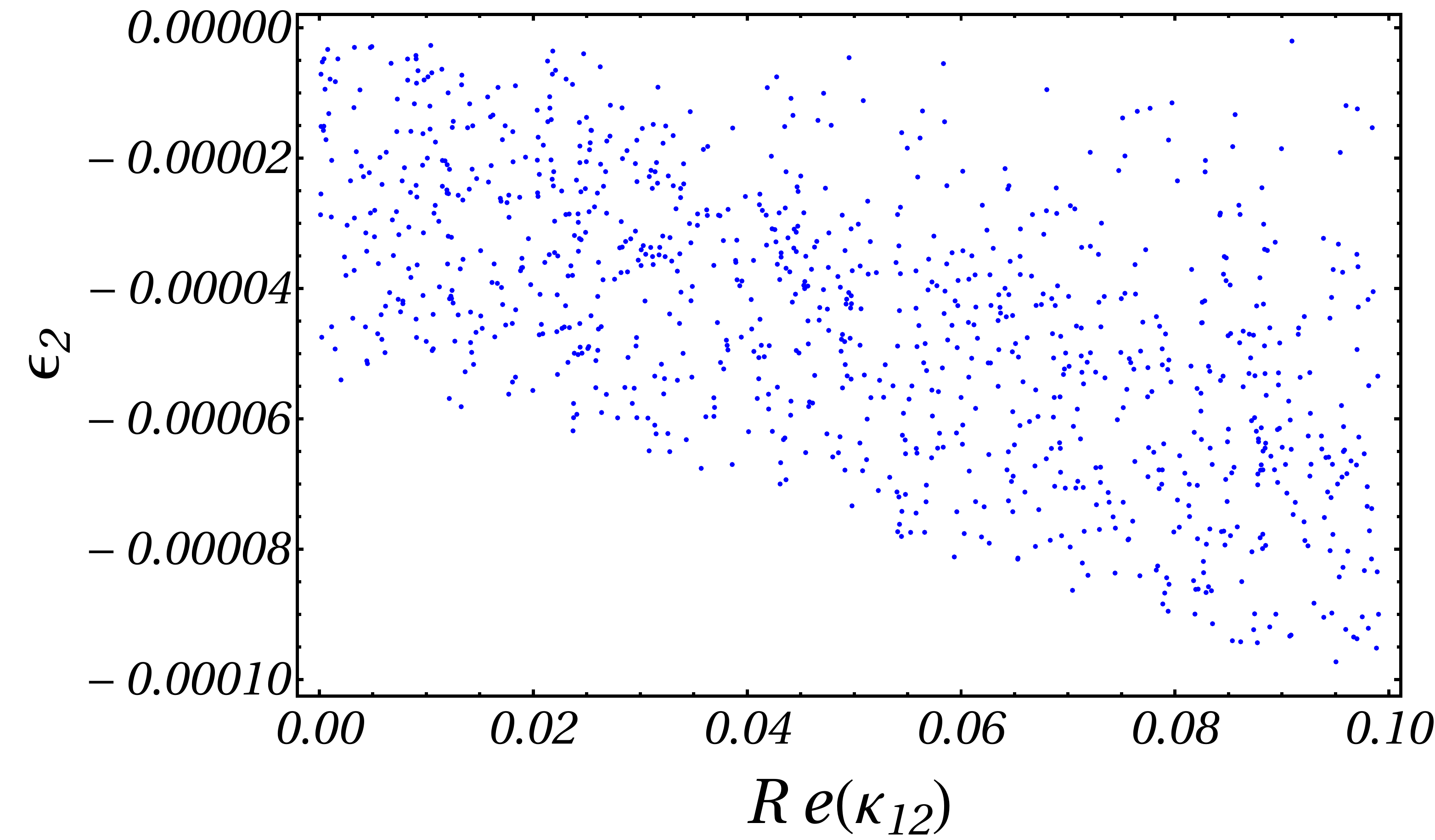}
		\includegraphics[width=1.60in]{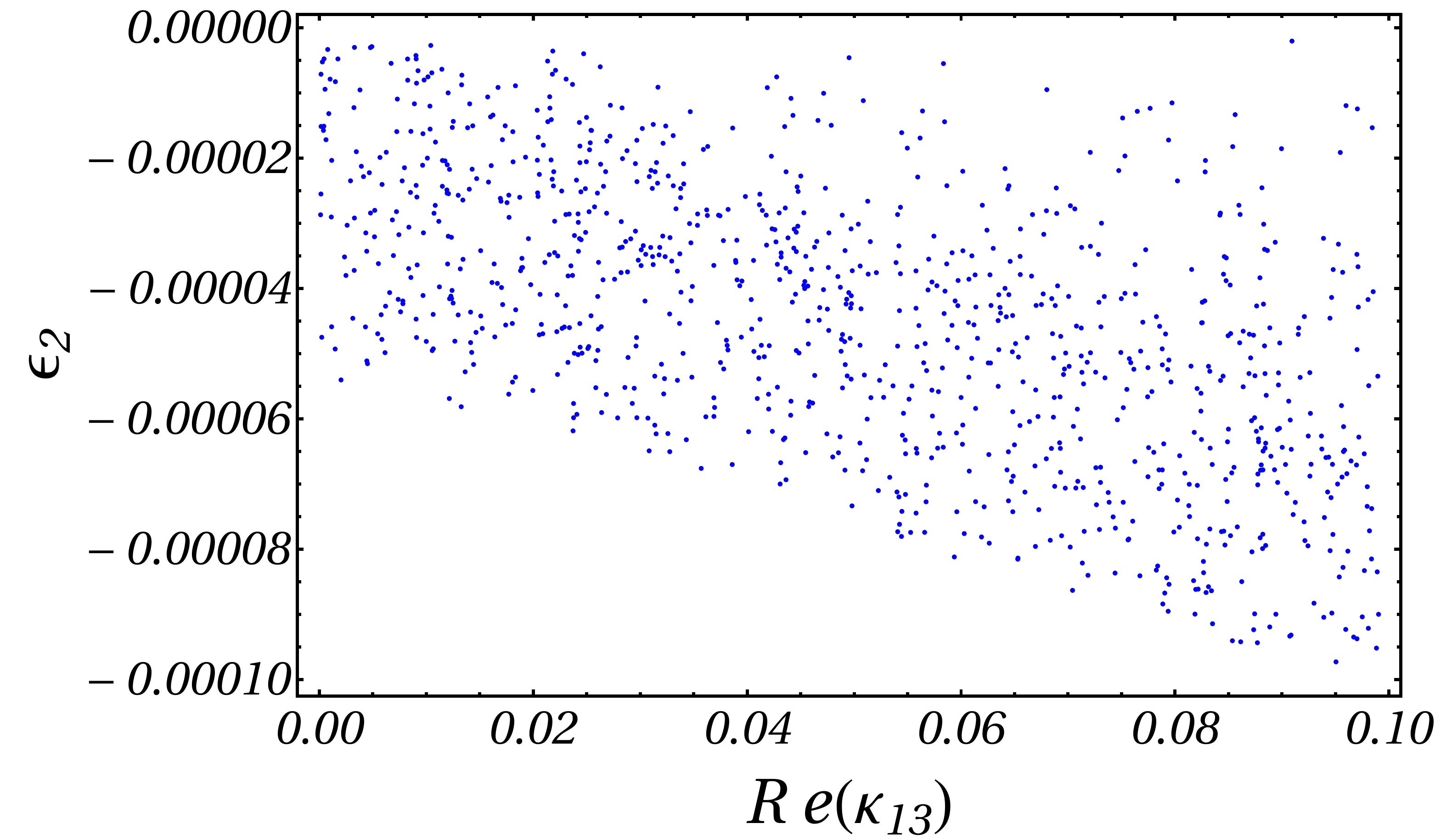}
		\caption{CP-asymmetry against the real and imaginary parts of the relevant coupling combination 
		for Case 4a, with complex $\kappa$ and constant $m_{N_j}$,   as given in Table \ref{tab:summary}.  }
			\label{fig:ECk}
	\end{center}
\end{figure}

  We now consider the case where $\kappa_{12}$ has both real and imaginary parts. The parameters $K_{12}$ and $K_{13}$ are complex but $K_{11}$ is real. The real part of $\kappa_{12}$ contributes to the  asymmetry, together with the combination of the imaginary part of $K_{12}$, the imaginary part of $\kappa_{12}$ and the real part of $K_{12}.$ There are two phases in this case: one in the  coupling constant $Y_N$ and  another in $\kappa_{12}$. We find that $\epsilon_{1}$ and $\epsilon_{2}$ take negative numerical values for complex $\kappa_{12}$ and $\kappa_{13}$.  Important in this case is that numerical values for  $\epsilon_{2}$ and $\epsilon_{1}$ are of the same order. Scan plots for $\epsilon_1$ and $\epsilon_2$ as functions of $\Re \kappa_{12}$ and $\Im \kappa_{12}$,  and $\Re \kappa_{13}$ and $\Im \kappa_{13}$, are shown in Fig. \ref{fig:ECk}.    Note the similarities between the dependence of $\epsilon_1$ and $\epsilon_2$ on $\Re \kappa_{12}$ and $\Re \kappa_{13}$; while the plots for $\epsilon_2$ versus $\Im \kappa_{12}$ and $\Im \kappa_{13}$ favor values close to 0, the variation of $\epsilon_1$ on the same parameters is slightly different.

\noindent
{\bf Case 4b: Complex $\kappa$ and variable mass}\\[5mm]

\begin{figure}[H]
	\begin{center}
		\includegraphics[height=1.2in,width=1.60in]{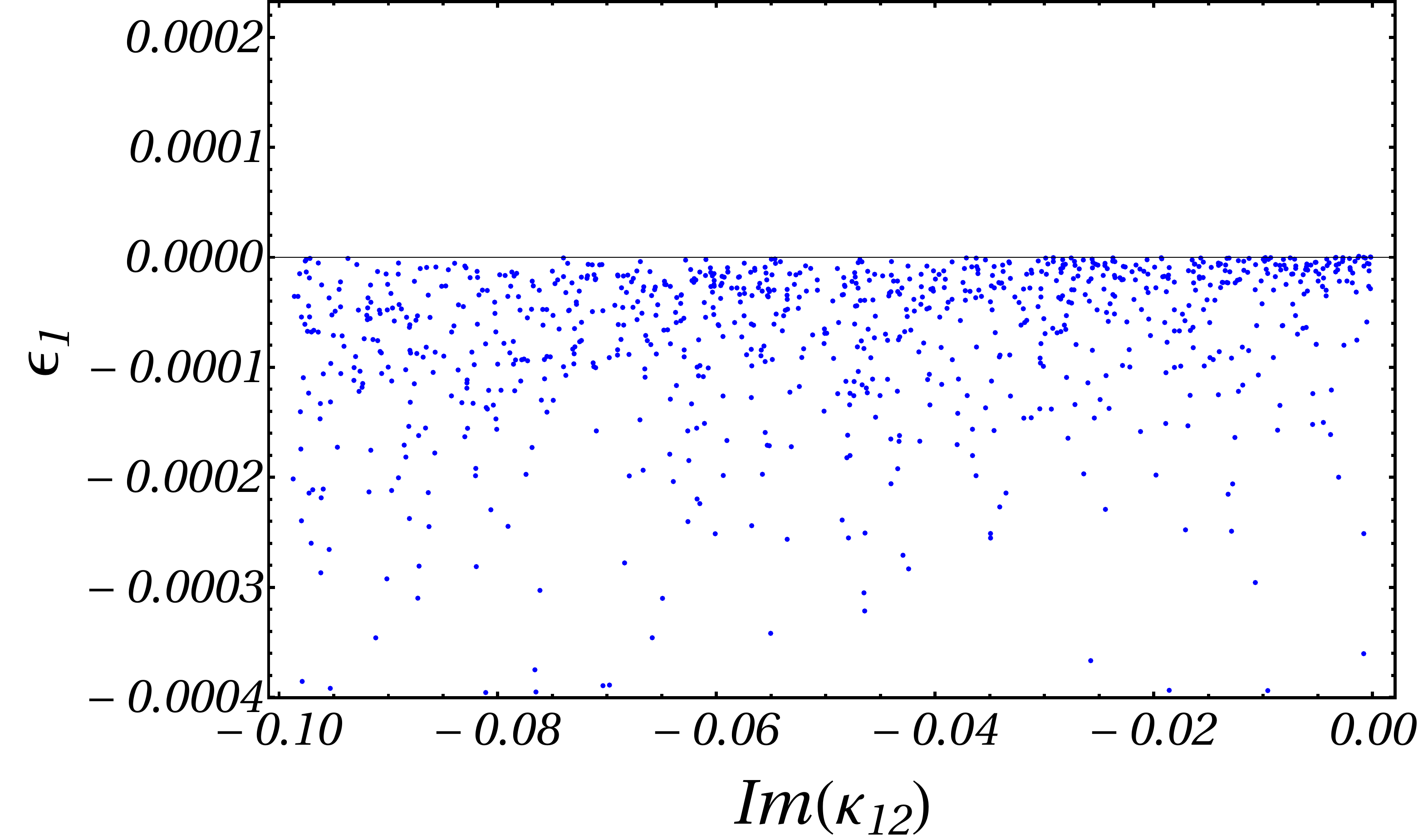}
		\includegraphics[height=1.2in,width=1.60in]{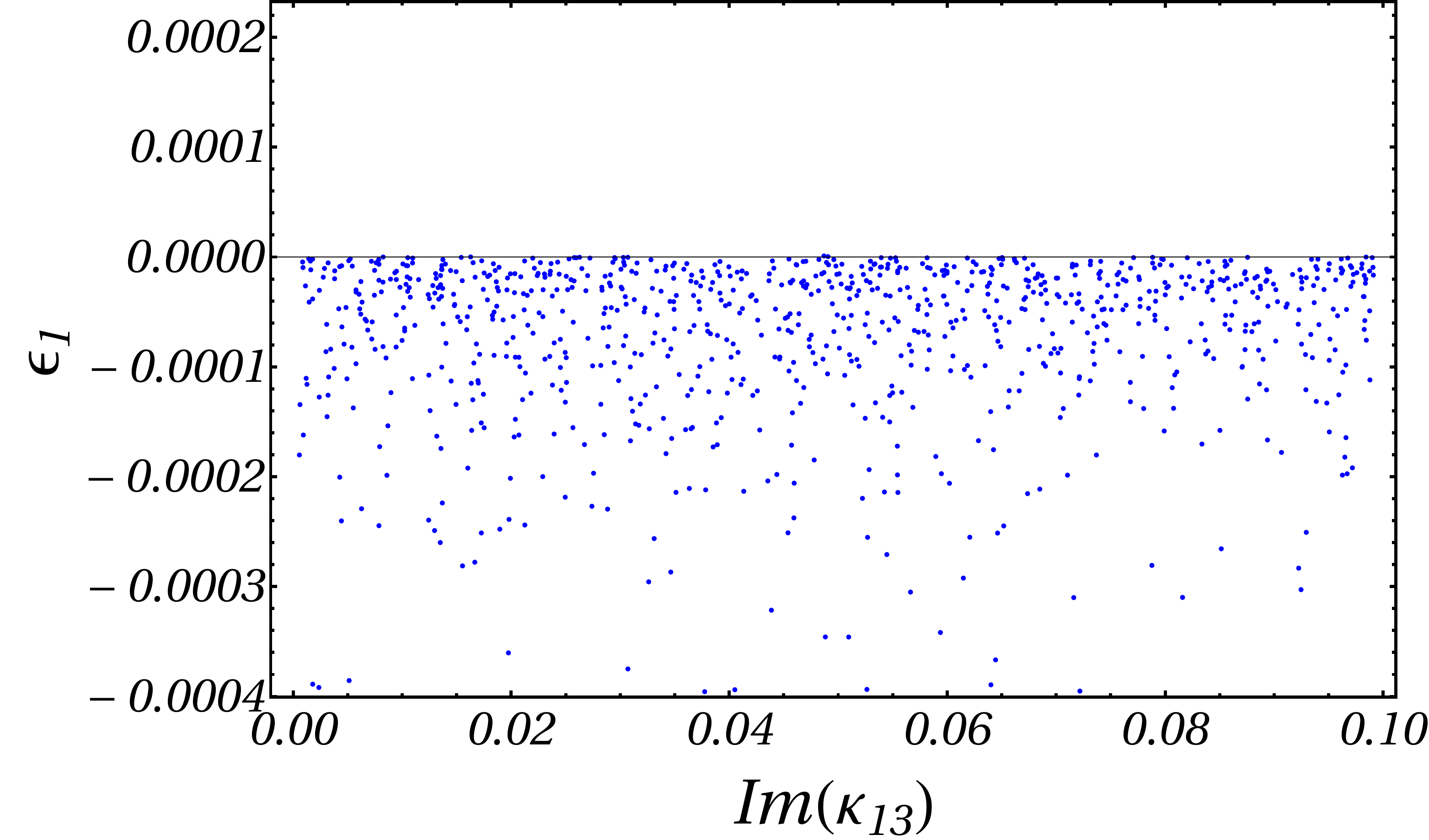}
		\includegraphics[height=1.2in,width=1.60in]{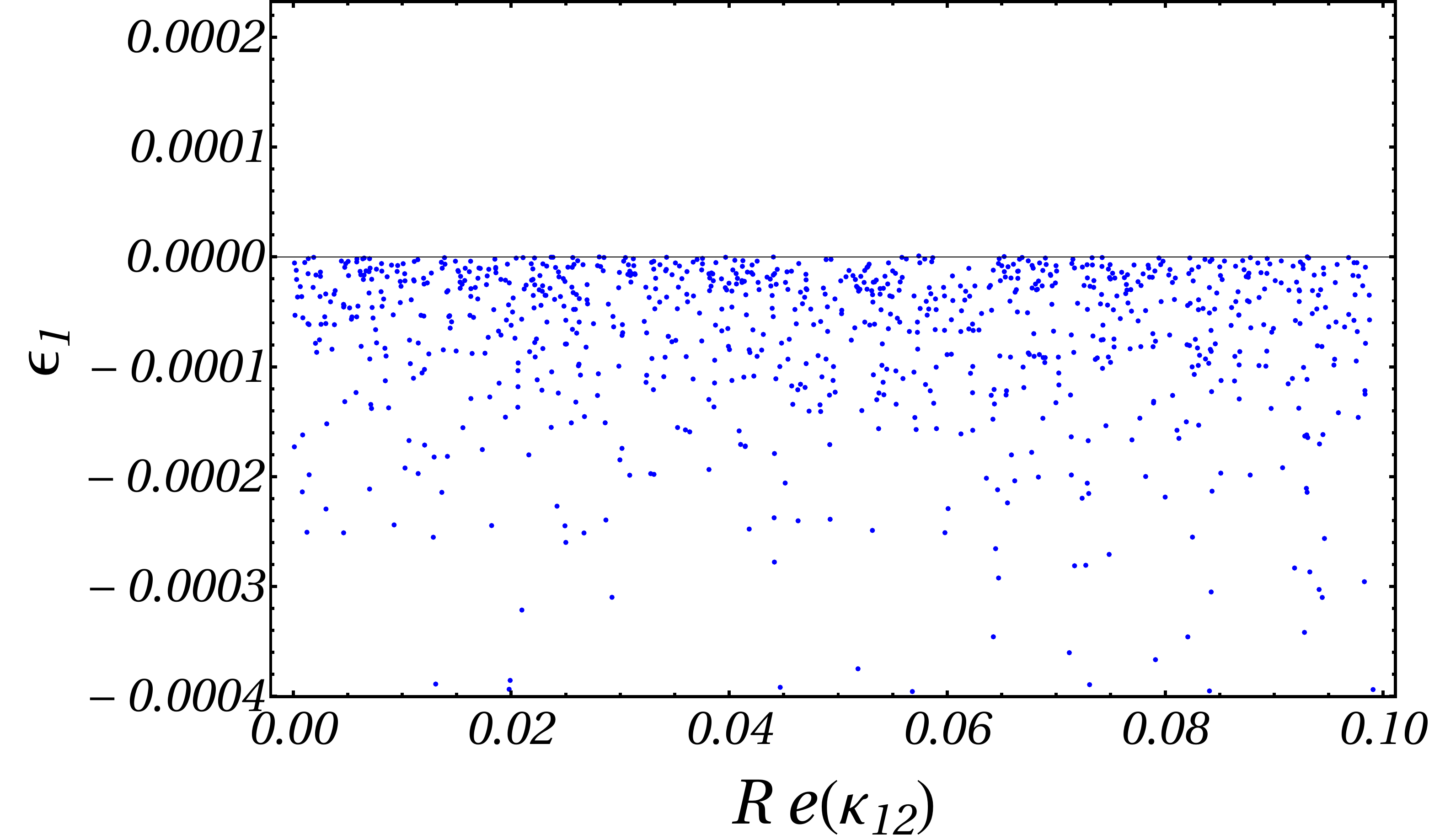}
		\includegraphics[height=1.2in,width=1.60in]{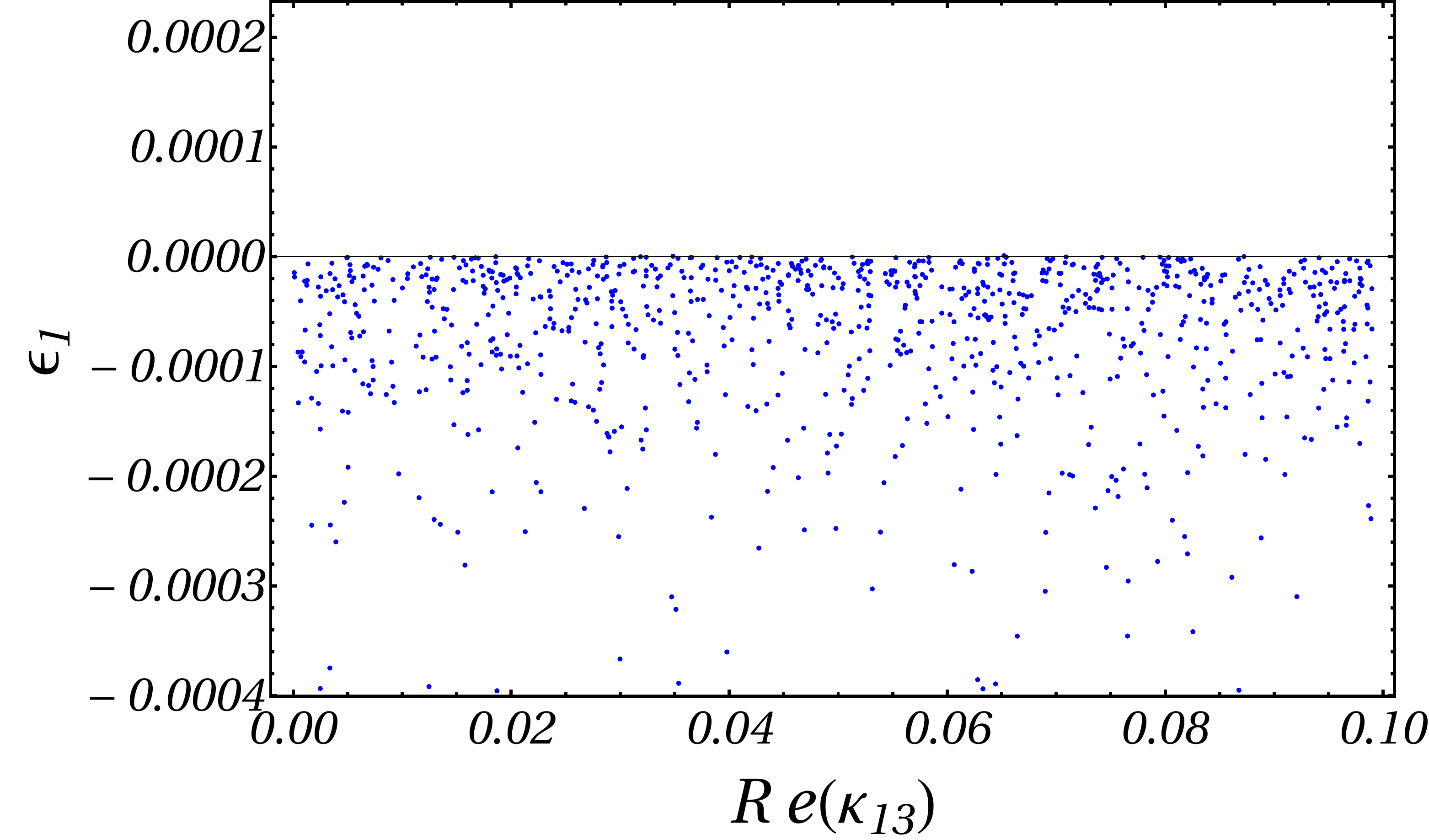}
		\includegraphics[height=1.2in,width=1.60in]{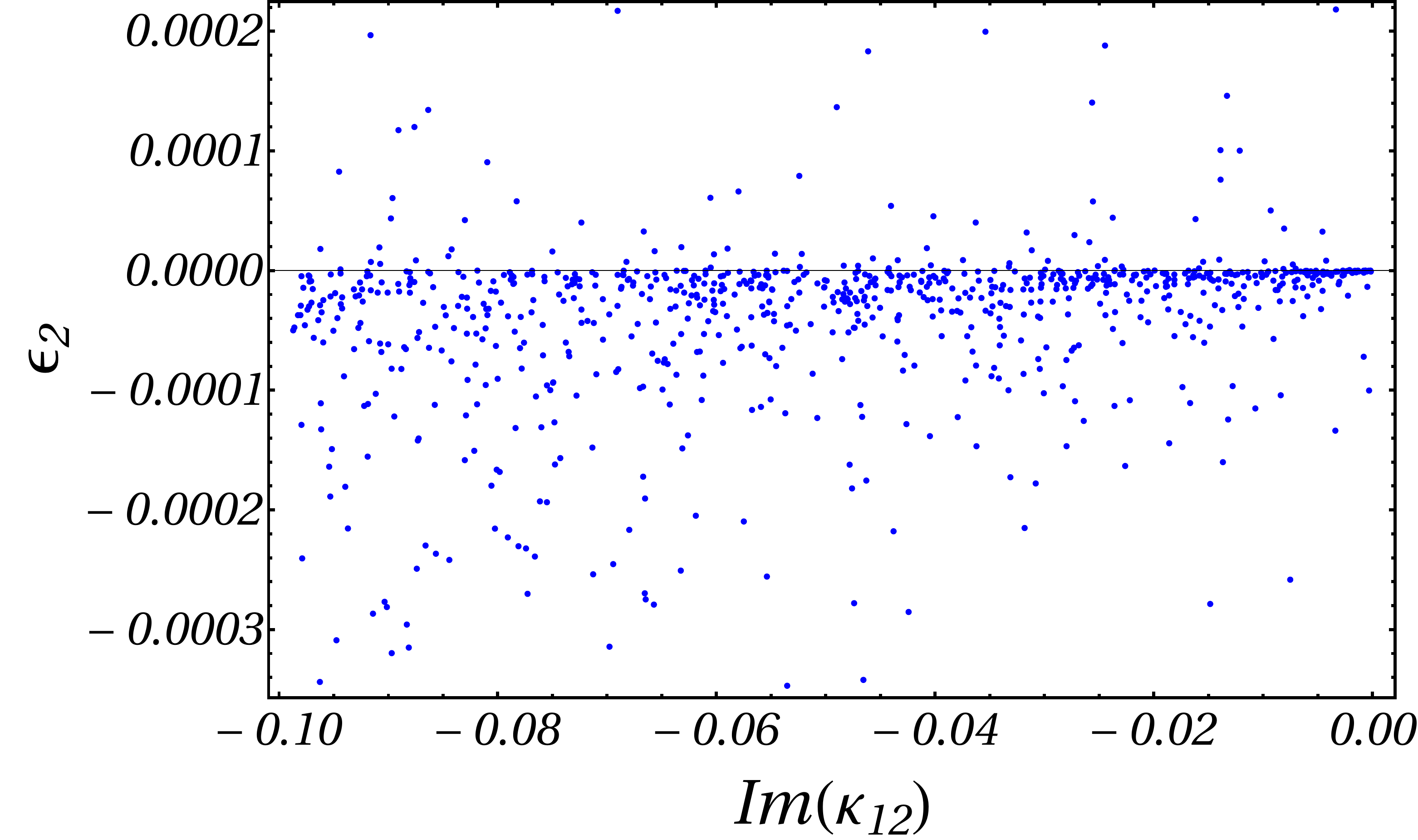}
		\includegraphics[height=1.2in,width=1.60in]{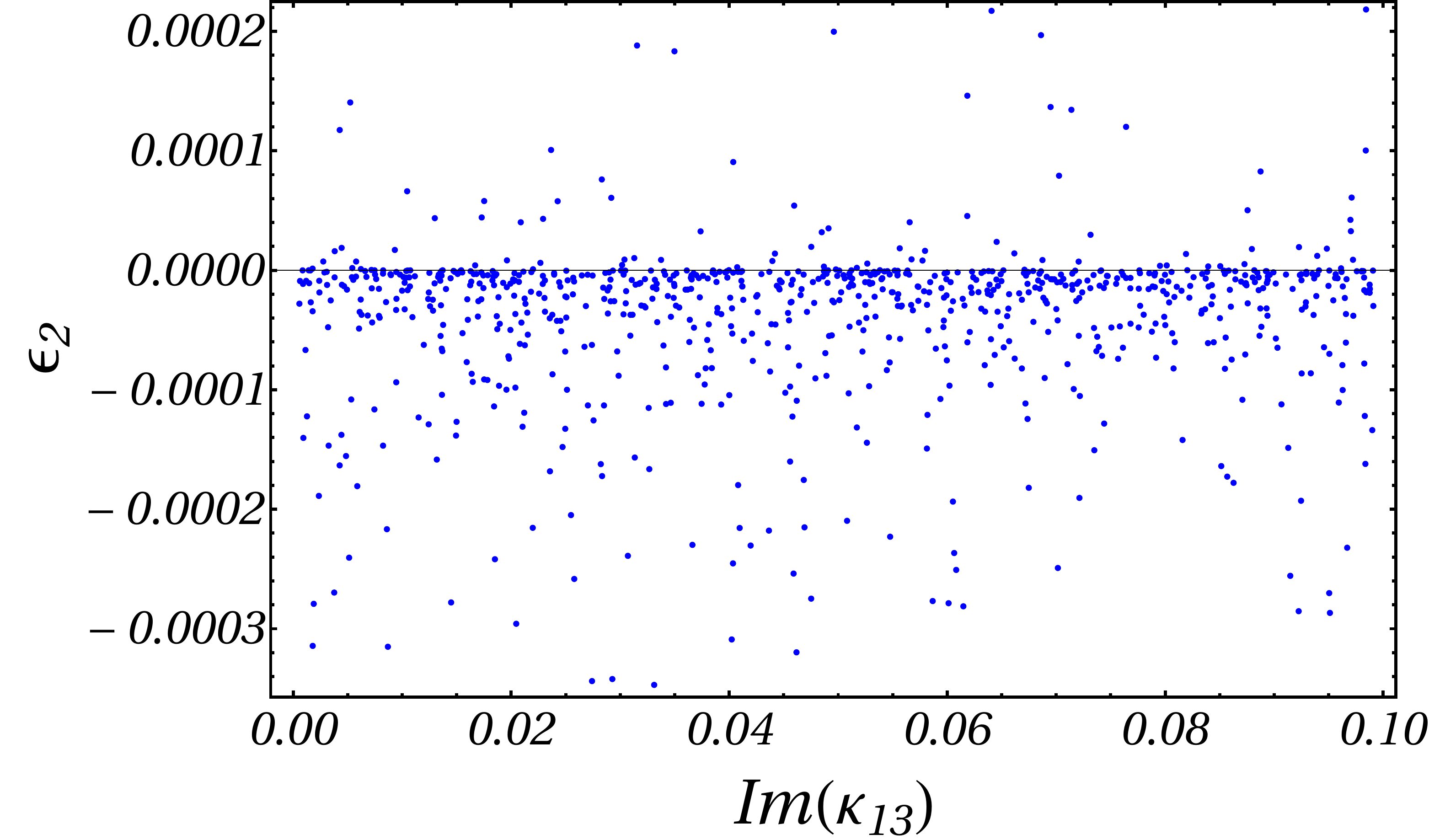}
		\includegraphics[height=1.2in,width=1.60in]{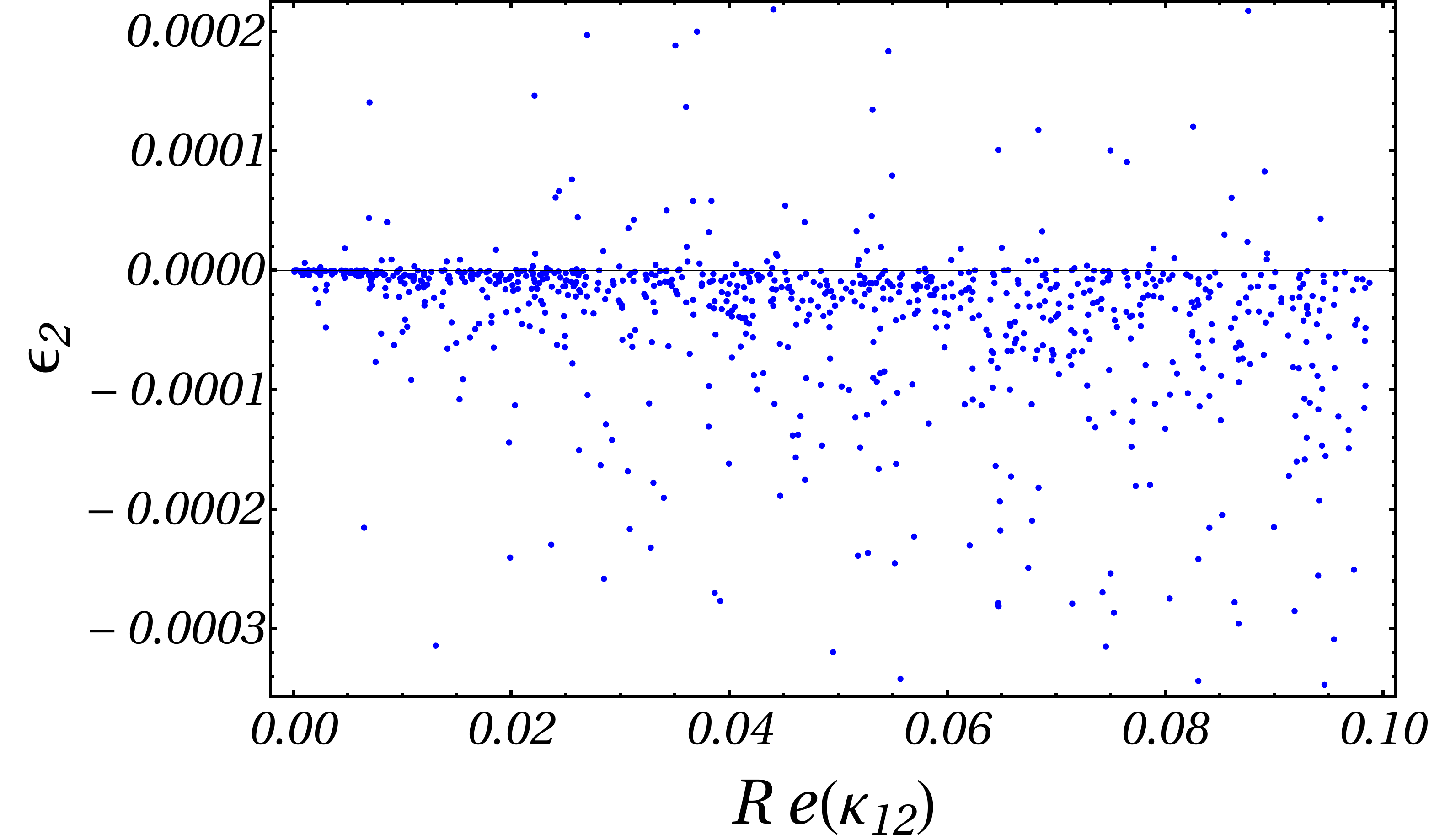}
		\includegraphics[height=1.2in,width=1.60in]{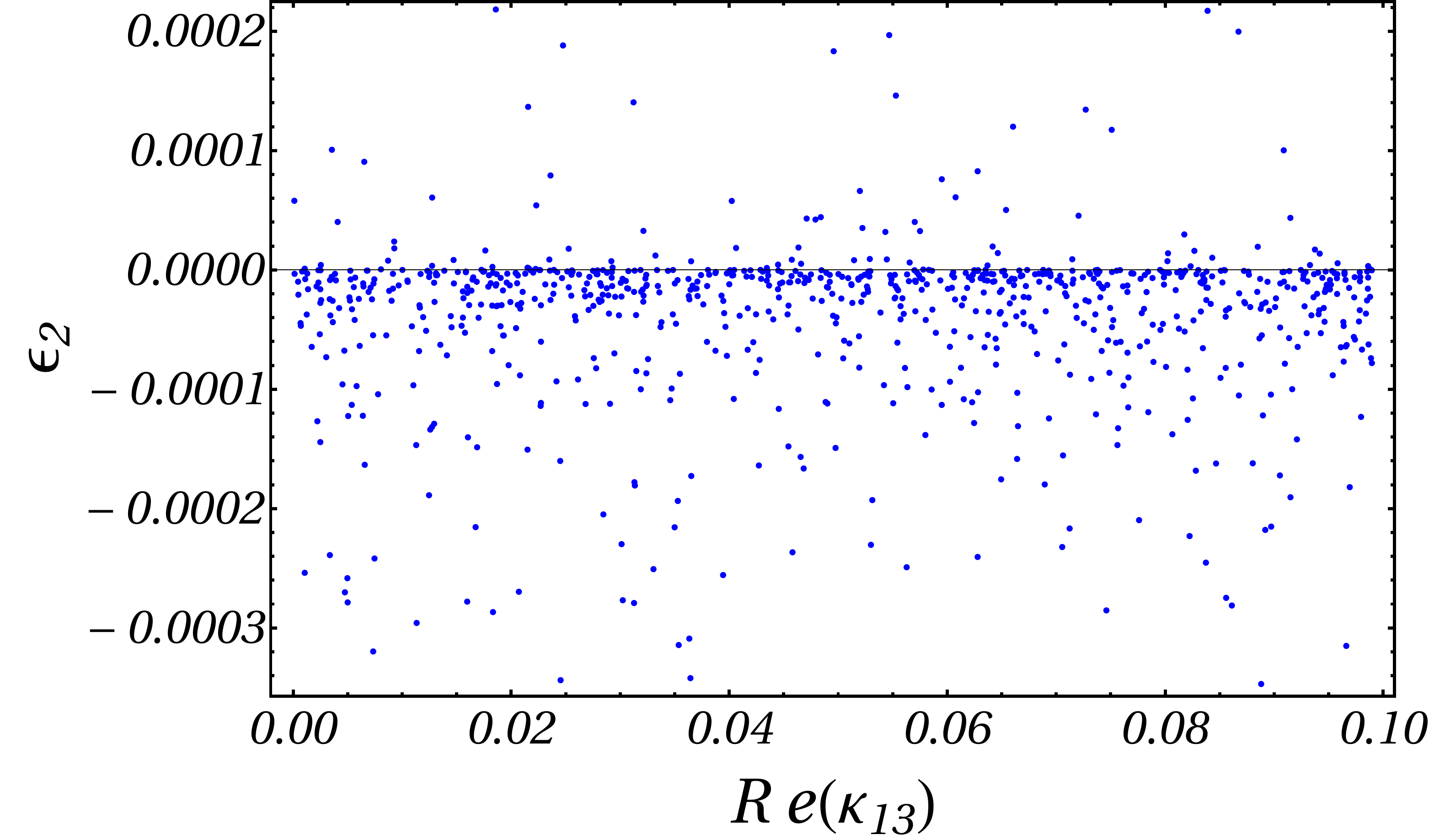}
		\caption{CP-asymmetry $\epsilon_1$ (top panels) and $\epsilon_2$ (bottom panels) plotted against the real and imaginary parts of the relevant coupling combination for Case 4b, with complex $\kappa$ and varying $m_{N_j}$. }	
	\label{fig:CkM1}
	\end{center}
\end{figure}

\begin{figure}[H]
	\begin{center}
		\includegraphics[width=2.2in]{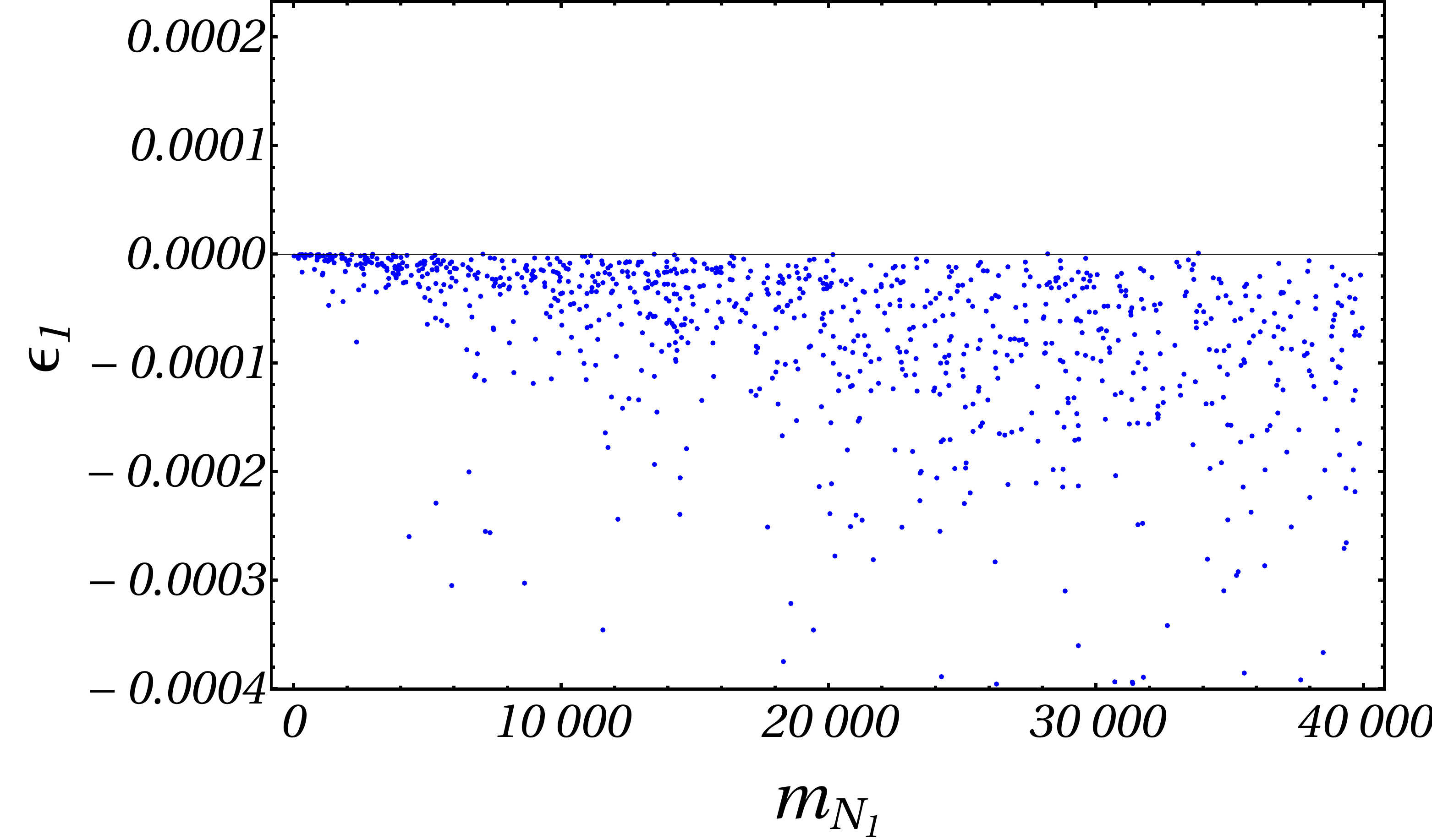}
		\includegraphics[width=2.2in]{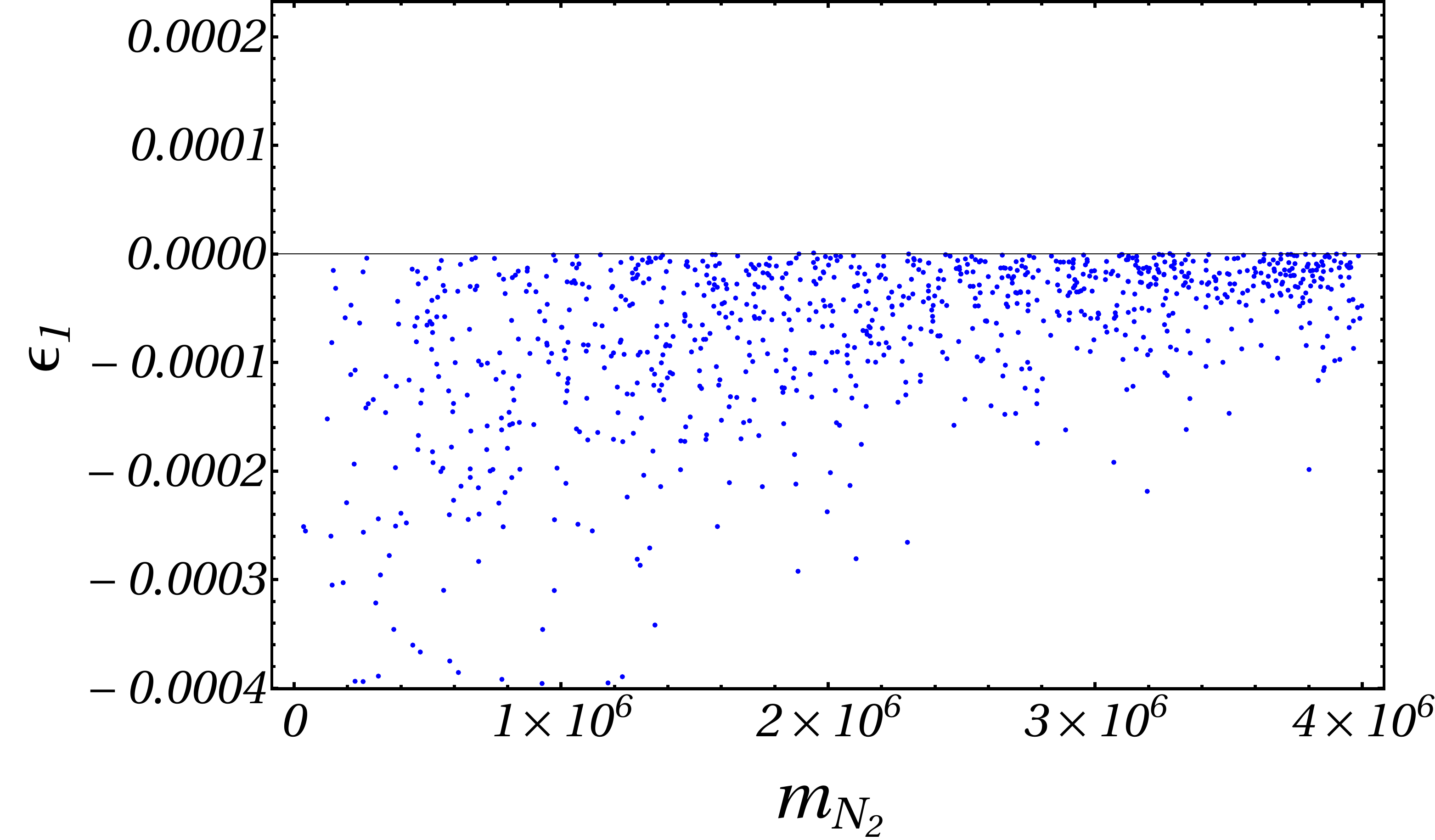}
		\includegraphics[width=2.2in]{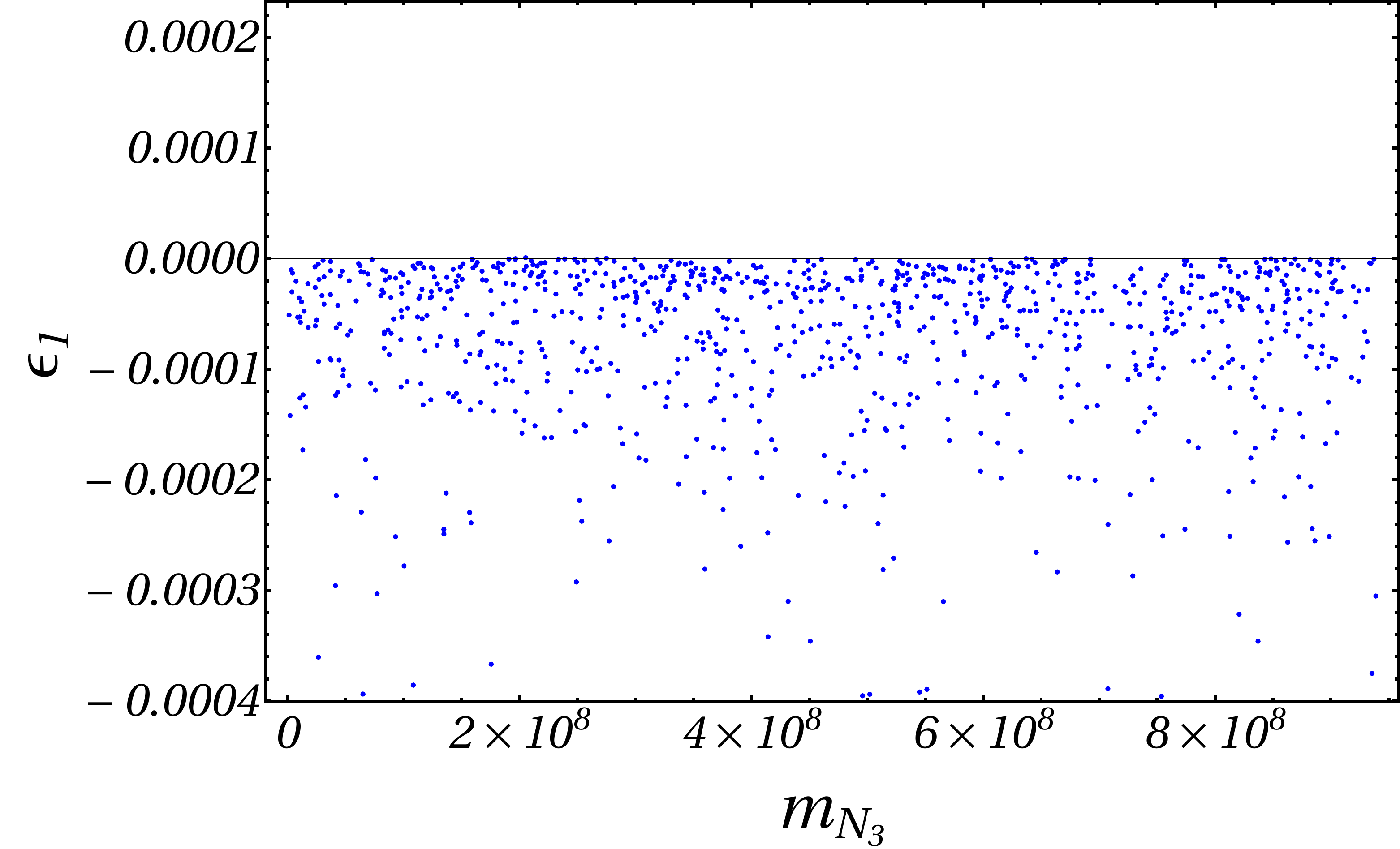}
		\includegraphics[width=2.2in]{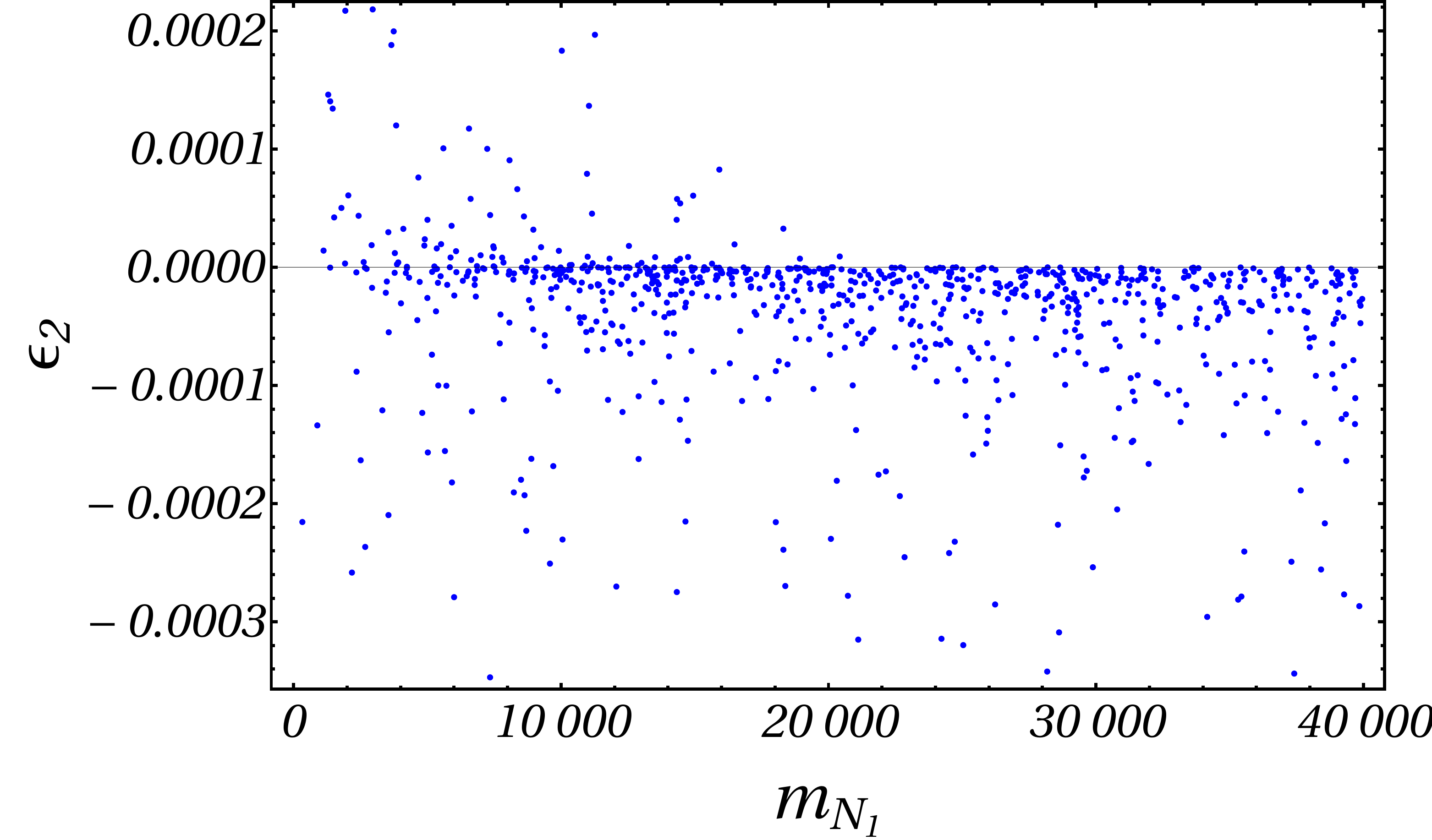}
		\includegraphics[width=2.2in]{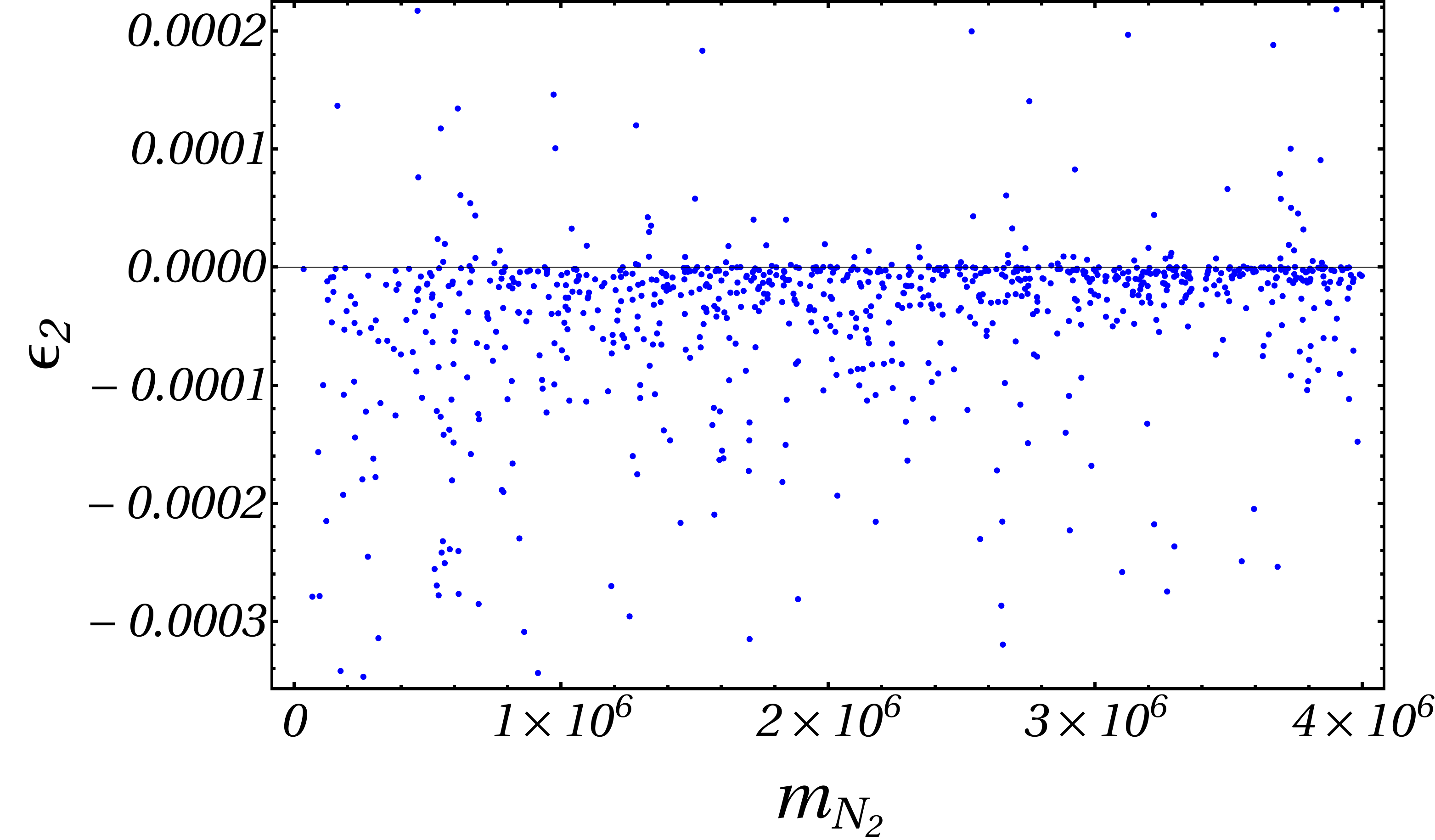}
		\includegraphics[width=2.2in]{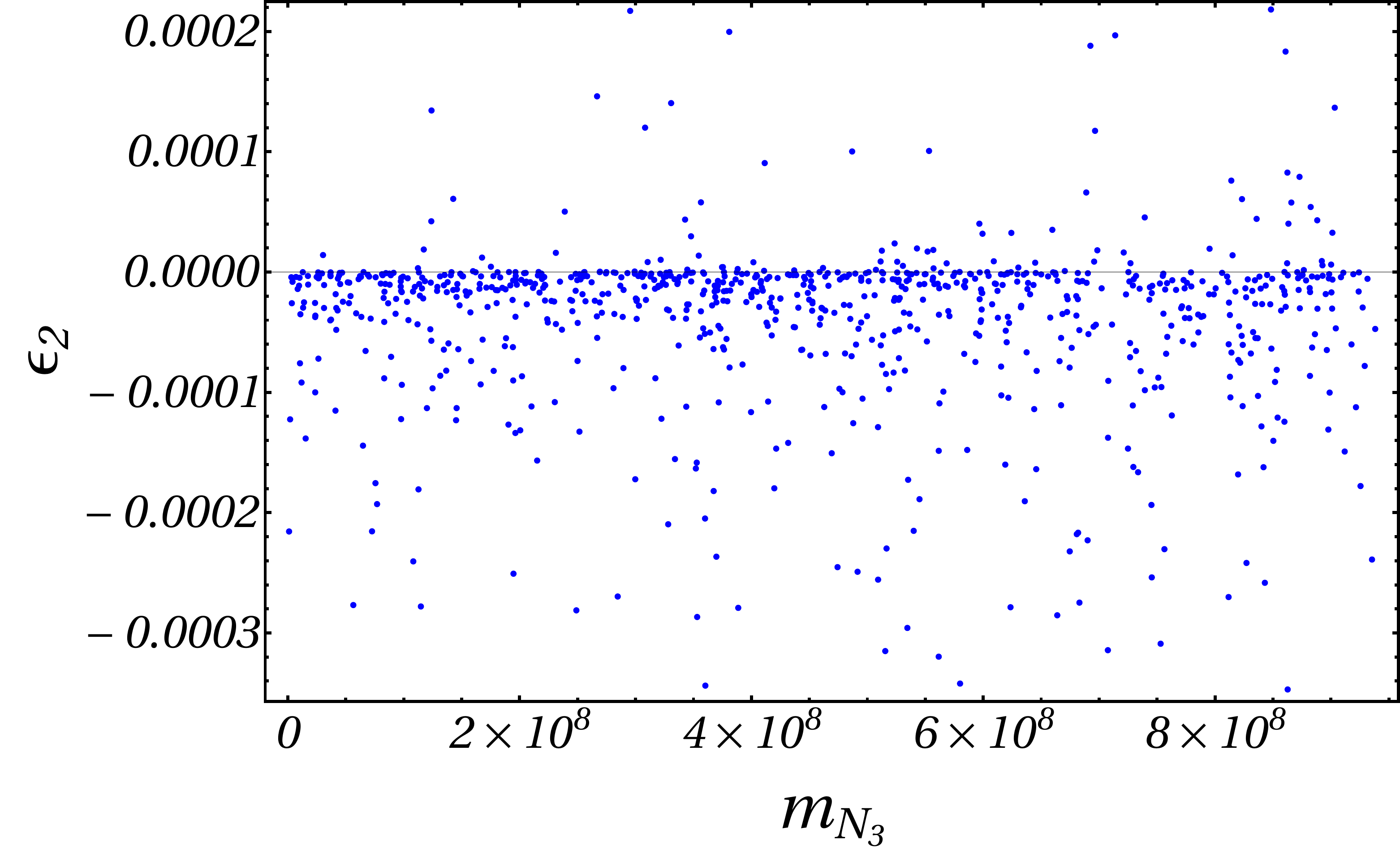}
		\caption{CP-asymmetry $\epsilon_1$ (top panels) and $\epsilon_2$ (bottom panels)  plotted against the masses, $m_{N_1},~m_{N_2}$ and $m_{N_3}$, 
		for Case 4b, with complex $\kappa$. }	
	\label{fig:CkM2}
	\end{center}
\end{figure}

\begin{figure}[H]
	\begin{center}
		\includegraphics[width=2.8in]{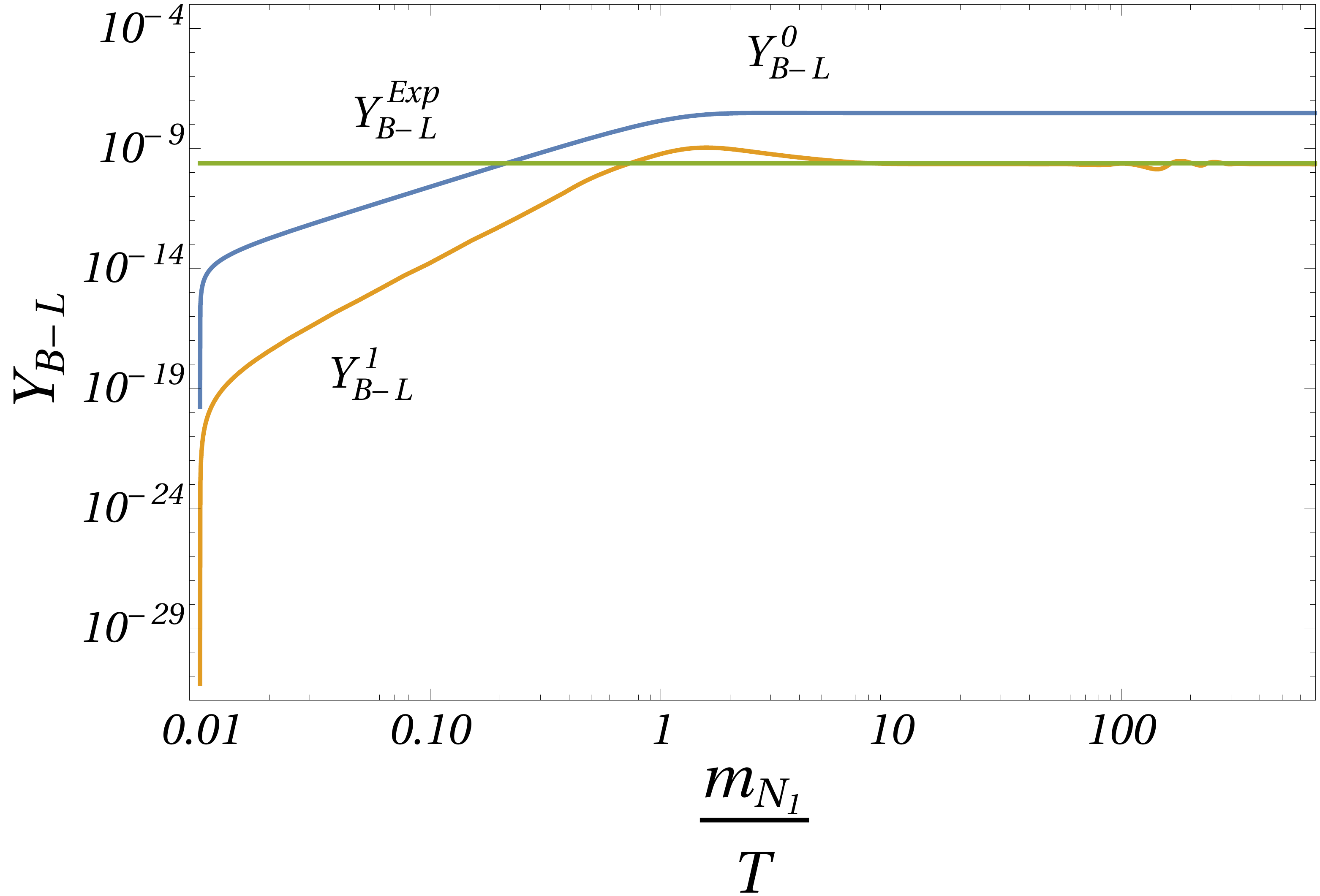}
		\includegraphics[width=2.8in]{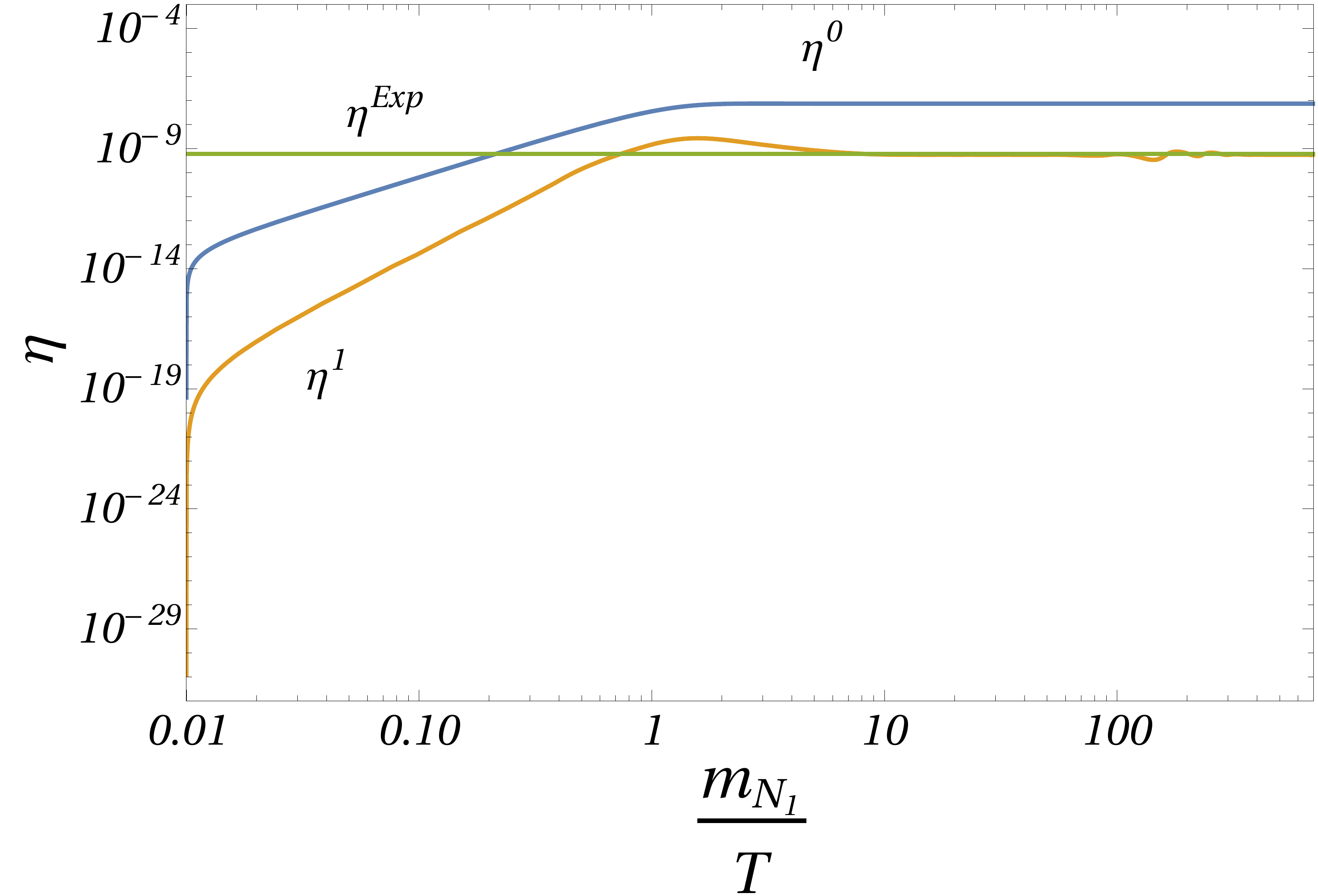}
		\caption{The lepton asymmetry, $Y_{B-L}$ (left) and the baryon asymmetry, ${\eta} $ (right) versus $z=\frac{m_{N_1}}{T}$ for the case with complex $\kappa$ and $m_{N_1}=10$ TeV, $m_{N_2}=10^3$ TeV,  $\kappa_{13} =  10^{-1} $,  $\kappa_{12} = 2(1-\iota)10^{-3} $.  The superscripts $0$ and $1$ correspond to the cases without and with the washout terms induced by the scattering processes, respectively.}
	\label{fig:YBLetaCk}
		\end{center}
\end{figure}

Finally in this section we allow  both $\kappa_{12}$ and $\kappa_{13}$ to be complex, while varying the right-handed neutrino masses.
In Fig. \ref{fig:CkM1} we plot the dependence of $\epsilon_1$ (top panels) and $\epsilon_2$ (bottom panels) with the parameter $\kappa_{12}$  and $\kappa_{13}$ separately for their real and imaginary parts, while in Fig.  \ref{fig:CkM2} we plot the dependence of $\epsilon_1$  and $\epsilon_2$ on the relevant right-handed neutrino masses $m_{N_1},~m_{N_2}$ and $m_{N_3}$. We note that by varying neutrino masses, regions of $\epsilon_1$ negative and close to 0 are evenly populated, while $\epsilon_2$ is allowed to be positive.
To study the asymmetry, in Fig. \ref{fig:YBLetaCk} we plot the dependence of the lepton asymmetry, $Y_{B-L}$ (left) and the baryon asymmetry, ${\eta},  $ with $z=\frac{m_{N_1}}{T}$ for the case with complex $\kappa$ for the specific set of parameter values, as indicated in the caption and in Table \ref{tab:summary}.  One can see that, unlike scenarios where we consider $\kappa_{12}$ purely real or purely imaginary, this case is compatible with the requirement of yielding sufficient baryonic asymmetry in the universe.

\noindent
{\bf Comparison of the  standard leptogenesis case with the new model}\\[5mm]

\begin{figure}[H]
	\begin{center}
		\includegraphics[width=2.80in]{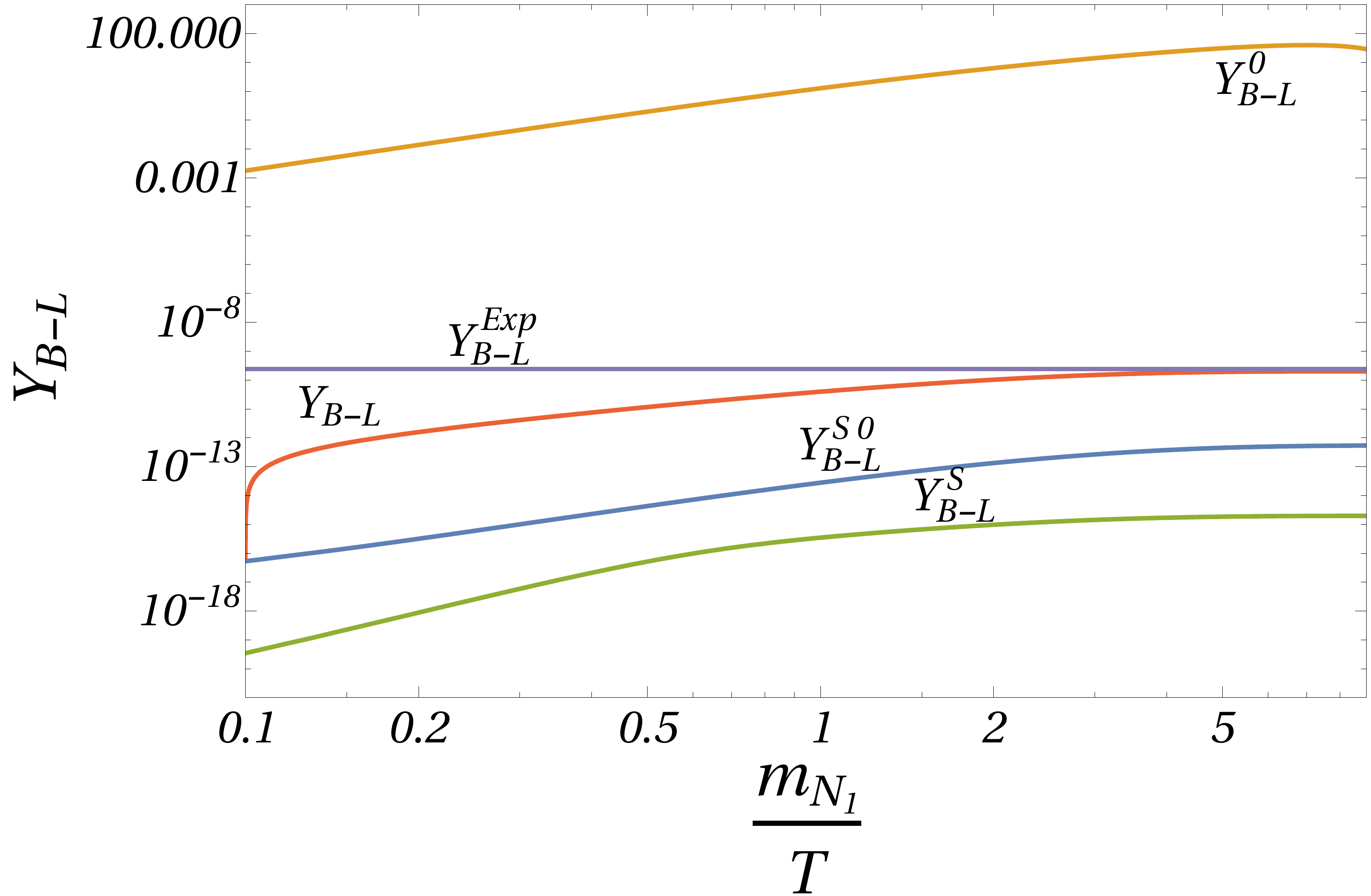}
		\includegraphics[width=2.80in]{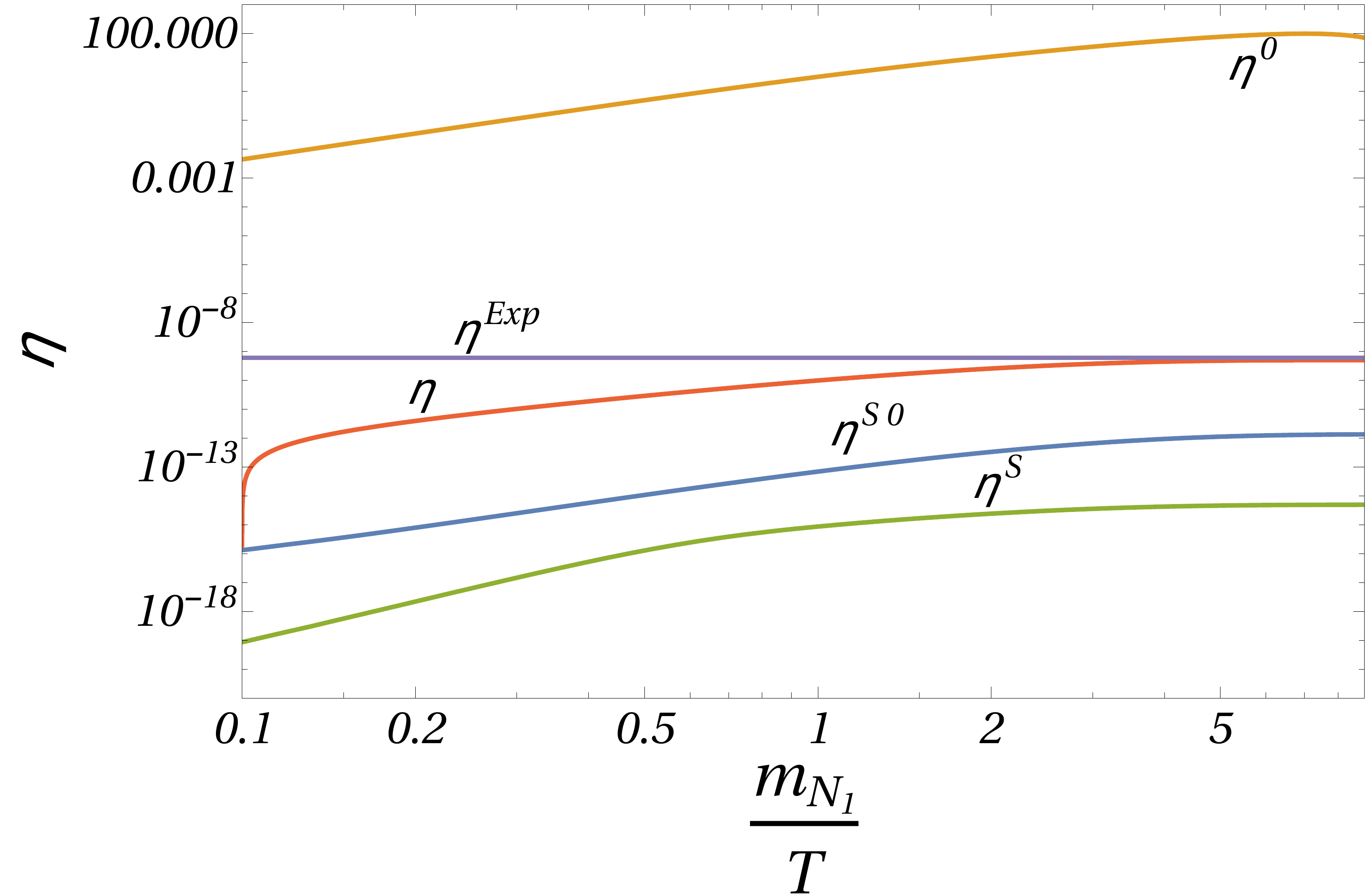}
		\caption{ Plot for $Y_{B-L}$ and $\eta$ showing the contribution of the standard case and the total contribution in the new scenario for the 
	benchmark point given in	 Table \ref{tab:YBLetaSum}.  Here the superscript $S$ indicates the standard contribution, while superscripts $0$ and $1$ represent contributions without and with scattering, respectively. }	
		\label{fig:YBLetaSum}
	\end{center}
\end{figure}

We demonstrated that our model with suitably chosen parameter values could explain the observed baryon asymmetry of the universe. The presence of the new charged scalar that interacts directly with the right-handed neutrinos boosts sufficiently  the CP asymmetry inducing the required baryon asymmetry We further illustrate this by comparing the standard leptogenesis case  (in the absence of the new degrees of freedom) with the scenario considered here, including all possibilities described above. To highlight the differences, 
we fix  the mass of the lightest right-handed neutrino, $m_{N_1}=10$ TeV. Choosing  $m_{N_1}$ so light means leptogenesis is not possible in seesaw Type 1 models. Whereas in our model  we have an additional channel that violates lepton number in the decay process. Therefore, both channels are important for our analysis. As a benchmark point, we consider the rest of the parameter values as given in Table \ref{tab:YBLetaSum}.
\begin{table}[H]
	\begin{center}
	\small
	\begin{tabular}
{p{0.3in} |p{0.6in} |p{0.6in} |p{0.3in} |p{0.3in}  |p{0.3in} |p{0.6in} |p{1in} |p{1.1in}  }
		\hline \hline
$m_{N_1}$ (TeV) & $m_{N_2}$ (TeV) & $m_{N_3}$ (TeV)& $m_\chi $ (GeV)& $m_\psi$ (GeV) & $m_S$ (GeV) & $\kappa_{11}$ & $\kappa_{12}$ &$\kappa_{13}$ 
\\[1mm] \hline 
10&$1.02\times10^{3}$ &$8.76\times10^{5}$ & 200 & 60&175 & 1.82$\times 10^{-2}$& $(2+.055 \iota)\times 10^{-3}$ & $(4.51-5.5 \iota )\times 10^{-4}$ \\
\hline \hline
		\end{tabular}
	\caption{Benchmark point  for the comparison of standard case with the new scenario.}
	\label{tab:YBLetaSum}
 \end{center}
\end{table}
Numerical solutions of Boltzmann equations for this case are show in Fig. \ref{fig:YBLetaSum}, with the left plot showing the evolution of the lepton asymmetry and the right plot showing with evolution of the corresponding baryon asymmetry.  The left side plot shows the lepton asymmetry $Y_{B-L}$, with  curves labelled  $Y_{B-L}^0$ and $Y_{B-L}^{S0}$ (for new leptogenesis contribution, and standard contribution, respectively, without scattering), and $Y_{B-L}^1$ and $Y_{B-L}^{S1}$ (for new and standard contribution with scattering). The curve $Y_{B-L}^{\rm Exp}$ represents the lepton asymmetry needed to generate the correct matter-antimatter asymmetry. On the right side plot,  the top curves
$\eta^0$ and $\eta^1$ correspond to  the new contribution  in our model and two bottom curves labelled $\eta^{S0}$ and $\eta^{S1}$ corresponding to standard leptogenesis. The top ones in each of these ($\eta^0$ and $\eta^{S0}$) are both generated without involving any scattering, meaning these curves represent the total asymmetry generated by the decay channels, while the bottom ones ($\eta^1$ and $\eta^{S1}$) correspond to the case including the effect of scattering processes. The flat curve labelled $\eta^{\rm Exp}$ represents the observed asymmetry.  
It is clear  that while the standard leptogenesis alone is not able to generate required matter-antimatter asymmetry, the addition of the new degrees of freedom makes up for the deficit. 
\section{Conclusion}
\label{sec:conclusions}

In this work  we extended the SM to allow for both a dark matter candidate and a mechanism for leptogenesis which could produce sufficient  asymmetry to account for the matter antimatter discrepancy in the Universe.  The extension includes, in addition to three Majorana right-handed neutrinos, a gauge singlet charged scalar field $S^{+}$, plus a charged $(\chi^+$) and a neutral ($\psi$) singlet fermions, the latter  two  which are odd under an additional $Z_2$ symmetry. In this scenario, $\psi$ is a stable dark matter candidate, interacting with the other particles through its Yukawa coupling only. When this coupling is very small, the abundance of dark matter is due to the slow two-body decay of $\chi^+ \to S^+ \psi $, if kinematically allowed, or the four-body decay $\chi^+ \to H \ell \nu \psi $, in the regions where the two-body decay is forbidden. We implement the freeze-in mechanism and show that, for a wide range of masses and couplings the relic abundance is consistent with the experimental data.

We then analyse this parameter space to find regions favorable to decays of heavy right-handed neutrinos into leptons (or anti-leptons) plus doublet Higgs bosons, or the new charged scalar $S^+$. Leptogenesis is generated by the CP-violating interference between the tree-level process and one-loop contributions from self-energy and vertex corrections, facilitated by the non-equilibrium conditions when the temperature of the Universe is of the same order as the mass of the decaying heavy neutrino.

In canonical thermal leptogenesis with hierarchal right-handed neutrinos, an upper limit on the CP asymmetry results in a lower limit on the mass of lightest right hand neutrino, but this is in conflict with the bounds from naturalness, limiting the radiative corrections to the Higgs boson mass from Yukawa couplings. In standard leptogenesis, a way out of this inconsistency is provided by 
a resonant enhancement of the CP-asymmetry,  possible when the mass difference between two of the  right-handed neutrinos is small and comparable to their decay widths, meaning the neutrinos are almost degenerate in mass.  This enables low energy leptogenesis with Majorana right-handed neutrino masses in the  $1-10$ TeV range.

In our model,  additional possibilities open through the Yukawa interaction of the additional  charged scalar $S^+$. We obtain contributions from both $N\to S\ell$ and its CP-conjugate process, providing new decay channels, as  well as  the self-energy and vertex corrections for each of the decays.

We derive the Boltzmann equations for our model, and proceed to a detailed analysis of the parameter space. The CP parameters $\epsilon_1$ and $\epsilon_2$ depend on the coupling constants of the heavy neutrinos with the usual Higgs boson ($Y_N$) and with $S$ ($y_2$), where the first is constrained by light neutrino masses and mixings, while the second is largely unconstrained. We isolate two relevant parameters $K_{ij}=\left(Y_{N}^\dagger Y_N\right)_{ij}$ and $\kappa_{ij}=\left(y^\dagger_{2}y_2\right)_{ij}$, where the second is essential for leptogenesis: the amount of CP-asymmetry is most sensitively dependent on it, and on the right-handed neutrino masses.

We look at cases where the mass is fixed, or it varies, and where $\kappa$ is either real, purely imaginary, or complex.
The results of our investigations are as follows.
\begin{itemize}
\item For the real $\kappa$ case only the imaginary $K_{12}$ contributes to the CP asymmetry, while the real part of $K_{12}$ contributes to CP asymmetry for imaginary values of $\kappa$. The real and imaginary parts of $K_{12}$ have same order. 	
	\item Both the real and imaginary parts of $K_{12}$ contribute to the CP asymmetry when $\kappa_{12}$ is complex. 
	\item  While the required matter antimatter asymmetry cannot be  generated for real or purely imaginary $\kappa$,  if $\kappa$ is complex we can generate the required leptogenesis for $m_{N_1}=10$ TeV, thus for low scale. 
		\item Smaller values of $\kappa_{12}$ give the required matter antimatter asymmetry at higher masses of right-handed neutrinos, while larger values of $\kappa_{12}$ give the required matter antimatter asymmetry for lower masses of right-handed neutrino.
\end{itemize}

In conclusion, we have shown that  a simple,  minimally extended SM scenario, can account for both dark matter though the freeze-in mechanism, and  provide sufficient CP-asymmetry through leptogenesis. This mechanism is capable of generating the required matter antimatter symmetry for relatively light right-handed neutrino masses (10 TeV) without resorting to resonant leptogenesis, that is, without requiring two of the right-handed  neutrinos to be degenerate in mass.

\section{Acknowledgments}
The work of MF  has been partly supported by NSERC through the grant number SAP105354.

\bibliography{Leptogenesis.bib}
\bibliographystyle{jhep}
\end{document}